\documentclass[11pt,titlepage,a4paper]{myarticle}

\usepackage[T1]{fontenc} 
\usepackage{comment} 

\usepackage{amsmath, mathrsfs, amsfonts, amssymb, amsthm, mathtools, graphicx, color, ucs, xparse, tikz, lmodern, physics, varioref, tensor, bm}
\usepackage[nosort]{cite}
\usepackage{youngtab}
\usepackage{float,multirow}
\newcommand{\ybar}[1]{\overline{\vphantom{1}#1}}
\usepackage{booktabs}
\usepackage{multirow}
\usepackage{placeins}

\newcommand{\centerresize}[2]{%
  \makebox[\textwidth][c]{\resizebox{#1\textwidth}{!}{#2}}}

\newcommand{\R}{\mathbb{R}}

\newcommand{\Z}{\mathbb{Z}}

\newenvironment{eq}
    {\begin{equation}
    \begin{aligned}
    }
    { 
    \end{aligned}
    \end{equation}
    \ignorespacesafterend
    }

\usepackage{pifont}

\RequirePackage[colorlinks=true
,urlcolor=blue
,anchorcolor=blue
,citecolor=blue
,filecolor=blue
,linkcolor=blue
,menucolor=blue
,linktocpage=true
,pdfproducer=medialab
,pdfa=true
]{hyperref}

\usepackage{cleveref}
\usepackage{tabularx}

\begin{document}

\preprint{{\tt IFT-UAM/CSIC-26-171}}

\title{Gauge groups of type 0 orientifolds}
\author{Vittorio Larotonda${}^{1,2}$ and Salvatore Raucci${}^{3,4}$
     \oneaddress{
     ${}^1$Dipartimento di Fisica e Astronomia, Universit\`{a} di Bologna, via Irnerio 46, Bologna, Italy\\
     ${}^2$INFN, Sezione di Bologna, viale Berti Pichat 6/2, Bologna, Italy\\
     ${}^3$Instituto de F\'{i}sica Te\'{o}rica IFT-UAM/CSIC, C/ Nicol\'{a}s Cabrera 13--15, Campus de Cantoblanco, 28049 Madrid, Spain \\
     ${}^4$Departamento de F\'{i}sica Te\'{o}rica, Universidad Aut\'{o}noma de Madrid, Cantoblanco, 28049 Madrid, Spain\\      
      {~}
}}

\Abstract{We determine the global form of the gauge groups of the orthogonal and symplectic type 0 orientifolds, using the constraints from perturbative and brane states. In the process, we correct some misprints in type 0 vacuum amplitudes that have propagated in the literature. For type 0A, our analysis reveals a parity condition on the numbers of D9-branes in the orthogonal orientifold and a puzzle concerning the decay of a tachyonic D0-brane in the symplectic orientifold.
}

\maketitle
\setcounter{page}{1}

\setcounter{tocdepth}{2}
\tableofcontents

\section{Introduction and main results}

Gauge sectors naturally occur in some of the ten-dimensional string theories. Besides their phenomenological relevance, there are more structural questions that one might ask, such as: which gauge groups can possibly arise? Determining the group, and not just the gauge algebra, is far from a technicality: type I string theory and one of the supersymmetric heterotic theories share the gauge group $Spin(32)/\Z_2$, and this has been one of the tests~\cite{Witten:1998cd,Sen:1998tt,Sen:1998ki,Sen:1999mg,Witten:2023snr} of their S-duality~\cite{Dabholkar:1995ep,Hull:1995nu,Polchinski:1995df}.

The theories containing gauge sectors can be divided into two classes: the heterotic theories and the orientifolds. Most of them are not spacetime supersymmetric; in particular, type I string theory is the only supersymmetric ten-dimensional orientifold. For the heterotic theories, the gauge groups have been extracted from the lattice formulation~\cite{Fraiman:2023cpa}, while only partial results are available for the orientifolds, namely for the Sugimoto string~\cite{Larotonda:2024thv} and for two special orientifolds of type 0 string theories~\cite{Fraiman:2026ltu}. In this work, we extend this analysis to the orthogonal and symplectic orientifolds of the type 0 theories.\footnote{Recent reviews describing the type 0 orientifolds in the context of non-supersymmetric string theories can be found, e.g., in~\cite{Mourad:2017rrl,Basile:2021mkd,Angelantonj:2024tns,Raucci:2024fnp,Leone:2025mwo,Dudas:2025ubq}.} These come in families, and we determine the gauge group topology for all of them. 

We extract the gauge groups from the perturbative spectra and the brane states, following the paradigm of type I string theory. Every object, perturbative or non-perturbative, must transform in a well-defined representation of the gauge group, which is then constrained by the available representations.
Specifically, for an orientifold with gauge algebra $\mathfrak{g}$, we consider the universal cover $\widetilde{G}$, with center $\mathcal{Z}(\widetilde{G})=\{z_i\}$. We then find the representation classes $[{\bf R}]$ arising from the quantization of the perturbative excitations and those arising from the branes, and determine the central elements $z_i$ that act trivially on the classes $[{\bf R}]$. The gauge group that we obtain is then the maximal central quotient compatible with the spectrum.

The relevant brane states come from open strings stretched between the branes and the background D9-branes, which generate the gauge sector.
For D0-branes, which are particles in spacetime, these open strings clearly give rise to states that must fit into representations of the gauge group.
However, one should also check the quantization of modes coming from the other $p$-branes,\footnote{If one insists on obtaining a particle, the $p$-branes can be compactified on $p$-cycles.} as emphasized in~\cite{Witten:2023snr,Larotonda:2024thv}. Assuming that the tachyonic branes decay into branes free of open-string tachyons and perturbative states, it suffices to consider the tachyon-free branes in each theory. This has been the approach of~\cite{Fraiman:2026ltu}, which we follow in this work. We will find one puzzling case that may evade this paradigm: a tachyonic D0-brane in the symplectic orientifold of type 0A; we comment on this in \cref{sec:0A/Omega_s_gauge_group}.
There may be other non-perturbative effects that further break the center of $\widetilde{G}$ (the tachyonic D0 above may be an example), so that, strictly speaking, our results should be interpreted as ``upper bounds'' on the invariant centers.

In addition to the orthogonal and symplectic orientifolds that we will discuss, type 0B string theory also has unitary orientifolds~\cite{Bianchi:1990yu,Sagnotti:1995ga,Sagnotti:1996qj}; these will be discussed in a separate work.

Our analysis in \cref{sec:groups_orthogonal,sec:groups_symplectic} relies on several known results and a few new ones that we collect in \cref{sec:type_0_review} and in two appendices on the brane spectra of type 0 orientifolds. In particular, in \cref{sec:type_0_review}, we display the relevant one-loop amplitudes obtained from~\cite{Dudas:2001wd}, correcting several misprints that have propagated in the literature (especially for the orientifolds of type 0A string theory).\footnote{The type 0A/$\Omega$ amplitudes of~\cite{Dudas:2001wd} containing misprints are eqs.~(7.8), (7.9), and (7.11).} We also use a more modern notation, providing a dictionary between the results of~\cite{Kaidi:2019tyf} and the amplitudes of~\cite{Dudas:2001wd}. 
Before delving into the detailed analysis, we summarize our main findings.

\subsection{The gauge groups}\label{sec:summary}

\paragraph{Type 0A orthogonal orientifold} The gauge algebra is $\mathfrak{so}(n_1)\oplus \mathfrak{so}(n_2)$. Our first result is that $n_2$ must be even to cancel a K-theory tadpole (equivalently, a worldvolume global anomaly) and $n_1$ must be even for a type of D0-brane to have gauge-invariant states. For $n_{1,2}>2$, the identity component of the gauge group is
\begin{eq}
  \frac{SO(n_1)\times Spin(n_2)}{\Z_2} \qquad &\text{if}\quad (n_1,n_2)\equiv (0,0)\pmod{4}\,,\\
    SO(n_1)\times Spin(n_2)\qquad &\text{if}\quad (n_1,n_2)\not\equiv (0,0)\pmod{4}\,.
\end{eq}
For the special cases when one of the two factors is absent (and the other is still $>2$), we obtain

\begin{eq}
\resizebox{0.92\linewidth}{!}{$
    \begin{aligned}
        n_2=0 \ &:
            \begin{array}{ll}
                \displaystyle\frac{SO(n_1)}{\Z_2} & \text{if } n_1\equiv 0\pmod{4}\,,\\[10pt]
                SO(n_1) & \text{if } n_1\equiv 2\pmod{4}\,;
            \end{array}\qquad
        n_1=0 \ &:
            \begin{array}{ll}
                \displaystyle\frac{Spin(n_2)}{\Z_2} & \text{if } n_2\equiv 0\pmod{4}\,,\\[10pt]
                Spin(n_2) & \text{if } n_2\equiv 2\pmod{4}\,.
            \end{array}
    \end{aligned}
    $}
\end{eq}
The discussion about disconnected components can be found in \cref{sec:0A/Omega_gauge_group}.

\paragraph{Type 0B orthogonal orientifold}
The gauge algebra is $\mathfrak{so}(k_1)^{\oplus 2}\oplus \mathfrak{so}(k_2)^{\oplus 2}$. We obtain the following identity components for non-vanishing $k_{1,2}>2$:
\begin{eq}
    \frac{Spin(k_1)^2\times Spin(k_2)^2}{\Z_2^3} \qquad &\text{if}\quad (k_1,k_2)\equiv (0,0) \pmod{4}\,,\\
    \frac{Spin(k_1)^2\times Spin(k_2)^2}{\Z_2^2} \qquad &\text{if} \quad (k_1,k_2) \not\equiv (0,0) \pmod{2}\,,\\
    \frac{Spin(k_1)^2\times Spin(k_2)^2}{ \Z_4\times \Z_2} \qquad &\text{otherwise}\,.
\end{eq}
When $k_1=0$ or $k_2=0$, if $k>2$ denotes the non-vanishing multiplicity, we obtain
\begin{eq}
    \frac{Spin(k)^2}{\Z_2^2} \qquad &\text{if}\quad k\equiv 0 \pmod{4}\,,\\
    \frac{Spin(k)^2}{\Z_4} \qquad &\text{if}\quad k\equiv 2 \pmod{4}\,,\\
    \frac{Spin(k)^2}{\Z_2} \qquad &\text{if}\quad k\equiv 1 \pmod{2}\,.
\end{eq}
The discussion about disconnected components can be found in \cref{sec:0B/Omega_gauge_group}.

\paragraph{Type 0A symplectic orientifold}
The gauge algebra is $\mathfrak{sp}\left(\frac{n_1}{2}\right)\oplus \mathfrak{sp}\left(\frac{n_2}{2}\right)$, and we obtain the groups\footnote{However, see \cref{sec:0A/Omega_s_gauge_group} for a subtlety in the case $(n_1,n_2)\equiv (2,0)\pmod{4}$.}
\begin{eq}
    \frac{Sp\left(\frac{n_1}{2}\right)\times Sp\left(\frac{n_2}{2}\right)}{\Z_2} \qquad &\text{if}\quad n_2\equiv 0 \pmod{4}\,,\\
    Sp\left(\frac{n_1}{2}\right)\times Sp\left(\frac{n_2}{2}\right) \qquad &\text{if}\quad n_2\equiv 2 \pmod{4}\,.
\end{eq}
The special cases with a vanishing $n_1$ or $n_2$ are obtained by removing one factor from the above groups.

\paragraph{Type 0B symplectic orientifold}
The gauge algebra is $\mathfrak{sp}\left(\frac{k_1}{2}\right)^{\oplus 2}\oplus \mathfrak{sp}\left(\frac{k_2}{2}\right)^{\oplus 2}$, and we obtain the group
\begin{eq}
    \frac{Sp\left(\frac{k_1}{2}\right)^2\times Sp\left(\frac{k_2}{2}\right)^2}{\Z_2} \,.
\end{eq}
The special cases with a vanishing $k_1$ or $k_2$ are obtained by removing one pair of factors.

\section{Type 0 strings and their branes}\label{sec:type_0_review}

Type 0 strings are built in the light-cone gauge from 8 transverse bosons and 8 Majorana--Weyl fermions on each side of the worldsheet. The fermionic contribution to their one-loop vacuum amplitudes can then be written in terms of the level-one $\mathfrak{so}(2r)_1$ characters for $2r$ free Majorana fermions, $O_{2r}$, $V_{2r}$, $S_{2r}$, and $C_{2r}$, as
\begin{eq}
    \mathcal{T}_{0A} & = O_8\bar{O}_8+V_8\bar{V}_8+S_8\bar{C}_8+C_8\bar{S}_8 \,, \\
    \mathcal{T}_{0B} & = O_8\bar{O}_8+V_8\bar{V}_8+S_8\bar{S}_8+C_8\bar{C}_8 \,.
\end{eq}
In these expressions, we are suppressing the modular integral with the measure and the contribution from the bosonic oscillators, following the conventions of~\cite{Angelantonj:2002ct}.

Starting from these two type 0 closed-string theories, one can impose different orientifold projections, following from four allowed Klein bottle amplitudes~\cite{Sagnotti:1987tw,Pradisi:1988xd,Bianchi:1990yu,Fioravanti:1993hf,Pradisi:1995pp,Sagnotti:1995ga,Sagnotti:1996qj}: one for type 0A string theory and three for type 0B string theory. In this work, we will be interested in two of them,
\begin{eq}\label{eq:0A-0B_Orientifolds}
    \mathcal{K}_{0A}=\frac{1}{2}\left(O_8+V_8\right) \,,\qquad \mathcal{K}_{0B}=\frac{1}{2}\left(O_8+V_8-S_8-C_8\right)\,.
\end{eq}
In the transverse channel, these become
\begin{eq}
    \tilde{\mathcal{K}}_{0A}=\frac{2^5}{2}\left(O_8+V_8\right) \,,\qquad \tilde{\mathcal{K}}_{0B}= 2^5 V_8\,,
\end{eq}
and signal that the type 0A orientifold planes couple to the closed-string tachyon and dilaton, while the type 0B orientifold planes only source the dilaton. 

In addition to the O9-planes, the type 0A orientifold contains two types of D9-branes, which couple to the closed-string tachyon with opposite signs. Following the literature, we denote those with a positive coupling to the closed tachyon by $n_1$, and those with a negative coupling to the tachyon by $n_2$. The annulus and M\"{o}bius amplitudes are
\begin{eq}\label{eq:0A_orientifold_A-M}
    \mathcal{A}_{99}&=\frac{n_1^2 + n_2^2}{2}\left(O_8+V_8\right)-n_1 n_2 \left(S_8+C_8\right)\,,\\
    \mathcal{M}_9&=\sigma\left(\frac{n_1-n_2}{2}\hat{O}_8-\frac{n_1+n_2}{2}\hat{V}_8\right)\,,
\end{eq}
where the use of hatted characters is explained in~\cite{Angelantonj:2002ct}. In \cref{eq:0A_orientifold_A-M}, $\sigma$ denotes a sign choice in the M\"obius amplitude that gives rise to two different orientifolds:\footnote{This choice is not present for all orientifold projections. In fact, it is not an available option for the other two orientifolds of type 0B string theory that we are not exploring in this work: in one case, $\sigma = -1$ would exchange the two unitary factors of the orientifold, while in the other, it would lead to an inconsistent model.} an orthogonal orientifold with gauge algebra $\mathfrak{so}(n_1)\oplus\mathfrak{so}(n_2)$ when $\sigma=+1$ and a symplectic orientifold with gauge algebra $\mathfrak{sp}(n_1/2)\oplus\mathfrak{sp}(n_2/2)$ when $\sigma=-1$. Note that, in this latter case, $n_1$ and $n_2$ must be even. 
In the transverse channel, the amplitudes of \cref{eq:0A_orientifold_A-M} become
\begin{eq}
    \widetilde{\mathcal{A}}_{99}&=\frac{2^{-5}}{2}\left[(n_1-n_2)^2 O_8+(n_1+n_2)^2 V_8\right]\,,\\
    \widetilde{\mathcal{M}}_9&=-\sigma\left[(n_1-n_2)\hat{O}_8+(n_1+n_2)\hat{V}_8\right]\,.
\end{eq}
Note that the two types of branes are not symmetric. For example, the open-string tachyons are in the symmetric and antisymmetric representations, depending on $\sigma$, as can be seen from the multiplicity of the $O_8$ character,
\begin{eq}\label{eq:0A_tachyons_D9}
    \#O_8=\frac{n_1(n_1+\sigma)}{2}+\frac{n_2(n_2-\sigma)}{2}\,.
\end{eq}
This has interesting consequences when $\sigma=+1$, which we will discuss in \cref{sec:0A_parity}.

The orientifold of type 0B in \cref{eq:0A-0B_Orientifolds} can have four types of D9-branes: two pairs of branes and antibranes. Denoting them by $n_1$, $n_2$, $n_3$, and $n_4$, the annulus and M\"obius amplitudes are
\begin{eq}\label{eq:0B_orientifold_A-M}
    \mathcal{A}_{99}& =\sum_{i=1}^4\frac{n_i^2}{2} V_8+(n_1n_2+n_3n_4)O_8-(n_1n_3+n_2n_4)S_8-(n_1n_4+n_2n_3)C_8\,,\\
    \mathcal{M}_{9}& =-\sigma\sum_{i=1}^4 \frac{n_i}{2}\hat{V}_8\,,
\end{eq}
where again, $\sigma=\pm 1$ selects the orthogonal or the symplectic version of the orientifolds, with gauge algebras $\bigoplus_{i}\mathfrak{so}(n_i)$ when $\sigma=+1$ and $\bigoplus_{i}\mathfrak{sp}(n_i/2)$ when $\sigma=-1$. RR tadpole cancellation requires that $n_1=n_2\equiv k_1$ and $n_3=n_4\equiv k_2$, which amounts to imposing an equal number of branes and antibranes for each family. From the amplitudes in \cref{eq:0B_orientifold_A-M}, one obtains that the open-string tachyons are in bifundamental representations associated with brane-antibrane pairs, while chiral fermions are in bifundamentals connecting the pairs; see~\cite[Fig.~2]{Altavista:2026evd} for a pictorial representation of these features. In the transverse channel, the amplitudes of \cref{eq:0B_orientifold_A-M} are
\begin{eq}
    \widetilde{\mathcal{A}}_{99}& =2^{-5}\left[(k_1-k_2)^2 O_8+ (k_1+k_2)^2 V_8\right] \,,\\
    \widetilde{\mathcal{M}}_9&=-2\sigma(k_1+k_2)\hat{V}_8\,.
\end{eq}

\subsection{Branes in orientifold models}\label{sec:branes_type_0}

The branes of the type 0 orientifolds descend from those of the parent type 0A and 0B models. There, the physical state of a charged D$p$-brane can be written in terms of the boundary states $\ket{\mathrm{B}p,\eta}$ as~\cite{Bergman:1997rf,Kaidi:2019tyf} 
\begin{eq}\label{eq:boundary_state}
    \ket{\mathrm{D}p,\eta}\propto\eta \ket{\mathrm{B}p,\eta}_{\mathrm{NSNS}}+\ket{\mathrm{B}p,\eta}_{\mathrm{RR}}\,,
\end{eq}
where $\eta=\pm1$ distinguishes two types of branes. This parameter will play a key role in our work. The uncharged branes of type 0 theories only have the NSNS boundary-state component.

The boundary state $\ket{\mathrm{B}p,\eta}_{\mathrm{NSNS}}$ contains the closed-string tachyon with a positive coefficient and the massless NSNS fields with a coefficient proportional to $\eta$. Therefore, taking into account the extra factor of $\eta$ in \cref{eq:boundary_state}, the coupling of a D$p$-brane to the closed tachyon is $\propto\eta$, while the tension is always positive. In this sense, $\eta$ denotes different ``types'' of branes, distinguished by the coupling to the closed tachyon. For example, with our convention for the D9-branes of type 0A, the index $n_1$ corresponds to branes of the type $\eta=+1$, while the index $n_2$ corresponds to branes of the type $\eta=-1$. For the D9-branes of type 0B, the index $k_1$ corresponds to branes of the type $\eta=+1$, while the index $k_2$ corresponds to branes of the type $\eta=-1$.
In the following, we will write that a brane is of type $+$ or $-$ depending on the corresponding value of $\eta$. A feature of type 0 strings that will be useful to us is that strings between branes of the same type give rise to bosons, while strings between branes of opposite types give rise to fermions. In particular, given a D$p$-brane of type $\eta$, massless fermions from D$p$--D9 strings come from D9-branes of the type $-\eta$. This will be used several times in \cref{sec:groups_orthogonal,sec:groups_symplectic}.

The above analysis applies to the branes of the parent type 0 models. To obtain those of the orientifolds, one must know how $\Omega$ acts on the boundary states. As shown in \cite[eq.~(5.3)]{Kaidi:2019tyf},
\begin{eq}\label{eq:branes_and_omega}
    \Omega\ket{\mathrm{B}p,\eta}_{\mathrm{NSNS}}=\ket{\mathrm{B}p,\eta}_{\mathrm{NSNS}}\,,\qquad \Omega\ket{\mathrm{B}p,\eta}_{\mathrm{RR}}=i^{\eta\nu}\ket{\mathrm{B}p,\eta}_{\mathrm{RR}}\,,
\end{eq}
where
\begin{eq}\label{eq:nu_definition}
    \nu\equiv\eta n +p-1 \pmod{8}
\end{eq}
and the values of $n$ correspond to the different orientifolds as follows:
\begin{eq}\label{eq:orientifold_types}
    \mathrm{0A/}\Omega \to n=1\,,\quad \mathrm{0A/}\Omega_s \to n=5\,,\quad \mathrm{0B/}\Omega \to n=0\,,\quad \mathrm{0B/}\Omega_s \to n=4\,.
\end{eq}
We have denoted the symplectic orientifolds by $\Omega_s$; this will be our convention for the rest of this work. From \cref{eq:branes_and_omega,eq:nu_definition}, one can extract all the properties of the branes. In particular, when $\nu\equiv0\pmod{4}$, each brane is mapped to itself by $\Omega$, thus leading to branes and antibranes with orthogonal or symplectic Chan--Paton factors. Following~\cite{Dudas:2001wd}, we call them \emph{charged branes}, because they are charged under RR fields. When $\nu\equiv2\pmod{4}$, $\Omega$ exchanges branes and antibranes, and the configuration that survives after the orientifold corresponds to $m$ branes and $m$ antibranes with unitary Chan--Paton factors. These are the \emph{uncharged branes} of~\cite{Dudas:2001wd}. Branes with $\nu$ odd only have NSNS components and are also uncharged, with real Chan--Paton labels.

In general, for each $p$, there are D$p$-branes of both types, with $\eta=+1$ and $\eta=-1$. In particular, let $\eta_d(p)$ be the type of charged D$p$-branes in type 0A/$\Omega$ (hence, $n=1$ in \cref{eq:orientifold_types}). Then,
\begin{eq}
    \eta_d(p)=\begin{cases}
        +1 \qquad p=0,4,8\,,\\
        -1 \qquad p=2,6\,,
    \end{cases}
\end{eq}
while the branes with the same dimensions but different types correspond to uncharged branes. Note that the charged branes of type 0A/$\Omega_s$ have the same $\eta$ types, $\eta_d(p)$, because replacing $n=1$ with $n=5$ does not change $\nu$ modulo $4$.

The full classification of branes in the type 0 orientifolds can be found in~\cite{Kaidi:2019tyf}, and the explicit one-loop amplitudes in~\cite{Dudas:2001wd}. They are fully characterized by the values of $n$ and $\eta$, which define $\nu$ as in \cref{eq:nu_definition}. These determine the worldvolume gauge symmetry, the representation of the open-string tachyons (and their possible absence), and the K-theory charges; we summarize these properties in \cref{tab:type0_orientifold_branes}.
For example, in type 0A/$\Omega$, this yields the spectrum of branes given in \cref{tab:type0A/Omega_branes}.

\begin{table}[ht]
    \centering
    \begin{tabular}{c|c|c|c}
        $\nu$ & {gauge algebra} & {open tachyon} & {K-charge}  \\ \hline
        0 & $\mathfrak{so}\oplus\mathfrak{so}$ & bifund & $\Z$ \\
        1 & $\mathfrak{so}$ & sym & $0$ \\
        2 & $\mathfrak{u}$ & sym & $0$ \\
        3 & $\mathfrak{sp}$ & sym & $0$ \\
        4 & $\mathfrak{sp}\oplus \mathfrak{sp}$ & bifund & $\Z$ \\
        5 & $\mathfrak{sp}$ & antisym & $0$ \\
        6 & $\mathfrak{u}$ & antisym & $\Z_2$ \\
        7 & $\mathfrak{so}$ & antisym & $\Z_2$
    \end{tabular}
    \caption{Classification of branes from~\cite{Kaidi:2019tyf} with the worldvolume gauge algebra, the representation of the open-string tachyon, and the K-theory charge. From this table, one obtains the branes of all the type 0 orientifolds (of types $\eta=\pm1$) by using the definition of $\nu$ in \cref{eq:nu_definition}.}
    \label{tab:type0_orientifold_branes}
\end{table}

\begin{table}[ht]
    \centering
    \begin{tabular}{c|c|c|c|c}
        $p$ & $\eta=+1$ & K-charge & $\eta=-1$ & K-charge \\ \hline
        9 & $n_1$ stack, $\mathfrak{so}$ & $0$ & $n_2$ stack, $\mathfrak{so}$ & $\Z_2$ \\
        8 & charged, $\mathfrak{so}\oplus\mathfrak{so}$ & $\Z$ & uncharged, $\mathfrak{u}$ & $\Z_2$ \\
        7 & $\mathfrak{so}$ & $\Z_2$ & $\mathfrak{sp}$ & $0$ \\
        6 & uncharged, $\mathfrak{u}$ & $\Z_2$ & charged, $\mathfrak{sp}\oplus \mathfrak{sp}$ & $\Z$ \\
        5 & $\mathfrak{sp}$ & $0$ & $\mathfrak{sp}$ & $0$ \\
        4 & charged, $\mathfrak{sp}\oplus\mathfrak{sp}$ & $\Z$ & uncharged, $\mathfrak{u}$ & $0$ \\
        3 & $\mathfrak{sp}$ & $0$ & $\mathfrak{so}$ & $0$ \\
        2 & uncharged, $\mathfrak{u}$ & $0$ & charged, $\mathfrak{so}\oplus\mathfrak{so}$ & $\Z$ \\
        1 & $\mathfrak{so}$ & $0$ & $\mathfrak{so}$ & $\Z_2$ \\
        0 & charged, $\mathfrak{so}\oplus\mathfrak{so}$ & $\Z$ & uncharged, $\mathfrak{u}$ & $\Z_2$ \\
        -1 & $\mathfrak{so}$ & $\Z_2$ & $\mathfrak{sp}$ & $0$
    \end{tabular}
    \caption{Branes of type 0A/$\Omega$ depending on their $\eta$ type. All the $p$-odd branes are uncharged. This has been derived by using \cref{tab:type0_orientifold_branes}.}
    \label{tab:type0A/Omega_branes}
\end{table}

In order to extract the brane spectra, we will use the one-loop amplitudes from~\cite{Dudas:2001wd}. We display them in this section explicitly, correcting some misprints that propagated in the literature, especially for the type 0A orientifolds. In these cases, we provide the amplitudes in both the direct and transverse channels for ease of comparison with the (conflicting) literature; for the type 0B orientifolds, we only provide the direct-channel amplitudes. We stress that the following discussion applies to both the orthogonal and the symplectic cases.

\subsubsection{Branes in 0A orientifolds}\label{sec:0A_branes_amplitudes}

Branes with $p$ even in the type 0A orientifolds are labeled by two real ($d_1,d_2$) and one complex ($m$) multiplicities, $(d_1,d_2,m)$. $d_1$ and $d_2$ denote the branes and antibranes of type $\eta_d(p)$ that carry RR charge, while $m$ denotes uncharged branes. Recall that $\eta_d(p)$ is the $\eta$-type of the charged D$p$-branes, and therefore, the type of the uncharged branes is $-\eta_d(p)$.
The annulus and M\"obius amplitudes are
\begin{eq}\label{eq:0A_orientifold_charged_direct}
    \mathcal{A}_{pp}&=\left(\frac{d_1^2+d_2^2}{2}+m\bar{m}\right)\left(O_{p-1}V_{9-p}+V_{p-1}O_{9-p}\right)\\ & +\left(d_1d_2+\frac{m^2 +\bar{m}^2}{2}\right)\left(O_{p-1}O_{9-p}+V_{p-1}V_{9-p}\right)-(d_1+d_2)(m+\bar{m})S_{p-1}'S_{9-p}'\,,\\
    \mathcal{M}_p&=\sigma\epsilon\left[\frac{d_1+d_2}{2}\left(\hat{O}_{p-1}\hat{V}_{9-p}-\hat{V}_{p-1}\hat{O}_{9-p}\right)-\eta_d(p)\frac{m+\bar{m}}{2}\left(\hat{O}_{p-1}\hat{O}_{9-p}+\hat{V}_{p-1}\hat{V}_{9-p}\right)\right]\,,
\end{eq}
where $\epsilon=1 $ for $p=0,2,8$ and $\epsilon=-1$ for $p=4,6$. In the transverse channel, these amplitudes become
\begin{eq}\label{eq:0A_orientifold_charged_transverse}
    \widetilde{\mathcal{A}}_{pp}&=\frac{2^{-\frac{p+1}{2}}}{4}\left[(d_1+d_2+m+\bar{m})^2 \left(V_{p-1}O_{9-p}+O_{p-1}V_{9-p}\right) \right.\\
    & +(d_1+d_2-m-\bar{m})^2\left(O_{p-1}O_{9-p}+V_{p-1}V_{9-p}\right)\\
    & \left. -2\left((d_1-d_2)^2-(m-\bar{m})^2\right) S_{p-1}' S_{9-p}'\right]\,,\\
    \widetilde{\mathcal{M}}_p&=\frac{\sigma}{\sqrt{2}}\left[-\eta_d(p)(d_1+d_2-m-\bar{m})\left(\hat{O}_{p-1}\hat{O}_{9-p}+\hat{V}_{p-1}\hat{V}_{9-p}\right)\right.\\
    & \left.+(d_1+d_2+m+\bar{m})\left(\hat{O}_{p-1}\hat{V}_{9-p}-\hat{V}_{p-1}\hat{O}_{9-p}\right)\right]\,.
\end{eq}
The multiplicities of the characters $V_{p-1}O_{9-p}$ and $O_{p-1}O_{9-p}$ in the direct channel fix the worldvolume gauge algebras and the open-string tachyons:
\begin{eq}
    \# V_{p-1}O_{9-p} & = \frac{d_1(d_1-\epsilon\sigma)}{2}+\frac{d_2(d_2-\epsilon\sigma)}{2}+m\bar{m}\,,\\
    \# O_{p-1}O_{9-p} & = d_1d_2+\frac{m^2+\bar{m}^2}{2}-\sigma\epsilon\eta_d(p)\frac{m+\bar{m}}{2}\,.
\end{eq}
An additional contribution at one loop, which is actually the most relevant one for our analysis, comes from D$p$--D9 strings. In the direct and transverse channels, these lead to
\begin{eq}\label{eq:0A_orientifold_charged_p9}
    \mathcal{A}_{p9}&=\left[n_{\eta_d(p)}(d_1+d_2)+n_{-\eta_d(p)}(m+\bar{m})\right]\left(O_{p-1}+V_{p-1}\right)S_{9-p}' \\
    & - \left[n_{-\eta_d(p)}(d_1+d_2)+n_{\eta_d(p)}(m+\bar{m})\right]S_{p-1}'\left(O_{9-p}+V_{9-p}\right)\,,\\
    \widetilde{\mathcal{A}}_{p9}&=\frac{2^{-5}}{\sqrt{2}}\left[\eta_d(p) (n_1-n_2)(d_1+d_2-m-\bar{m})\left(O_{p-1}O_{9-p}-V_{p-1}V_{9-p}\right)\right. \\ 
    &\left. +(n_1+n_2)(d_1+d_2+m+\bar{m})\left(V_{p-1}O_{9-p}-O_{p-1}V_{9-p}\right)\right]\,,
\end{eq}
where in $\mathcal{A}_{p9}$ we used the notation $n_+\equiv n_1$ and $n_-\equiv n_2$.

Branes with $p$ odd in the type 0A orientifolds are labeled by two real multiplicities, $(d_1,d_2)$, which denote branes of types $+$ and $-$. The corresponding amplitudes are 
\begin{eq}\label{eq:0A_orientifold_uncharged_direct}
    \mathcal{A}_{pp}&=\frac{d_1^2+d_2^2}{2}\left(O_{p-1}+V_{p-1}\right)\left(O_{9-p}+V_{9-p}\right)-d_1d_2 \left(S_{p-1}+C_{p-1}\right)\left(S_{9-p}+C_{9-p}\right)\,,\\
    \mathcal{M}_p&=\sigma\epsilon\left[\frac{d_1-\epsilon' d_2}{2}\left(\hat{O}_{p-1}\hat{O}_{9-p}+\hat{V}_{p-1}\hat{V}_{9-p}\right)+\frac{\epsilon'd_1+d_2}{2}\left(\hat{O}_{p-1}\hat{V}_{9-p}-\hat{V}_{p-1}\hat{O}_{9-p}\right)\right]\,,\\
    \mathcal{A}_{p9}&=(n_1d_1+n_2d_2)\left(O_{p-1}+V_{p-1}\right)\left(S_{9-p}+C_{9-p}\right)\\
    & -(n_1d_2+n_2d_1)\left(S_{p-1}+C_{p-1}\right)\left(O_{9-p}+V_{9-p}\right)\,,
\end{eq}
where $(\epsilon,\epsilon')=(-1,-1),(+1,+1),(+1,-1),(-1,+1),(-1,-1)$ for the D$(-1)$, D1, D3, D5, and D7.
In the transverse channel,
\begin{eq}\label{eq:0A_orientifold_uncharged_transverse}
    \widetilde{\mathcal{A}}_{pp}&=\frac{2^{-\frac{p+1}{2}}}{2}\left[(d_1-d_2)^2 \left(O_{p-1}O_{9-p}+V_{p-1}V_{9-p}\right)+(d_1+d_2)^2\left(O_{p-1}V_{9-p}+V_{p-1}O_{9-p}\right)\right]\,,\\
    \widetilde{\mathcal{M}}_p&=\sigma\left[-(d_1-d_2)\left(\hat{O}_{p-1}\hat{O}_{9-p}+\hat{V}_{p-1}\hat{V}_{9-p}\right)+(d_1+d_2)\left(\hat{O}_{p-1}\hat{V}_{9-p}-\hat{V}_{p-1}\hat{O}_{9-p}\right)\right]\,,\\
    \widetilde{\mathcal{A}}_{p9}&=2^{-5}\left[(n_1-n_2)(d_1-d_2)\left(O_{p-1}O_{9-p}-V_{p-1}V_{9-p}\right)\right.\\
    &\left. +(n_1+n_2)(d_1+d_2)\left(V_{p-1}O_{9-p}-O_{p-1}V_{9-p}\right)\right]\,.
\end{eq}

From these amplitudes, one can obtain the branes that are free of open-string tachyons in the 0A/$\Omega$ orientifold, which will be used in \cref{sec:0A/Omega_gauge_group}. We list them in \cref{tab:OA/Omega_tachyon_free_branes}, where we also display the branes that become tachyon-free for the special configurations with $n_1=0$ or $n_2=0$.
\begin{table}[H]
    \centering
    \begin{minipage}{0.86\linewidth}
        \renewcommand{\arraystretch}{1.20}
        \setlength{\tabcolsep}{0pt}
        \begin{tabular*}{\linewidth}{@{\extracolsep{\fill}}llc@{}}
            \toprule[0.8pt]
            & Brane & Configurations \\[3pt]
            \midrule[0.4pt]
            \addlinespace[4pt]
            \multirow[t]{5}{*}{$n_1n_2\neq0$}
                & D4      & $(2l,0,0)$ \\
                & D2      & $(d,0,0)$  \\
                & D1      & $(0,1)$    \\
                & D0      & $(d,0,m)$  \\
                & D$(-1)$ & $(1,0)$    \\
            \addlinespace[5pt]
            \midrule[0.4pt]
            \addlinespace[4pt]
            \multirow[t]{3}{*}{$n_1=0$}
                & D8 & $(d,0,0)$ \\
                & D7 & $(1,0)$   \\
                & D6 & $(0,0,1)$ \\
            \addlinespace[4pt]
            \cmidrule[0.3pt]{2-3}
            \addlinespace[4pt]
            \multirow[t]{2}{*}{$n_2=0$}
                & D8 & $(0,0,1)$  \\
                & D6 & $(2l,0,0)$ \\
            \addlinespace[4pt]
            \bottomrule[0.8pt]
        \end{tabular*}
\end{minipage}
        \caption{D-branes of type 0A/$\Omega$ that are free of open-string tachyons. The Chan--Paton multiplicities are $(d_1,d_2,m)$ for $p$ even and $(d_1,d_2)$ for $p$ odd. For the D0, $m$ can be either $0$ or $1$. In the lower blocks, we list the additional branes in the special cases with $n_1=0$ or $n_2=0$.}
        \label{tab:OA/Omega_tachyon_free_branes}   
\end{table}

For the symplectic type 0A/$\Omega_s$ orientifold, the analogous list can be found in \cref{tab:sOA/Omega_tachyon_free_branes} and will be used in \cref{sec:s0A_gauge_group}.
\begin{table}[H]
    \centering
    \begin{minipage}{0.86\linewidth}
        \renewcommand{\arraystretch}{1.20}
        \setlength{\tabcolsep}{0pt}
        \begin{tabular*}{\linewidth}{@{\extracolsep{\fill}}llc@{}}
            \toprule[0.8pt]          
            & Brane & Configurations \\[3pt]
            \midrule[0.4pt]
            \addlinespace[4pt]
            \multirow[t]{5}{*}{$n_1n_2\neq0$}
                & D5 & $(0,1)$    \\
                & D4 & $(d,0,m)$  \\
                & D3 & $(1,0)$    \\
                & D2 & $(2l,0,m)$ \\
                & D0 & $(2l,0,0)$ \\
            \addlinespace[5pt]
            \midrule[0.4pt]
            \addlinespace[4pt]
            $n_1=0$ & D8 & $(2l,0,0)$ \\
            \addlinespace[4pt]
            \cmidrule[0.3pt]{2-3}
            \addlinespace[4pt]
            $n_2=0$ & D6 & $(d,0,0)$ \\
            \addlinespace[4pt]
            \bottomrule[0.8pt]
        \end{tabular*}
 \end{minipage}
        \caption{D-branes of type 0A/$\Omega_s$ that are free of open-string tachyons. The Chan--Paton multiplicities are $(d_1,d_2,m)$ for $p$ even and $(d_1,d_2)$ for $p$ odd. For D2- and D4-branes, $m$ can be either 0 or 1. In the lower blocks, we list the branes in the special cases with $n_1=0$ or $n_2=0$.}
        \label{tab:sOA/Omega_tachyon_free_branes}
\end{table}

\subsubsection{Branes in 0B orientifolds}\label{sec:0B_branes_amplitudes}

In the type 0B orthogonal and symplectic orientifolds, the charged D1- and D5-branes are labeled by four real multiplicities $d_i$. The direct-channel amplitudes read
\begin{eq}\label{eq:0B_orientifold_charged_direct}
    \mathcal{A}_{pp}& = \frac{1}{2}\sum_{i=1}^4 d_i^2 \left(O_{p-1}V_{9-p}+V_{p-1}O_{9-p}\right)+(d_1d_2+d_3d_4)\left(O_{p-1}O_{9-p}+V_{p-1}V_{9-p}\right)\\
    & - (d_1d_3+d_2d_4)\left(S_{p-1}S_{9-p}+C_{p-1}C_{9-p}\right)-(d_1d_4+d_2d_3)\left(S_{p-1}C_{9-p}+C_{p-1}S_{9-p}\right)\,,\\
    \mathcal{M}_p&=-\sigma\frac{\epsilon_p}{2}\sum_{i=1}^4 d_i \left(\hat{O}_{p-1}\hat{V}_{9-p}-\hat{V}_{p-1}\hat{O}_{9-p}\right) \,,\\
    \mathcal{A}_{p9}&=\left[k_1(d_1+d_2)+k_2(d_3+d_4)\right]\left(O_{p-1}+V_{p-1}\right)\left(S_{9-p}+C_{9-p}\right)\\
    & -\left[k_1(d_3+d_4)+k_2(d_1+d_2)\right]\left(S_{p-1}+C_{p-1}\right)\left(O_{9-p}+V_{9-p}\right)\,,
\end{eq}
where $\epsilon_1=-1$ and $\epsilon_5=+1$, and where $k_{1,2}$ have been defined after \cref{eq:0B_orientifold_A-M}. To extract the spectrum, it is also useful to have the last amplitude $\mathcal{A}_{p9}$ in terms of the multiplicities $n_i$: for the D5,
\begin{eq}
    \mathcal{A}_{59}&=\left(n_1d_1+n_2d_2+n_3d_3+n_4d_4\right)\left(O_{p-1}C_{9-p}+V_{p-1}S_{9-p}\right)\\
    &+\left(n_1d_2+n_2d_1+n_3d_4+n_4d_3\right)\left(O_{p-1}S_{9-p}+V_{p-1}C_{9-p}\right)\\
    &-\left(n_1d_4+n_2d_3+n_3d_2+n_4d_1\right)\left(S_{p-1}V_{9-p}+C_{p-1}O_{9-p}\right)\\
    &-\left(n_1d_3+n_2d_4+n_3d_1+n_4d_2\right)\left(S_{p-1}O_{9-p}+C_{p-1}V_{9-p}\right)\,,
\end{eq}
while for the D1, $\mathcal{A}_{19}$ is obtained by exchanging all the $S$ and $C$ characters in $\mathcal{A}_{59}$.

Uncharged branes with $p=-1,3,7$ are labeled by two complex multiplicities, $d_1$ and $d_2$, and the direct-channel amplitudes are
\begin{eq}\label{eq:0B_orientifold_uncharged_odd}
    \mathcal{A}_{pp}&=(d_1\bar{d}_1+d_2\bar{d}_2)\left(O_{p-1}V_{9-p}+V_{p-1}O_{9-p}\right)+\frac{d_1^2+\bar{d}_1^2+d_2^2+\bar{d}_2^2}{2}\left(O_{p-1}O_{9-p}+V_{p-1}V_{9-p}\right)\\
    &- (d_1\bar{d}_2+\bar{d}_1 d_2)\left(S_{p-1}S_{9-p}+C_{p-1}C_{9-p}\right)-(d_1d_2+\bar{d}_1\bar{d}_2)\left(S_{p-1}C_{9-p}+C_{p-1}S_{9-p}\right)\,,\\
    \mathcal{M}_p&=-\sigma\frac{\epsilon_p}{2}(d_1+\bar{d}_1+d_2+\bar{d}_2)\left(\hat{O}_{p-1}\hat{O}_{9-p}+\hat{V}_{p-1}\hat{V}_{9-p}\right)\,,\\
    \mathcal{A}_{p9}&=\left[k_1(d_1+\bar{d}_1)+k_2(d_2+\bar{d}_2)\right]\left(O_{p-1}+V_{p-1}\right)\left(S_{9-p}+C_{9-p}\right)\\
    & - \left[k_1(d_2+\bar{d}_2)+k_2(d_1+\bar{d}_1)\right]\left(S_{p-1}+C_{p-1}\right)\left(O_{9-p}+V_{9-p}\right)\,,
\end{eq}
where $\epsilon_{-1}=\epsilon_7=+1$ and $\epsilon_3=-1$. As for the D1- and D5-branes, to extract the spectrum, one needs $\mathcal{A}_{p9}$ in terms of the multiplicities $n_i$. For $p=-1$ and $7$, this is
\begin{eq}
    \mathcal{A}_{p9}&=\left(n_1d_1+n_2\bar{d}_1+n_3d_2+n_4\bar{d}_2\right)\left(O_{p-1}S_{9-p}+V_{p-1}C_{9-p}\right)\\
    & +\left(n_1\bar{d}_1+n_2{d}_1+n_3\bar{d}_2+n_4{d}_2\right)\left(O_{p-1}C_{9-p}+V_{p-1}S_{9-p}\right)\\
    & -\left(n_1\bar{d}_2+n_2{d}_2+n_3\bar{d}_1+n_4{d}_1\right)\left(S_{p-1}V_{9-p}+C_{p-1}O_{9-p}\right)\\
    & -\left(n_1{d}_2+n_2\bar{d}_2+n_3{d}_1+n_4\bar{d}_1\right)\left(S_{p-1}O_{9-p}+C_{p-1}V_{9-p}\right)\,,
\end{eq}
while for $p=3$, all the characters $S$ and $C$ are exchanged.

Finally, uncharged branes with $p$ even come with two real labels, $(d_1,d_2)$, and their one-loop amplitudes read
\begin{eq}\label{eq:0B_orientifold_uncharged_even}
    \mathcal{A}_{pp}&=\frac{d_1^2+d_2^2}{2}\left(O_{p-1}+V_{p-1}\right)\left(O_{9-p}+V_{9-p}\right)-2 d_1d_2 S_{p-1}'S_{9-p}'\,,\\
    \mathcal{M}_p&=-\sigma\frac{d_1+d_2}{\sqrt{2}}\left[\sin\left(\frac{p-5}{4}\pi\right)\left(\hat{O}_{p-1}\hat{O}_{9-p}+\hat{V}_{p-1}\hat{V}_{9-p}\right)\right. \\
    & \left. +\cos\left(\frac{p-5}{4}\pi\right)\left(\hat{O}_{p-1}\hat{V}_{9-p}-\hat{V}_{p-1}\hat{O}_{9-p}\right)\right]\,,\\
    \mathcal{A}_{p9}
    &=\left[\left(n_1+n_2\right)d_1+\left(n_3+n_4\right)d_2\right]\left(O_{p-1}+V_{p-1}\right)S_{9-p}'\\
    & -\left[\left(n_1+n_2\right)d_2+\left(n_3+n_4\right)d_1\right]S_{p-1}'\left(O_{9-p}+V_{9-p}\right)\,.
\end{eq}

From all these amplitudes, we obtain the branes that are free of open-string tachyons in \cref{tab:0B/Omega_tachyon_free_branes,tab:0B/Omega_prime_tachyon_free_branes} for the orthogonal and symplectic orientifolds.
\begin{table}[!htbp]
    \centering
    \begin{minipage}{0.86\linewidth}
        \renewcommand{\arraystretch}{1.20}
        \setlength{\tabcolsep}{0pt}
        \begin{tabular*}{\linewidth}{@{\extracolsep{\fill}}llc@{}}
            \toprule[0.8pt]
            & Brane & Configurations \\[3pt]
            \midrule[0.4pt]
            \addlinespace[4pt]
            \multirow[t]{4}{*}{$k_1k_2\neq0$}
                & D5 & $(d_1,d_2,d_3,d_4)$ \text{ with } $d_1d_2=d_3d_4=0$ \\
                & D1 & $(d_1,d_2,d_3,d_4)$ \text{ with } $d_1d_2=d_3d_4=0$ \\
                & D0 & $(d_1,d_2)$ \text{ with } $d_{1,2}\leq1$ \\
                & D$(-1)$ & $(d_1,d_2)$ \text{ with } $d_{1,2}\leq1$ \\
            \addlinespace[5pt]
            \midrule[0.4pt]
            \addlinespace[4pt]
            \multirow[t]{2}{*}{$k_1=0$}
                & D8 & $(d_1,d_2) = (1,0)$ \\
                & D7 & $(d_1,d_2) = (1,0)$ \\
            \addlinespace[4pt]
            \cmidrule[0.3pt]{2-3}
            \addlinespace[4pt]
            \multirow[t]{2}{*}{$k_2=0$}
                & D8 & $(d_1,d_2) = (0,1)$ \\
                & D7 & $(d_1,d_2) = (0,1)$ \\
            \addlinespace[4pt]
            \bottomrule[0.8pt]
        \end{tabular*}
    \end{minipage}
    \caption{D-branes of type 0B/$\Omega$ that are free of open-string tachyons. For each entry, we specify the corresponding Chan--Paton multiplicity: for D5- and D1-branes, the configurations correspond to branes or antibranes of the two types, while the D0 and D$(-1)$ entries each describe two families. In the lower blocks, we list the additional branes in the special cases $k_1=0$ or $k_2=0$.}
    \label{tab:0B/Omega_tachyon_free_branes}
\end{table}

\begin{table}[!htbp]
    \centering
    \begin{minipage}{0.86\linewidth}
        \renewcommand{\arraystretch}{1.20}
        \setlength{\tabcolsep}{0pt}
        \begin{tabular*}{\linewidth}{@{\extracolsep{\fill}}llc@{}}
            \toprule[0.8pt]
            & Brane & Configurations \\[3pt]
            \midrule[0.4pt]
            \addlinespace[4pt]
            \multirow[t]{4}{*}{$k_1k_2\neq0$}
                & D5 & $(d_1,d_2,d_3,d_4)$ \text{ with } $d_1d_2=d_3d_4=0$ \\
                & D4 & $(d_1,d_2)$ \text{ with } $d_{1,2}\leq1$ \\
                & D3 & $(d_1,d_2)$ \text{ with } $d_{1,2}\leq1$ \\
                & D1 & $(d_1,d_2,d_3,d_4)$ \text{ with } $d_1d_2=d_3d_4=0$ \\
            \addlinespace[4pt]
            \bottomrule[0.8pt]
        \end{tabular*}
    \end{minipage}
    \caption{D-branes of type 0B/$\Omega_s$ that are free of open-string tachyons. For each entry, we specify the corresponding Chan--Paton multiplicity: for D5- and D1-branes, the configurations correspond to branes or antibranes of the two types, while the D4 and D3 entries each describe two families.}
    \label{tab:0B/Omega_prime_tachyon_free_branes}
\end{table}

\FloatBarrier

\subsection{D9-branes in type 0A/\texorpdfstring{$\Omega$}{Omega}}\label{sec:0A_parity}

In the orientifold procedure, one imposes the cancellation of RR tadpoles for consistency. This is what fixes the $\mathfrak{so}(32)$ gauge algebra in type I string theory and $\mathfrak{sp}(16)$ in the Sugimoto string. For the type 0 orientifolds, we have seen that RR tadpole cancellation is not equally restrictive: there are no constraints on the number of D9-branes in type 0A/$\Omega$ and type 0A/$\Omega_s$, while in type 0B/$\Omega$ and type 0B/$\Omega_s$, the four initial real multiplicities $n_i$ are restricted to $n_1=n_2\equiv k_1$ and $n_3=n_4\equiv k_2$, and the values of $k_{1,2}$ are not further constrained by consistency. Additional conditions can arise by requiring the cancellation of tachyon or dilaton tadpoles, but these are not required by consistency, and their presence is a signal of dynamical effects of supersymmetry breaking in string theory; see~\cite{Fischler:1986ci,Fischler:1986tb,Callan:1986bc,Tseytlin:1990mv,Dudas:2004nd,Raucci:2024fnp,Leone:2025mwo} for relevant literature on this topic.

However, the brane scan that we have provided, and in particular \cref{tab:type0_orientifold_branes}, reveals a peculiar feature that holds for type 0A/$\Omega$. As we show explicitly in \cref{tab:type0A/Omega_branes}, the D9-branes of type $-$, i.e., those labeled by $n_2$, carry a $\Z_2$ K-theory charge. This means that when $n_2$ is odd, there would be a space-filling $\Z_2$ charge: a K-theory tadpole. This must cancel, just like the RR tadpoles in more familiar setups~\cite{Uranga:2000xp}, which fixes the value of $n_2$ to be even. 
We saw a hint of the existence of this $\Z_2$ when extracting the multiplicity of tachyons in \cref{eq:0A_tachyons_D9}. There, we could have inferred that a single D9 of type $-$ is a stable configuration, while here we are finding that such a configuration would lead to a non-vanishing K-theory tadpole. 

One can see the inconsistency of $n_2$ odd from a probe argument, following~\cite{Uranga:2000xp}. Consider a charged D0-brane of labels $(d_1,d_2,m)=(1,0,0)$. This is of type $+$, and therefore it leads to massless fermions from the open strings connecting it to the D9-branes of type $-$, labeled by $n_2$. These give rise to $n_2$ real zero modes
\begin{eq}
    \acomm{\psi_a}{\psi_b}=\delta_{ab}\,,\qquad a=1,\dots,n_2\,,
\end{eq}
which are all odd under the worldvolume $O(1)\equiv \Z_2$ gauge symmetry living on the D0. The inconsistency of $n_2$ odd can be seen from this $\Z_2$ action: no operator that implements it can exist when $n_2$ is odd. In fact, the $\Z_2$ should correspond to an operator $U$ such that $U\psi_a U^{-1}=-\psi_a$ for every $a$, and therefore $U$ should anticommute with $\prod_a \psi_a$. However, when $n_2$ is odd, $\prod_a \psi_a$ commutes with all the $\psi_a$ and is thus a scalar on the irreducible module, which then implies that no operator $U$ can anticommute with it. Note that when $n_2$ is even, $U$ is the chirality operator and, as usual, the worldvolume $O(1)$ projects out one chirality of the Clifford module.

While this proves that $n_2$ must be even, there is another, more hidden condition that holds for type 0A/$\Omega$. In fact, the other D9 label, $n_1$, must also be even. 
This is not because of a K-theory charge---the D9-branes of type $+$ have no such charge, as can be seen from \cref{tab:type0A/Omega_branes}---but is due to an issue with the uncharged D0-branes. A single uncharged D0-brane, of type $-$, has a worldvolume $U(1)$ gauge symmetry and complex fermions in the vector representation of $\mathfrak{so}(n_1)$. 
In particular, since the uncharged D0 comes from a D0 and an anti-D0 of the same type in 0A, exchanged by the orientifold operator, strings from the $n_1$ D9-branes to the D0 and the anti-D0 give complex zero modes $\phi_i$ and $\chi_i$ respectively, with opposite charges under the $U(1)$. These are then exchanged by the orientifold operator, with the orientation reversal that identifies $\chi_i=\phi^\dagger_i$, so that $\chi^\dagger_i\chi_i=1-\phi^\dagger_i\phi_i$. The $U(1)$ charge operator must then be symmetric in both $\phi_i$ and $\chi_i$\footnote{This can change in the presence of a worldvolume Chern--Simons term; this can happen, for instance, in the presence of a non-vanishing Romans mass, for which the D0-branes have a worldvolume tadpole.} and, letting $\psi_i\propto \phi_i$, and normalizing so that $\psi_i^\dagger$ have charge $+1$,
\begin{eq}
    Q=\frac{1}{2}\sum_{i=1}^{n_1} \left(\phi_i^\dagger\phi_i-\chi^\dagger_i\chi_i\right)=\sum_{i=1}^{n_1}\left(\psi_i^\dagger \psi_i -\frac{1}{2}\right)=N-\frac{n_1}{2}\,.
\end{eq}
This means that one obtains $n_1$ complex zero modes $\psi_i$, such that $\acomm{\psi_i}{\psi_j^\dagger}=\delta_{ij}$.
Therefore, gauge-invariant states under the worldvolume symmetry only come from $\Lambda^{n_1/2}\mathbf{n}_1$. In particular, no gauge-invariant state exists when $n_1$ is odd. One is then forced to choose between imposing $n_1$ even and removing the uncharged D0 from the brane spectrum of the theory. We will choose the former option, since otherwise there would be a K-theory $\Z_2$ charge without a charged source in the type 0A classification (see~\cref{tab:type0A/Omega_branes}).

Note that another possible source of restrictions could be global anomalies on the worldvolumes of the branes, in the same way as the charged D0 constrains $n_2$. However, using the brane scan that we provided, one finds no other constraints of this type. For example, the D1-brane of type 0A/$\Omega$ has Majorana--Weyl fermions of both chiralities in the $\mathbf{n}_1$ of $\mathfrak{so}(n_1)$, which are odd under the worldvolume $O(1)$. Since these are $2n_1$ real modes, their parity leads to no non-trivial phase from their Pfaffians. The same holds in all other cases; in particular, this holds in the type 0B orientifolds because of the doubling of the D9-branes in each pair $\{1,2\}$ and $\{3,4\}$.

To summarize, we have obtained the constraint that the numbers of D9-branes in type 0A/$\Omega$ must be even,
\begin{eq}
    n_1\equiv n_2\equiv0\pmod{2}\,.
\end{eq}

\section{Gauge groups of type 0 orthogonal orientifolds}\label{sec:groups_orthogonal}

In this section, we extract the global forms of the gauge groups that are compatible with the perturbative-string states and the non-perturbative brane states for the orthogonal orientifolds of type 0A and 0B string theory. Some of the results of this section have already appeared in~\cite{Fraiman:2026ltu} for special cases,\footnote{In an early version of~\cite{Fraiman:2026ltu}, the authors did not notice the typos in the type 0A/$\Omega$ amplitudes of~\cite{Dudas:2001wd}. Their results for this orientifold are consistent with~\cite{Dudas:2001wd}, but not accurate in light of our corrected amplitudes. The type 0B/$\Omega$ result of~\cite{Fraiman:2026ltu} remains unchanged for the identity component.} and we skip some details that are already explained at length in that work.
We follow the general approach of~\cite{Witten:2023snr,Larotonda:2024thv,Fraiman:2026ltu}, and we fix our conventions about the centers of the simply connected covers in appendix~\ref{app:centers}. We begin with type 0A/$\Omega$ in \cref{sec:0A/Omega_gauge_group} and then proceed to the case of type 0B/$\Omega$ in \cref{sec:0B/Omega_gauge_group}. 

\subsection{Type \texorpdfstring{0A/$\Omega$}{0A/Omega}}\label{sec:0A/Omega_gauge_group}

The spacetime gauge algebra of this model is $\mathfrak{so}(n_1)\oplus\mathfrak{so}(n_2)$ with $n_1$ and $n_2$ even, as we discussed in \cref{sec:0A_parity}.
States from D9-branes give the two classes
\begin{equation}
    (o,o) \quad \text{and}\quad (v,v)\,.
\end{equation}
For the brane states, non-trivial classes can only come from quantizing fermion zero modes in vector representations of the spacetime gauge algebra. 

In general, quantizing $r$ fermion zero modes gives rise to a Clifford module. When $r$ is even, if there is a worldvolume $O(1)\cong\Z_2$ that flips all the modes, imposing gauge invariance under it projects out one chiral half. These considerations suffice to cover all cases for this type 0A orthogonal orientifold, with some care regarding the number of zero modes. In fact, while one real mode per spacetime vector index gives rise to a spacetime spinor, two real modes assemble into a complex creation operator, for which the Fock space is $\Lambda^\bullet \mathbf{n}$. The worldvolume $O(1)$ projection then selects forms of even degree; therefore, only the trivial class $o$ is generated. The remaining cases can be built from copies of these two basic ones. 

The above arguments apply to the D$p$-branes with $p\geq 1$ of type 0A/$\Omega$. For example, the relevant states for the tachyon-free uncharged D1-brane are $n_1$ left-moving and $n_1$ right-moving real fermions, all odd under the worldvolume $O(1)$, as can be seen in~\cref{tab:0A/Omega_D1_spectrum}. These can be combined into complex creation operators so that the Hilbert space is 
\begin{eq}
    \mathcal{H}_{\mathrm{D1}}=\bigoplus_{r\text{ even}}\Lambda^r \mathbf{n}_1\,,
\end{eq}
thus yielding the class $(o,o)$. Similarly, the D2 has two real zero modes for each pair of $n_1$ and worldvolume indices (see~\cref{tab:0A/Omega_D2_spectrum}), which gives rise to complex creation operators in the representation $\mathbf{n}_1\otimes \mathbf{d}$, and hence to the trivial class again. The D4 with $d=2l$ has worldvolume gauge algebra $\mathfrak{sp}(l)$ (see~\cref{tab:0A/Omega_D4_spectrum}), with four real modes for each Chan--Paton pair of indices, which again gives the trivial class. The additional tachyon-free branes for the special cases $n_1=0$ or $n_2=0$ are analogous and only give rise to the trivial classes.

The only branes that break central elements are the D0s. A charged D0 has real fermion zero modes $\psi_a$ with $a=1,\dots,n_2$. We have already seen them in \cref{sec:0A_parity}, where they force $n_2$ to be even. Since $n_2$ is even, quantizing them gives the Clifford module $\text{\bf{S}}\oplus\text{\bf{C}}$, and the worldvolume $O(1)$ that flips all the modes keeps the spinorial class (of the second $\mathfrak{so}$ factor). Therefore, a single brane gives rise to the class $(o,s)$, and stacking more branes produces no further classes.

The other tachyon-free D0 is the uncharged one, of type $-$. This also played a role in \cref{sec:0A_parity}, constraining $n_1$ to be even, and we have discussed its quantization already in that section: there are complex modes $\psi_i$ charged under the worldvolume $U(1)$ gauge symmetry (see~\cref{tab:0A/Omega_D0_spectrum}), and the charge operator is 
\begin{eq}\label{eq:charge_operator}
    Q=N-\frac{n_1}{2}\,,
\end{eq}
where $N$ is the number operator. 
Since $n_1$ is even, the Hilbert space is 
\begin{eq}\label{eq:0A_D0_Hilbert_Space}
    \mathcal{H}_{\text{D0, uncharged}}=\Lambda^{n_1/2}\mathbf{n}_1  \,,   
\end{eq}
giving rise to the class $\left(\frac{n_1}{2}v,o\right)$, which is thus determined by the value of $n_1$ modulo $4$.

In addition to the brane states, this type 0A orientifold contains a tachyon-free D-instanton. As usual, it does not break central elements, but it can constrain the disconnected components of the gauge groups. This D$(-1)$ has $n_2$ real zero modes $\eta_a$, so that the measure transforms as 
\begin{eq}
    \prod_{a=1}^{n_2}\dd \eta_a\to (\det g_2)\prod_{a=1}^{n_2}\dd \eta_a\,,
\end{eq}
where, as in~\cite{Fraiman:2026ltu}, $g_2\in O(n_2)$ and $\det g_2$ is the determinant. This has the effect of breaking the reflection of the second factor. Note that this reflection is also broken by the charged D0, and the whole mechanism is analogous to type I string theory, where both the D0 and the D$(-1)$ break the reflection of $O(32)$.

To summarize, perturbative states belong to the classes
\begin{eq}
    (o,o)\quad\text{and}\quad(v,v)\,,
\end{eq}
while brane states add the classes
\begin{eq}
    (o,s)\quad\text{and}\quad\left(\frac{n_1}{2}v,o\right)
\end{eq}
under the spacetime gauge algebra $\mathfrak{so}(n_1)\oplus\mathfrak{so}(n_2)$, together with classes built by combining them. If one considers the perturbative spectrum alone, the subset of the center that acts trivially is
\begin{equation}
    \{(1,1),(z_v,1),(1,z_v),(z_v,z_v),(z_c,z_s),(z_s,z_c),(z_s,z_s),(z_c,z_c)\}\,.
\end{equation}
We now include the brane states for all possible combinations of the values of $n_1$ and $n_2$. We consider generic cases and special ones in which one of the two values vanishes, but our notation does not include the cases $n_{1,2}=2$, for which the spacetime algebra would contain a $\mathfrak{u}(1)$ factor. This last case will be the subject of a separate work, since it is similar to the two type 0B orientifolds that we are not considering here.

\subsubsection{The case $\mathfrak{so}(n_1=4n)\oplus\mathfrak{so}(n_2=4k)$}

In this case, the only additional class from the quantization of the brane modes is $(o,s)$. 
The subset of the center that acts trivially on the full spectrum is then
\begin{eq}
    \{(1,1),(z_v,1),(z_c,z_s),(z_s,z_s)\}\,,
\end{eq}
which is isomorphic to a $\Z_2\times \Z_2$ generated by $(z_v,1)$ and $(z_s,z_s)$.
The gauge group is then
\begin{eq}
    \frac{O(4n)\times Spin(4k)}{\langle(-1,z_s)\rangle}\,,
\end{eq}
where we use that $Spin(n_1)/{\langle z_v\rangle}\, \cong SO(n_1)$, and where $-1$ is the image of $z_s$ under the quotient.

\subsubsection{The case $\mathfrak{so}(n_1=4n+2)\oplus\mathfrak{so}(n_2=4k)$}

When $n_1\equiv 2\pmod{4}$ and $n_2\equiv 0\pmod{4}$, the D0-branes provide the classes $(v,o)$ and $(o,s)$.
Therefore, the unbroken subgroup of the center is
\begin{equation}
    \{(1,1),(z_v,1)\}\,,
\end{equation}
which is the $\Z_2$ such that $Spin(n_1)/\Z_2 \cong SO(n_1)$.
Taking into account the disconnected component, the gauge group is
\begin{eq}
    O(4n+2)\times Spin(4k)\,.
\end{eq}

\subsubsection{The case $\mathfrak{so}(n_1=2n)\oplus\mathfrak{so}(n_2=4k+2)$}

When $n_2\equiv 2\pmod{4}$, we have the class $(o,s)$. An interesting feature of this case is that the class $(v,o)$ is always present, regardless of the presence of $(v,o)$ from the uncharged D0. Indeed, we have that
\begin{eq}
    2(o,s) = (o,v) \quad\text{and}\quad (v,v)+(o,v) = (v,o)\,.
\end{eq}
The unbroken subgroup of the center is again
\begin{eq}
    \{(1,1),(z_v,1)\}\,,
\end{eq}
and for all the cases with $n_2\equiv 2\pmod{4}$, the gauge group is 
\begin{eq}
    O(2n)\times Spin(4k+2)\,.
\end{eq}

\subsubsection{The special case $\mathfrak{so}(n_1=2n)$} 

Consider now the case of vanishing $n_2$, so that we only have the gauge algebra $\mathfrak{so}(n_1)$. This means that only D9-branes of the $+$ type are present, $n_1=2n$, and all the states from these D9-branes are in the class $o$.
The quantization of the brane modes is much simpler since we only have D9s of one type, and we get the class $nv$. This corresponds to
\begin{eq}
    & o \qquad & \text{if }n\equiv0\pmod{2} \,,\\
    & v\qquad & \text{if }n\equiv1\pmod{2}\,.
\end{eq}
In the first case, we can take the quotient by the full center; in the second case, the only elements that act trivially are $\{1,z_v\}$. The D-instanton does not break the reflection element of the first factor, as in the general case above, and the global form of the gauge group is
\begin{eq}\label{eq:type0A_orthogonal_special_n20}
    & O(2n)/\{\pm1\} \qquad & \text{if }n\equiv0\pmod{2} \,,\\
    & O(2n)\qquad & \text{if }n\equiv1\pmod{2}\,.
\end{eq}

\subsubsection{The special case $\mathfrak{so}(n_2=2k)$} 

The other possibility is $n_1=0$ and $n_2=2k$. States from the D9s lie in the trivial class $o$, while the quantization of the brane modes adds the class $s$. Therefore, when $k\equiv0\pmod{2}$, the subgroup of the center that acts trivially is $\Z_2 =\{1,z_s\}$; when $k\equiv1\pmod{2}$, the center is completely broken.
The gauge groups are then
\begin{eq}
    & Spin(2k)/\langle z_s \rangle \qquad & \text{if }k\equiv0\pmod{2} \,,\\
    & Spin(2k)\qquad & \text{if }k\equiv1\pmod{2}\,.
\end{eq}

\subsection{Type \texorpdfstring{0B/$\Omega$}{0B/Omega}}\label{sec:0B/Omega_gauge_group}

The spacetime gauge algebra of this orientifold is $\mathfrak{so}(n_1)\oplus\mathfrak{so}(n_2)\oplus\mathfrak{so}(n_3)\oplus\mathfrak{so}(n_4)$ and, as we have discussed in \cref{sec:type_0_review}, RR tadpole cancellation fixes the Chan--Paton multiplicities to be $(k_1,k_1,k_2,k_2)$. We will sometimes denote the four factors by $n_i$, when it helps improve the presentation, and denote the two families of brane-antibrane pairs as $\{1,2\}$ and $\{3,4\}$. As in the previous type 0A orientifold, our formulas will not always be applicable for small values of $k_i$, and the cases $k_i=2$ will be the subject of a separate work.   

Perturbative states from the D9-branes generate the classes $(\varepsilon_1,\varepsilon_2,\varepsilon_3,\varepsilon_4)$, where two $\varepsilon_i$ are equal to $v$ and the other two are equal to $o$, with the understanding that $v\to o$ when the corresponding $k_i$ is odd (the vector representation is in the trivial class for odd $k_i$). To find the brane states, we have to scan the list in \cref{tab:0B/Omega_tachyon_free_branes}. Note that the system is symmetric in $k_1\leftrightarrow k_2$, so that every result that applies to $k_1$ also holds for $k_2$, allowing one to focus on a single brane-antibrane type.

It turns out that, also for this orientifold, the D0-branes suffice to fix the (identity component of the) gauge group. There are two types of D0s. The first, of type $+$, has a real fermion zero mode for each vector index of $\mathfrak{so}(n_3)\oplus\mathfrak{so}(n_4)=\mathfrak{so}(k_2)^{\oplus 2} $, which means $2k_2$ modes, and the worldvolume $O(1)\cong\Z_2$ keeps one chiral spinor of $Spin(2k_2)$; see \cref{tab:0B/Omega_D0_spectrum}. Hence,
\begin{eq}
    \mathcal{H}_{\mathrm{D0}_{(1,0)}}=\begin{cases}
        (\mathbf{S}\otimes\mathbf{S})\oplus(\mathbf{C}\otimes\mathbf{C}) \qquad & k_2 \text{ even}\,,\\
        \mathbf{S'}\otimes\mathbf{S'}\qquad & k_2 \text{ odd}\,.
    \end{cases}
\end{eq}
This expression also holds for $k_2=1$, with the understanding that $\mathbf{S'}$ is the one-dimensional singlet. The other D0-brane, of type $-$, leads to the same results for $k_1$, by the symmetry $k_1\leftrightarrow k_2$. To summarize the resulting classes, denoting both $k_{1,2}$ as $k$, for even $k\geq4$ we obtain the classes $(s,s,o,o)$ and $(c,c,o,o)$, which differ by $(v,v,o,o)$ (already present from the perturbative states), and $(o,o,s,s)$ and $(o,o,c,c)$, which differ by $(o,o,v,v)$ (already present from the perturbative states). For odd $k$, we obtain the classes $(s',s',o,o)$ and $(o,o,s',s')$. Therefore, the entire contribution of the two D0-branes can be summarized as adding to the perturbative states the classes\footnote{This holds for $k>2$. When $k=1$, one obtains $(o,o,o,o)$. Here, we are using $s$ to denote $s'$ when $k$ is odd to make the expressions clearer.}
\begin{eq}
    (s,s,o,o)\,,\qquad (o,o,s,s)\,.
\end{eq}
Note that a transformation from one $\mathfrak{so}$ factor with $\det=-1$ flips an odd number of modes and is thus excluded by these classes. Instead, a product of two transformations with determinant $\det=-1$ from the pairs $\{1,2\}$ or $\{3,4\}$ is allowed. For $k$ even, this can be represented by a reflection of one coordinate in each sector, $(R,R)$, while for $k$ odd, one can use the transformation $(-1,-1)$.

Consider now the charged D1- and D5-branes. For the D1s, one can take only branes; namely, D1s with labels $(d_1,0,d_3,0)$ (see \cref{tab:0B/Omega_D1_spectrum}). Those that correspond to $d_1$ have one real left-moving fermion in the $\mathbf{n}_4$ vector representation and one real right-moving fermion in the $\mathbf{n}_3$ vector representation; analogous fermions arise from branes that correspond to $d_3$ in $\mathbf{n}_{1,2}$ vector representations. From this, one obtains that a single D1 has the same Clifford algebra as the D0 of the same type,\footnote{This can be seen, for example, from the realization of the D0 as a brane-antibrane kink, which ultimately is the K-theory classification of branes.} with the $2k$ modes, and no further classes arise. Extending this to more D1-branes does not add classes.
The charged D5-branes lead to an even simpler spectrum. Consider D5-branes of one type, with worldvolume $Sp(l)$ gauge symmetry and with 6d Weyl fermions of opposite chirality in the two factors of the other type of D9 (see \cref{tab:0B/Omega_D5_spectrum}). Using the symplectic-Majorana condition, one obtains four real zero modes per Chan--Paton pair and, as in~\cite{Fraiman:2026ltu}, only classes $(o,o,o,o)$, $(v,v,o,o)$, and $(o,o,v,v)$.

For the special case with $k_1=0$ (recall the symmetry $k_1\leftrightarrow k_2$), we must also consider the single D7 and D8, which give rise to fermions charged under the pair $\{3,4\}$. The D8 has a worldvolume $O(1)$ gauge symmetry and 16 real zero modes per vector index. The $\Z_2$ projection keeps states with an even number of excitations, giving only the classes $(o,o)$ and $(v,v)$. The D7 has a worldvolume $U(1)$ gauge symmetry, with complex fermion zero modes whose creation operators carry charge $+1$ in $\mathbf{n}_3$ and $-1$ in $\mathbf{n}_4$. This is analogous to the uncharged D0-brane of type 0A/$\Omega$: each complex mode contributes to the charge operator as in \cref{eq:charge_operator}, but since $n_3=n_4=k_2$, the two sets have the same multiplicity and the constant contributions cancel,
\begin{eq}
    Q=N_3-N_4\,.
\end{eq}
Worldvolume gauge invariance requires $N_3=N_4$, thus yielding the classes $(o,o)$ and $(v,v)$.

Finally, one must include the effects of the two types of D-instantons, with $2k_2$ zero modes charged under the D9 pair $\{3,4\}$ and $2k_1$ zero modes charged under the D9 pair $\{1,2\}$. This imposes two conditions on the determinants of allowed transformations,\footnote{Note that the $U(1)$ on the D-instanton worldvolume adds no further phases to the measure. For example, take the D-instanton with $2k_2$ zero modes: the $k_2$ modes in $\mathbf{n}_3$ and the $k_2$ modes in $\mathbf{n}_4$ have opposite $U(1)$ charges, and the measure is $U(1)$-invariant.}
\begin{eq}\label{eq:0B_det_condition}
    \det g_1 \, \det g_2=1 \qquad \text{and} \qquad  \det g_3 \, \det g_4=1\,.
\end{eq}

In summary, for the type 0B/$\Omega$ orientifold, we have the classes $(s,s,o,o)$ and $(o,o,s,s)$ in addition to the perturbative ones from D9-branes. The perturbative classes imply that a central element $(a_1,a_2,a_3,a_4)$ must act with the same sign on all four vector representations: if we denote by $\chi_r$ the action of the central elements on the class $r$ (see \cref{tab:actions_of_Z_2k+1,tab:actions_of_Z_4k,tab:actions_of_Z_4k+2}), the perturbative spectrum implies that
\begin{eq}
    \chi_v(a_1)=\chi_v(a_2)=\chi_v(a_3)=\chi_v(a_4)\in\{\pm1\}\,,
\end{eq}
while from the D0 states we obtain the conditions
\begin{eq}
    \chi_s(a_1)\chi_s(a_2)=1\,,\qquad \chi_s(a_3)\chi_s(a_4)=1\,.
\end{eq}
If $\chi_v(a_i)=1$, this leaves only the subgroup generated by the central elements $(z_v,z_v,1,1)$ and $(1,1,z_v,z_v)$, and we denote the resulting connected component of the group by
\begin{eq}
    H_{k_1}^\circ\times H_{k_2}^\circ\,,
\end{eq}
where
\begin{eq}
    H_k^\circ=\frac{Spin(k)\times Spin(k)}{\langle(z_v,z_v)\rangle}\,.
\end{eq}
If instead $\chi_v(a_i)=-1$, which can only happen when $k_1$ and $k_2$ are both even, there is a further quotient to be imposed. To characterize this, embed $H_k^\circ$ in $Spin(2k)$ and let $z_s^{(2k)}$ be the central element that acts trivially on $\mathbf{S}_{2k}$. When $k\equiv 0\pmod{4}$, this is $(z_s,z_s)$, while when $k\equiv 2\pmod{4}$, this is $(z_s,z_c)$. With this definition, the additional quotient is generated by $\delta=(z_s^{(2k_1)},z_s^{(2k_2)})$, and the identity component of the resulting group is
\begin{eq}
    \frac{H_{k_1}^\circ\times H_{k_2}^\circ}{\langle\delta\rangle}\,.
\end{eq}

Explicitly,
\begin{itemize}
    \item when $k_1\equiv k_2\equiv 0\pmod{4}$, the identity component of the group is
    \begin{eq}
        G^\circ_{0,0}=\frac{Spin(k_1)^2\times Spin(k_2)^2}{\Z_2^3}\,,
    \end{eq}
    with generators $(z_v,z_v,1,1)$, $(1,1,z_v,z_v)$, and $(z_s,z_s,z_s,z_s)$;
    \item when $k_1\equiv0\pmod{4}$ and $k_2\equiv 2\pmod{4}$, the identity component of the group is
    \begin{eq}
        G^\circ_{0,2}=\frac{Spin(k_1)^2\times Spin(k_2)^2}{\Z_4\times \Z_2}\,,
    \end{eq}
    with generators $(z_s,z_s,z_s,z_c)$ and $(z_v,z_v,1,1)$;
    \item when $k_1\equiv k_2\equiv 2\pmod{4}$, the identity component of the group is
    \begin{eq}
        G^\circ_{2,2}=\frac{Spin(k_1)^2\times Spin(k_2)^2}{\Z_4\times \Z_2}\,,
    \end{eq}
    with generators $(z_s,z_c,z_s,z_c)$ and $(z_v,z_v,1,1)$;
    \item when at least one of $k_1$ and $k_2$ is odd, the identity component of the group is
    \begin{eq}
        G^\circ_{\text{odd}}=\frac{Spin(k_1)^2\times Spin(k_2)^2}{\Z_2\times \Z_2}\,,
    \end{eq}
    with generators $(z_v,z_v,1,1)$ and $(1,1,z_v,z_v)$;
    \item when one of the two vanishes, the identity components of the groups in the three cases $k\equiv0\pmod4$, $k\equiv2\pmod4$, and $k $ odd are
    \begin{eq}
        \frac{Spin(k)^2}{\Z_2\times \Z_2}\,,\qquad \frac{Spin(k)^2}{\Z_4}\,,\qquad \frac{Spin(k)^2}{\Z_2}\,,
    \end{eq}
    with generators $(z_v,z_v)$ and $(z_s,z_s)$, $(z_s,z_c)$, and $(z_v,z_v)$ respectively.
\end{itemize}

In general, the groups will have other disconnected components, but obtaining them is more involved. This is the topic of the next subsection, in which we provide a partial understanding of these groups. 

\subsubsection{The disconnected component}

The subtlety in the analysis of the disconnected components is the action of the reflections on the D0 states. In fact, an order-two transformation in the orthogonal groups can lift to an order-four operator on the spinors. To see this, consider the splitting $\R^{2k}=\R^k\oplus \R^k$. A transformation that preserves each factor is $\mathrm{diag}(A,B)$ with $A,B\in O(k)$. From the condition on the determinants in \cref{eq:0B_det_condition}, the allowed transformations are 
\begin{eq}
    S\left(O(k)\times O(k)\right)=\{\mathrm{diag}(A,B) \ : \ A,B\in O(k),\, \det A \, \det B=1\}\,,
\end{eq}
which is a subgroup of $SO(2k)$. In our case, the $2k$ zero modes from the D0-branes generate one Clifford algebra, as we have discussed. Denoting by $\gamma_i$ and $\gamma_{i}'$ the generators for the two blocks, normalized as $\acomm{\gamma_i}{\gamma_j}=\acomm{\gamma_i'}{\gamma_j'}=2\delta_{ij}$---they anticommute because there is a single Clifford algebra---a reflection in the $i=1$ direction admits the spin lift 
\begin{eq}
    \rho=\gamma_1\gamma_1'\,.
\end{eq}
Since the two generators anticommute, $\rho^2=-1=z_v^{(2k)}$, which means that $\rho$ has order four on the spinors. This fact has a simple geometric interpretation in the $\R^{2k}$ space, since $(R,R)$ is a $\pi$ rotation in the plane spanned by the two reflected coordinates, and squaring it gives the spin lift of a $2\pi$ rotation.

Given this fact, one would like to add the component of the group containing $\rho$ to the identity components of the previous section. A naive guess would be to consider the double-cover map $\pi:Spin(2k)\to SO(2k)$ and define the group
\begin{eq}\label{eq:definition_of_H}
    H_k=\pi^{-1}\left(S\left(O(k)\times O(k)\right)\right)\,,
\end{eq}
which explicitly reads
\begin{eq}
    H_k=\{g\in Spin(2k) \ : \ \pi(g)=\mathrm{diag}(A,B), \, A,B\in O(k), \, \det A \, \det B = 1 \}\,.
\end{eq}
Note the chosen convention: $H_k^\circ$ is the identity component of $H_k$. In fact, the group of the type 0B orientifold relevant for the Bergman--Gaberdiel duality has been expressed in terms of these $H_k$ groups in~\cite{Fraiman:2026ltu}.

However, it is not clear whether $H_k$ is the correct extension in general. The issue is that $O(k)$ has double covers $Pin^+(k)$ and $Pin^-(k)$, and our analysis does not distinguish between the two. In the following, we will arrive at an expression for the group without further assumptions on the branes, and additional information is needed to select a specific form of the groups---for example, since the D0-brane is its own antibrane, one could try to investigate the effects of some reality conditions on its worldvolume.

To illustrate where the subtlety lies, consider $H_k^\circ$. The kernel of its covering map is $\{1,z\}$, with 
\begin{eq}
    z=(z_v,1)=(1,z_v)\in H_k^\circ\,.
\end{eq}
To specify how $(R,R)$ is implemented, one must know $\rho^2$. For example, in $H_k$,
\begin{eq}
    \rho^2=\gamma_1\gamma_1'\gamma_1\gamma_1'=-\gamma_1^2{\gamma_1'}^2=-1=z\,
\end{eq}
from the anticommutation relations. However, one could multiply the operator representing the two reflections by a phase, so that its square would be $\rho^2=1$. This corresponds to the two choices $Pin^\pm$ as the double covers of $O$. Therefore, in the definition of the complete groups with the disconnected components included, one must specify which, among $Pin^\pm$, has to be chosen. Actually, there is some correlation among the choices because of the $\langle (z_v,z_v)\rangle$ quotient: if the lift of $(R,R)$ is $g=(e,f)$ with $e^2=\epsilon_e$ and $f^2=\epsilon_f$, where $-1$ corresponds to $z_v$, then $g^2=(\epsilon_e,\epsilon_f)$. However, $(1, z_v\equiv -1)\sim (z_v,1)$ and $(1,1)\sim (z_v,z_v)$, so that only a single relative sign remains meaningful. Therefore, we can define the groups
\begin{eq}
    \mathcal{E}_k^\epsilon=\frac{\{(u,v)\in Pin^+(k)\times Pin^\epsilon(k) \ : \ \det \pi(u)=\det \pi(v) \}}{\langle( z_v,z_v)\rangle}\,,
\end{eq}
where $\epsilon=\pm1$ and $\pi:Pin^\pm(k)\to O(k)$. The gauge groups of the type 0B orthogonal orientifold are then possible quotients of 
\begin{eq}\label{eq:0B_orthogonal_groups}
    G=\begin{cases}
        \left(\mathcal{E}_{k_1}^{\epsilon_1}\times \mathcal{E}_{k_2}^{\epsilon_2}\right)/{\langle\delta\rangle} \qquad &\text{if }k_{1,2} \text{ are even}\,,\\
        \mathcal{E}_{k_1}^{\epsilon_1}\times \mathcal{E}_{k_2}^{\epsilon_2}\qquad &\text{otherwise}\,,
    \end{cases}
\end{eq}
where $\epsilon_{1,2}$ are unspecified signs that are to be fixed by further conditions not included in our analysis. We stress that the actual groups may involve additional quotients involving the disconnected components, and our result of \cref{eq:0B_orthogonal_groups} should be interpreted as an ``upper bound'' for invariant centers.

\section{Gauge groups of type 0 symplectic orientifolds}\label{sec:groups_symplectic}

In this section, we discuss the gauge groups of the symplectic orientifolds of type 0 strings. All the amplitudes that will be used in the following can be obtained from \cref{sec:type_0_review} after setting $\sigma=-1$. Since these orientifold models are not often discussed in the literature, we collect the relevant information about the low-lying spectra in appendix~\ref{app:0A/Omega_s_branes} for the type 0A orientifold and in appendix~\ref{app:0B/Omega_s_branes} for the type 0B orientifold. As we mentioned in \cref{sec:branes_type_0}, we denote the symplectic orientifolds by $\Omega_s$.

\subsection{Type \texorpdfstring{0A/$\Omega_s$}{0A/Omega s}}\label{sec:s0A_gauge_group}

For the symplectic orientifold of type 0A, the spacetime gauge algebra is $\mathfrak{sp}(n_1/2)\oplus\mathfrak{sp}(n_2/2)$, where the Chan--Paton labels $n_1$ and $n_2$ of the background D9-branes must be even by construction. As in type 0A/$\Omega$, the two types of D9-branes are not symmetric: there are tachyons in the antisymmetric representation of the first factor and in the symmetric (adjoint) representation of the second. The perturbative spectrum includes fermions in the bifundamental representation, so we obtain the class $(v,v)$ from this sector. To analyze the brane states, we now consider the branes of \cref{tab:sOA/Omega_tachyon_free_branes}. We are not aware of a similar discussion in the literature, and therefore we will discuss each case individually, including more details than in the corresponding orthogonal setups.

\subsubsection{D0 quantization}\label{sec:s0A_D0}

\begin{table}[H]
\centering
\renewcommand{\arraystretch}{1.2}
\begin{tabular}{l|c|c|c}
Worldvolume state  & $\mathfrak{so}(9)_T$ & $\mathfrak{sp}(\tfrac{n_1}{2})\oplus\mathfrak{sp}(\tfrac{n_2}{2})$ & worldvolume $\mathfrak{sp}(k)$ \\
\hline
Fermion $\psi$ & $\mathbf{1}$ & $ (\mathbf{1}, \mathbf{n_2}) $ & $ \mathbf{2k} $
\end{tabular}
\caption{Open-string light spectrum charged under the gauge symmetry from a D0-brane stack with $d_1=2k$ and $d_2=m=0$ in the symplectic orientifold of type 0A string theory.}
\label{tab:s0A/Omega_chargedD0_relevant_spectrum}
\end{table}

Consider first charged D0-branes. A single such brane has multiplicity two ($k=1$ in \cref{tab:s0A/Omega_chargedD0_relevant_spectrum}) and a worldvolume $Sp(1)$ gauge symmetry. The fermion zero modes charged under the spacetime gauge sector are in the bifundamental of the worldvolume $\mathfrak{sp}(1)$ and spacetime $\mathfrak{sp}\left(\frac{n_2}{2}\right)$, thus only transforming under the second factor. This is consistent with the fact that the charged D0 is a brane of type $+$; see \cref{sec:type_0_review}.
For the quantization of the fermion zero modes, we use the following result, which we prove in appendix~\ref{app:symplectic_proof}: given $4mn$ real fermionic modes in the $(\mathbf{2m},\mathbf{2n})$ of $\mathfrak{sp}(m)\oplus \mathfrak{sp}(n)$, and given the Clifford module $M$ they generate, every weight $\lambda$ of $M$ is such that 
\begin{eq}\label{eq:weights_result_spsp}
    [\lambda]_{\mathfrak{sp}(m)}+[\lambda]_{\mathfrak{sp}(n)}\equiv mn\pmod{2}\,,
\end{eq}
where the classes $o$ and $v$ are represented as $o\to 0$ and $v\to 1$.
From this result, it follows that states transforming trivially under the worldvolume gauge symmetry generate the spacetime class 
\begin{eq}
    \left(o,\frac{n_2}{2}v\right)=\begin{cases}
        (o,o)\qquad  &\text{if }n_2\equiv 0\pmod{4}\,,\\
        (o,v)\qquad  &\text{if }n_2 \equiv 2\pmod{4}\,.
    \end{cases}
\end{eq}
With a stack of D0-branes, we can still use \cref{eq:weights_result_spsp}, which leads to no additional center classes beyond those of the single D0.

\subsubsection{D2 quantization}\label{sec:s0A_D2}

There are two types of D2-branes in this model. Consider first the charged D2 with symplectic worldvolume gauge symmetry, whose relevant charged fermion spectrum is in \cref{tab:s0A/Omega_chargedD2_relevant_spectrum}.

\begin{table}[H]
\centering
\renewcommand{\arraystretch}{1.2}
\begin{tabular}{l|c|c|c}
Worldvolume state  & $\mathfrak{so}(7)_T$ & $\mathfrak{sp}(\tfrac{n_1}{2})\oplus\mathfrak{sp}(\tfrac{n_2}{2})$ & worldvolume $\mathfrak{sp}(k)$ \\
\hline
Fermion $\psi$ & $\mathbf{1}$ & $ (\mathbf{n_1},\mathbf{1}) $ & $ \mathbf{2k} $
\end{tabular}
\caption{Open-string light spectrum charged under the gauge symmetry from a D2-brane stack with $d_1=2k$ and $d_2=m=0$ in the symplectic orientifold of type 0A string theory.
}
\label{tab:s0A/Omega_chargedD2_relevant_spectrum}
\end{table}

The quantization of the fermion zero modes proceeds as in the charged D0 case, with the difference that in three dimensions there is an extra factor of two in the spinor multiplicity. The version of the theorem in appendix~\ref{app:symplectic_proof} with two copies, now applied to the first spacetime gauge factor, leads to the trivial class only, $(o,o)$.

The second type of D2-brane is a single uncharged D2. It is of type $+$, and thus it gives rise to fermions charged under the second spacetime $\mathfrak{sp}$ factor. It has a worldvolume $U(1)$ gauge symmetry.

\begin{table}[H]
\centering
\renewcommand{\arraystretch}{1.2}
\begin{tabular}{l|c|c|c}
Worldvolume state  & $\mathfrak{so}(7)_T$ & $\mathfrak{sp}(\tfrac{n_1}{2})\oplus\mathfrak{sp}(\tfrac{n_2}{2})$ & worldvolume $\mathfrak{u}(1)$ \\
\hline
Fermion $\psi$ & $\mathbf{1}$ & $ (\mathbf{1},\mathbf{n_2}) $ & $\bar{q}$ \\
Fermion $\chi$ & $\mathbf{1}$ & $ (\mathbf{1},\mathbf{n_2}) $ & $q$
\end{tabular}
\caption{Open-string light spectrum charged under the gauge symmetry from a D2-brane stack with $d_1 = d_2 = 0$ and $m=1$ in the symplectic orientifold of type 0A string theory.}
\label{tab:s0A/Omega_chargedD2_second_type_relevant_spectrum}
\end{table}

The quantization proceeds as in the other cases with unitary worldvolume groups (see, e.g., the D0 of type 0A/$\Omega$), but since there are two complex zero modes for each index of $\mathbf{n}_2$, the charge operator is
\begin{eq}
    Q=N-n_2\,.
\end{eq}
The only resulting class is then the trivial one, $(o,o)$.

\subsubsection{D3 quantization}

The single D3 from \cref{tab:sOA/Omega_tachyon_free_branes} has a worldvolume $O(1) \equiv \Z_2$ gauge symmetry with charged fermion modes under the spacetime $\mathfrak{sp}(n_2/2)$.

\begin{table}[H]
\centering
\renewcommand{\arraystretch}{1.2}
\begin{tabular}{l|c|c|c}
Worldvolume state  & $\mathfrak{so}(6)_T$ & $\mathfrak{sp}(\tfrac{n_1}{2})\oplus\mathfrak{sp}(\tfrac{n_2}{2})$ & worldvolume $O(1)$ \\
\hline
Fermion $\psi$ & $\mathbf{1}$ & $ (\mathbf{1},\mathbf{n_2}) $ & odd
\end{tabular}
\caption{Open-string light spectrum charged under the gauge symmetry from a D3-brane stack with $d_1 = 1$ and $d_2=0$ in the symplectic orientifold of type 0A string theory.}
\label{tab:s0A/Omega_chargedD3_relevant_spectrum}
\end{table}

The quantization proceeds as in the D5-brane case of \cite{Larotonda:2024thv}, thus giving only the trivial class $(o,o)$.

\subsubsection{D4 quantization}

The quantization of fermion zero modes in the case of the charged D4-brane proceeds as for the D5 of \cite{Larotonda:2024thv}. We have to quantize fermions in the bifundamental $(\mathbf{n_2}, \mathbf{d_1})$ of $\mathfrak{sp}(\tfrac{n_2}{2})\oplus\mathfrak{so}(d_1)$. Imposing the symplectic-Majorana condition on the fermions reduces the degrees of freedom of the representation $(\mathbf{n_2},\mathbf{d_1})$ of $ \mathfrak{sp}(\tfrac{n_2}{2}) \oplus \mathfrak{so}(d_1)$ to the $(\mathbf{\tfrac{n_2}{2}},\mathbf{d_1})$ of $\mathfrak{so}(\tfrac{n_2}{2}) \oplus \mathfrak{so}(d_1)$. Indeed, the special embedding
\begin{equation}
    \mathfrak{sp}\left({\tfrac{n_2}{2}}\right) \supset \mathfrak{sp}(1) \oplus \mathfrak{so}\left({\tfrac{n_2}{2}}\right) 
\end{equation}
gives the $\mathfrak{sp}(1) \cong \mathfrak{su}(2)$ factor that maps the two real parts of the fermion into each other, and the symplectic tensor is given by $\omega = \epsilon \otimes \delta $, where $\epsilon$ is the invariant tensor of $\mathfrak{su}(2)$ and $\delta$ of $\mathfrak{so}\left({\tfrac{n_2}{2}}\right) $. Therefore, we must quantize the $(\mathbf{\tfrac{n_2}{2}},\mathbf{d_1})$ of $\mathfrak{so}({\tfrac{n_2}{2}}) \oplus \mathfrak{so}(d_1)$, imposing 
\begin{equation}
    \{\psi_{i,\alpha}, {\psi}^\dagger_{j,\beta}\} = \delta_{ij}\delta_{\alpha\beta}\,,
\end{equation}
where $i,j = 1,\dots,\tfrac{n_2}{2}$ are the $\mathfrak{so}(\tfrac{n_2}{2})$ indices and $\alpha,\beta = 1,\dots,d_1$ are the $\mathfrak{so}(d_1)$ ones. In total, this gives $4 \frac{n_2}{2} d_1$ independent complex modes after taking into account all the spinor components.
Following \cite{Larotonda:2024thv}, one obtains the class $(o,o)$.

The uncharged D4-brane is similar to the uncharged D2, with a worldvolume $U(1)$ gauge symmetry; see \cref{tab:s0A/Omega_chargedD4_second_type_relevant_spectrum}.
\begin{table}[H]
\centering
\renewcommand{\arraystretch}{1.2}
\begin{tabular}{l|c|c|c}
Worldvolume state  & $\mathfrak{so}(5)_T$ & $\mathfrak{sp}(\tfrac{n_1}{2})\oplus\mathfrak{sp}(\tfrac{n_2}{2})$ & worldvolume $\mathfrak{u}(1)$ \\
\hline
Fermion $\psi$ & $\mathbf{1}$ & $ (\mathbf{n_1},\mathbf{1}) $ & $\bar{q}$ \\
Fermion $\chi$ & $\mathbf{1}$ & $ (\mathbf{n_1}, \mathbf{1}) $ & $q$
\end{tabular}
\caption{Open-string light spectrum charged under the gauge symmetry from a D4-brane stack with $d_1 = d_2 = 0$ and $m=1$ in the symplectic orientifold of type 0A string theory.}
\label{tab:s0A/Omega_chargedD4_second_type_relevant_spectrum}
\end{table}
Quantizing the fermion zero modes gives the class $(o,o)$.

\subsubsection{D5 quantization}

The single D5-brane that is free of open-string tachyons is analogous to the D3, with an $O(1) \equiv \Z_2$ gauge symmetry on its worldvolume; see \cref{tab:s0A/Omega_chargedD5_relevant_spectrum}.
\begin{table}[H]
\centering
\renewcommand{\arraystretch}{1.2}
\begin{tabular}{l|c|c|c}
Worldvolume state  & $\mathfrak{so}(4)_T$ & $\mathfrak{sp}(\tfrac{n_1}{2})\oplus\mathfrak{sp}(\tfrac{n_2}{2})$ & worldvolume $O(1)$ \\
\hline
Fermion $\psi$ & $\mathbf{1}$ & $ (\mathbf{n_1}, \mathbf{1}) $ & odd
\end{tabular}
\caption{Open-string light spectrum charged under the gauge symmetry from a D5-brane stack with $d_1 = 0$ and $d_2=1$ in the symplectic orientifold of type 0A string theory.}
\label{tab:s0A/Omega_chargedD5_relevant_spectrum}
\end{table}
The quantization proceeds as for the D5 of \cite{Larotonda:2024thv} with $4 \frac{n_1}{2}$ complex modes, only giving rise to states in the trivial class $(o,o)$.

\subsubsection{D6 and D8 quantization}

When $n_2=0$, a charged D6 has a worldvolume $O(1)$ symmetry and provides charged fermion zero modes under the first spacetime factor. Its modes resemble the D4 case, and their quantization leads to the same class $(o,o)$.

When $n_1=0$, a charged D8 has a worldvolume $Sp(1)$ gauge symmetry, behaving like the charged D2 (with 16 real modes per pair of Chan--Paton indices). This also gives the class $(o,o)$.

\subsubsection{The gauge group}\label{sec:0A/Omega_s_gauge_group}

Taking into account the perturbative states and the fermion zero modes from the branes, we always have the class $(v,v)$. Then, when $n_2\equiv2\pmod{4}$, we also have the class $(o,v)$. Therefore, the gauge group is
\begin{eq}\label{eq:0A/Omega_s_group}
    G=\begin{cases}
        \left[Sp\left(\frac{n_1}{2}\right)\times Sp\left(\frac{n_2}{2}\right)\right]/{\Z_2} \qquad & \text{if } n_2\equiv0\pmod{4}\,,\\
        Sp\left(\frac{n_1}{2}\right)\times Sp\left(\frac{n_2}{2}\right) \qquad & \text{if } n_2\equiv2\pmod{4}\,,
    \end{cases}
\end{eq}
where the $\Z_2$ quotient is the diagonal $\Z_2$ generated by $(z_v,z_v)$. Note that the group is connected because $Sp$ is connected, which matches nicely with the absence of tachyon-free D-instantons in this symplectic model.

However, there is a subtlety in this orientifold model. In our analysis, we have only considered branes free of open-string tachyons, assuming that the tachyonic ones decay into combinations of them and perturbative states. In type 0A/$\Omega_s$, there is one case in which this fact does not seem to hold: the uncharged D0-brane with labels $(d_1,d_2,m)=(0,0,1)$ and a worldvolume $U(1)$ gauge group. This brane contains complex fermion zero modes in the fundamental representation of $\mathfrak{sp}\left(\frac{n_1}{2}\right)$. They can be quantized as in the uncharged D0-brane of type 0A/$\Omega$ (see \cref{sec:0A/Omega_gauge_group}): the Hilbert space is the same as in \cref{eq:0A_D0_Hilbert_Space}, thus generating the class $\left(\frac{n_1}{2}v,o\right)$. This turns out to be an interesting case since, for $n_1\equiv2\pmod{4}$, there is no combination of uncharged tachyon-free branes and perturbative states that can produce this class. Including this D0, the invariant center changes only in the case $(n_1,n_2)\equiv(2,0)\pmod{4}$, so that the gauge group becomes
\begin{eq}\label{eq:0A/Omega_s_group_with_D0}
    G_{\text{uncharged D0}}=\begin{cases}
        \left[Sp\left(\frac{n_1}{2}\right)\times Sp\left(\frac{n_2}{2}\right)\right]/{\Z_2} \qquad & \text{if } (n_1, n_2)\equiv(0,0)\pmod{4}\,,\\
        Sp\left(\frac{n_1}{2}\right)\times Sp\left(\frac{n_2}{2}\right) \qquad & \text{if } (n_1, n_2)\not\equiv(0,0)\pmod{4}\,.
    \end{cases}
\end{eq}
Beyond the study of the global structure of the gauge group, this peculiar feature of the uncharged D0 requires further investigation. In particular, since it carries no K-theory charge, it is not clear if there is any reason that prevents it from decaying into the vacuum. Alternatively, if a non-trivial decay channel exists, one has to identify its endpoint, which evades our brane scan. It is therefore not clear whether one should include the $\left(\frac{n_1}{2}v,o\right)$ class and what the fate of this tachyonic D0 is.

Note that, among all the orientifolds considered in this work, this uncharged D0 is the only case in which a tachyonic brane adds a center class that is not already present in the perturbative spectrum and the tachyon-free branes. As we have already mentioned in the Introduction, our analysis gives an upper bound on the invariant center, and it is possible that further non-perturbative effects will break some central elements. The uncharged D0 of this symplectic orientifold may be an example of such a mechanism.
Given the scope of this work, we present \cref{eq:0A/Omega_s_group} as the main result, and leave the analysis of the object leading to \cref{eq:0A/Omega_s_group_with_D0} as an open question. 

\subsection{Type \texorpdfstring{0B/$\Omega_s$}{0B/Omega s}}\label{sec:s0B_gauge_group}

In this type 0B orientifold, RR tadpole cancellation fixes again the four Chan--Paton labels of the spacetime gauge symmetry to be $(k_1,k_1,k_2,k_2)$, with the additional constraint from the symplectic nature that the $k_i$ are both even. The spacetime gauge algebra is then $\mathfrak{sp}(k_1/2)^{\oplus 2}\oplus \mathfrak{sp}(k_2/2)^{\oplus 2}$.
The two sectors, labeled by $k_1$ and $k_2$, are symmetric as in type 0B/$\Omega$.

As in the corresponding orthogonal orientifold, the perturbative spectrum gives rise to the classes $(\varepsilon_1,\varepsilon_2,\varepsilon_3,\varepsilon_4)$ with two $\varepsilon_i$ equal to $v$ and the other two equal to $o$. This already leaves only the diagonal $\Z_2$ as the unbroken center.

Another similarity with type 0B/$\Omega$ is the doubled spectrum of branes: as is clear from \eqref{eq:nu_definition} with $n=4$ and \cref{tab:type0_orientifold_branes}, for each brane dimension, both branes of type $+$ and $-$ are present and give rise to fermions charged under the two pairs $\{3,4\}$ and $\{1,2\}$ respectively.

\subsubsection{D1 quantization}

The two types of charged D1-branes, of type $+$ and $-$ in the conventions of \cref{sec:type_0_review}, have a worldvolume $Sp(1)$ gauge symmetry. As usual, fermion zero modes arise from D1--D9 strings between branes of different types.

\begin{table}[H]
\centering
\renewcommand{\arraystretch}{1.2}
\centerresize{}{
\begin{tabular}{l|c|c|c}
Worldvolume state  & $\mathfrak{so}(8)_T$ & $\mathfrak{sp}(\tfrac{k_1}{2})\oplus \mathfrak{sp}(\tfrac{k_1}{2})\oplus\mathfrak{sp}(\tfrac{k_2}{2})\oplus \mathfrak{sp}(\tfrac{k_2}{2})$ & worldvolume $\mathfrak{sp}(\tfrac{d_1}{2})\oplus \mathfrak{sp}(\tfrac{d_3}{2})$ \\
\hline
Fermions $\phi$ & $\mathbf{1}$ & $ (\mathbf{1}, \mathbf{k_1}, \mathbf{1}, \mathbf{1}) $ & $(\mathbf{1},\mathbf{d_3})$ \\
Fermions $\psi$ & $\mathbf{1}$ & $ (\mathbf{1}, \mathbf{1}, \mathbf{1}, \mathbf{k_2}) $ & $(\mathbf{d_1},\mathbf{1})$ \\
Fermions $\zeta$ & $\mathbf{1}$ & $ (\mathbf{k_1}, \mathbf{1}, \mathbf{1}, \mathbf{1}) $ & $(\mathbf{1},\mathbf{d_3})$ \\
Fermions $\chi$ & $\mathbf{1}$ & $ (\mathbf{1}, \mathbf{1}, \mathbf{k_2}, \mathbf{1}) $ & $(\mathbf{d_1},\mathbf{1})$
\end{tabular}}
\caption{Open-string light spectrum charged under the gauge symmetry from a D1-brane stack with $d_2 = d_4 = 0$ in the symplectic orientifold of type 0B string theory.
}
\label{tab:s0B/Omega_chargedD1_relevant_spectrum}
\end{table}

Consider a D1 with real modes in the $(\mathbf{k}_2\oplus \mathbf{k}_2)\otimes \mathbf{2}$ of $\mathfrak{sp}(k_2/2)^{\oplus 2}\oplus \mathfrak{sp}(1)$. These can be quantized like the fermion zero modes from the D0-brane of type 0A/$\Omega_s$, with the only difference that there is an extra $\mathfrak{sp}$ representation in this case; see appendix~\ref{app:symplectic_proof}. Equivalently, one can consider the branching
\begin{equation}
    \mathfrak{sp}(k_2) \longrightarrow \mathfrak{sp}(k_2/2) \oplus \mathfrak{sp}(k_2/2)\,.
\end{equation}
Since $k_2$ is even, the result is that only the class $o$ of $\mathfrak{sp}(k_2)$ is present, which branches as 
\begin{equation}
    o \longrightarrow (o,o) \oplus (v,v)
\end{equation}
for the two subalgebras. Therefore, we obtain the classes
\begin{equation}
    (o,o,o,o) \quad \text{and} \quad (o,o,v,v)
\end{equation}
from the quantization of fermion zero modes on a D1-brane of the first type, and
\begin{equation}
    (o,o,o,o) \quad \text{and} \quad (v,v,o,o)
\end{equation}
from the quantization of fermion zero modes on a D1-brane of the other type. A stack of branes simply adds copies of the zero modes in the language of appendix~\ref{app:symplectic_proof}, so that D1-branes do not add further classes in addition to the perturbative ones.

\subsubsection{D3 quantization}

The two types of D3-branes have worldvolume $U(1)$ gauge symmetries. Given the usual symmetry under $k_1\leftrightarrow k_2$, one can consider a single type of D3. In \cref{tab:s0B/Omega_chargedD3_relevant_spectrum}, we show the relevant fermion modes from the D3 of type $-$.

\begin{table}[H]
\centering
\renewcommand{\arraystretch}{1.2}
\begin{tabular}{l|c|c|c}
Worldvolume state  & $\mathfrak{so}(6)_T$ & $\mathfrak{sp}(\tfrac{k_1}{2})\oplus \mathfrak{sp}(\tfrac{k_1}{2})\oplus\mathfrak{sp}(\tfrac{k_2}{2})\oplus \mathfrak{sp}(\tfrac{k_2}{2})$ & worldvolume $\mathfrak{u}(1)$ \\
\hline
Fermion $\psi$ & $\mathbf{1}$ & $ (\mathbf{k_1}, \mathbf{1}, \mathbf{1}, \mathbf{1}) $ & $\bar{q}$ \\
Fermion $\chi$ & $\mathbf{1}$ & $ (\mathbf{1}, \mathbf{k_1}, \mathbf{1}, \mathbf{1}) $ & $q$
\end{tabular}
\caption{Open-string light spectrum charged under the gauge symmetry from a D3-brane stack with $d_1 = 0$ and $d_2=1$ in the symplectic orientifold of type 0B string theory.
}
\label{tab:s0B/Omega_chargedD3_relevant_spectrum}
\end{table}

The quantization of these states proceeds as for the D7-brane of type 0B/$\Omega$: the charge operator is $Q=N_\psi-N_\chi$ and the only classes that are present are $(o,o,o,o)$ and $(v,v,o,o)$. Then, the other type of D3-brane adds the class $(o,o,v,v)$, with the overall result that no further classes are added to the perturbative ones.

\subsubsection{D4 quantization}

The two types of D4-branes have a worldvolume $O(1)\cong \Z_2$ gauge symmetry. Consider the D4 of type $-$, with the charged fermion zero modes from \cref{tab:s0B/Omega_chargedD4_relevant_spectrum}.

\begin{table}[H]
\centering
\renewcommand{\arraystretch}{1.2}
\begin{tabular}{l|c|c|c}
Worldvolume state  & $\mathfrak{so}(5)_T$ & $\mathfrak{sp}(\tfrac{k_1}{2})\oplus \mathfrak{sp}(\tfrac{k_1}{2})\oplus\mathfrak{sp}(\tfrac{k_2}{2})\oplus \mathfrak{sp}(\tfrac{k_2}{2})$ & worldvolume $O(1)$ \\
\hline
Fermion $\psi$ & $\mathbf{1}$ & $ (\mathbf{k_1}, \mathbf{1}, \mathbf{1}, \mathbf{1}) $ & odd \\
Fermion $\chi$ & $\mathbf{1}$ & $ (\mathbf{1}, \mathbf{k_1}, \mathbf{1}, \mathbf{1}) $ & odd
\end{tabular}
\caption{Open-string light spectrum charged under the gauge symmetry from a D4-brane stack with $d_1 = 0$ and $d_2=1$ in the symplectic orientifold of type 0B string theory.
}
\label{tab:s0B/Omega_chargedD4_relevant_spectrum}
\end{table}

Quantizing these zero modes works as for the D4-brane of the type 0A/$\Omega_s$ orientifold, and taking both types of D4-branes into account, one recovers the same classes again: $(o,o,o,o)$, $(v,v,o,o)$, and $(o,o,v,v)$.

\subsubsection{D5 quantization}

The two types of charged D5-branes have worldvolume orthogonal gauge symmetries.

\begin{table}[H]
\centering
\renewcommand{\arraystretch}{1.2}
\centerresize{}{
\begin{tabular}{l|c|c|c}
Worldvolume state  & $\mathfrak{so}(4)_T$ & $\mathfrak{sp}(\tfrac{k_1}{2})\oplus \mathfrak{sp}(\tfrac{k_1}{2})\oplus\mathfrak{sp}(\tfrac{k_2}{2})\oplus \mathfrak{sp}(\tfrac{k_2}{2})$ & worldvolume $\mathfrak{so}(d_1)\oplus \mathfrak{so}(d_3)$ \\
\hline
Fermions $\phi$ & $\mathbf{1}$ & $ (\mathbf{1}, \mathbf{k_1}, \mathbf{1}, \mathbf{1}) $ & $(\mathbf{1},\mathbf{d_3})$ \\
Fermions $\psi$ & $\mathbf{1}$ & $ (\mathbf{1}, \mathbf{1}, \mathbf{1}, \mathbf{k_2}) $ & $(\mathbf{d_1},\mathbf{1})$ \\
Fermions $\zeta$ & $\mathbf{1}$ & $ (\mathbf{k_1}, \mathbf{1}, \mathbf{1}, \mathbf{1}) $ & $(\mathbf{1},\mathbf{d_3})$ \\
Fermions $\chi$ & $\mathbf{1}$ & $ (\mathbf{1}, \mathbf{1}, \mathbf{k_2}, \mathbf{1}) $ & $(\mathbf{d_1},\mathbf{1})$
\end{tabular}
}
\caption{Open-string light spectrum charged under the gauge symmetry from a D5-brane stack with $d_2 = d_4 = 0$ in the symplectic orientifold of type 0B string theory.
}
\label{tab:s0B/Omega_chargedD5_relevant_spectrum}
\end{table}

Focus on the fermions $\psi$ and $\chi$ living on the branes of the first kind, with a $\mathfrak{so}(d_1)$ gauge algebra (see \cref{tab:s0B/Omega_chargedD5_relevant_spectrum}). These are symplectic-Majorana--Weyl modes, and we can consider the decomposition
\begin{equation}
    \mathfrak{sp}\left(\tfrac{k_2}{2}\right)^{\oplus 2} \oplus\mathfrak{so}(d_1) \longrightarrow \mathfrak{sp}(1)^{\oplus 2} \oplus \mathfrak{so}\left(\tfrac{k_2}{2}\right)^{\oplus 2} \oplus\mathfrak{so}(d_1)\,,
\end{equation}
under which the fundamental representation of each symplectic factor decomposes as
\begin{equation}
    \mathbf{k_2} \longrightarrow
    \left(\mathbf{2},\mathbf{\tfrac{k_2}{2}}\right)
\end{equation}
under $\mathfrak{sp}(1)\oplus \mathfrak{so}(k_2/2)$.
This means that we have to quantize complex modes in the $\left(\mathbf{\frac{k_2}{2}}\oplus \mathbf{\frac{k_2}{2}}\right)\otimes \mathbf{d_1}$ of $\mathfrak{so}(k_2/2)^{\oplus 2}\oplus \mathfrak{so}(d_1)$, thus leading to
\begin{equation}
     \{\psi_{i\alpha}, \psi^\dagger_{j \beta}\} = \delta_{ij}\delta_{\alpha\beta} \,, \qquad \{\chi_{a\alpha}, \chi^\dagger_{b \beta}\} = \delta_{ab} \delta_{\alpha\beta}\,, \qquad \{\psi_{i\alpha}, \chi_{a \beta}\} = 0 \,,
\end{equation}
where $i,j$ are the indices of the first spacetime $\mathfrak{so}\left(\tfrac{k_2}{2}\right)$ factor, $a,b$ the indices of the second $\mathfrak{so}\left(\tfrac{k_2}{2}\right)$ factor, and $\alpha,\beta$ those of the worldvolume $\mathfrak{so}(d_1)$ gauge algebra. Therefore, when $\frac{k_2}{2}$ is even, one can follow \cite{Larotonda:2024thv}: the Fock space invariant under the worldvolume gauge symmetry contains only states in the class $o$ of $\mathfrak{so}(k_2)$, which branches as
\begin{equation}
    o \longrightarrow (o,o) \oplus (v,v)\qquad\text{under}\qquad \mathfrak{so}(k_2) \longrightarrow \mathfrak{so}\left(\tfrac{k_2}{2}\right) \oplus \mathfrak{so}\left(\tfrac{k_2}{2}\right)\,.
\end{equation}
These recombine with the representation $(v,v)$ of $\mathfrak{sp}(1)\oplus \mathfrak{sp}(1)$, thus reconstructing the initial symplectic representations from
\begin{equation}
    \mathfrak{so}\left(\frac{k_2}{2}\right)\oplus\mathfrak{so}\left(\frac{k_2}{2}\right) \oplus \mathfrak{sp}(1)^{\oplus 2} \subset \mathfrak{sp}\left(\frac{k_2}{2}\right)\oplus\mathfrak{sp}\left(\frac{k_2}{2}\right)\,.
\end{equation}
The only non-trivial class that we obtain from this procedure is $(v,v)$ of $\mathfrak{sp}\left(\frac{k_2}{2}\right)\oplus\mathfrak{sp}\left(\frac{k_2}{2}\right)$, which, for the D5-branes of the two types, gives the class $(o,o,v,v)$ for the branes of the first type and the class $(v,v,o,o)$ for the branes of the second type.
When $\frac{k_2}{2}$ is odd, we only get the class $(o,o)$ of the two $\mathfrak{so}\left(\tfrac{k_2}{2}\right)$ factors, which recombines with the bi-vector class $(v,v)$ of $\mathfrak{sp}(1) \oplus \mathfrak{sp}(1)$, leading to the same result.

In general, for any value of $k_2$, one can directly consider the action of the central elements of the two symplectic factors on the modes: the central element of one $\mathfrak{sp}$ factor acts as $\psi\to-\psi$, while the central element of the other acts as $\chi\to-\chi$. Therefore, on the Fock space, they act as\footnote{An additional overall sign in these operators is excluded because it would enter as $(-1)^{k_2 d_1}$, but $k_2 d_1$ is even; see the reasoning in appendix~\ref{app:symplectic_proof}.} $(-1)^{N_\psi}$ and $(-1)^{N_\chi}$.
This has to be compared with the action of the worldvolume $\Z_2$ (or the element $-1\in O(d_1)$ in general), which flips all the modes and acts as $(-1)^{N_\psi+N_\chi}$. Then, gauge invariance requires $N_\psi+N_\chi\equiv0\pmod{2}$, which leads to the two classes $(o,o)$ or $(v,v)$ of the pair $\{3,4\}$. The same reasoning applies to the D5 of the other type. 

\subsubsection{The gauge group}

Considering the perturbative states and the brane modes that we have discussed, the only central elements that act trivially are then $(1,1,1,1)$ and $(z_v,z_v,z_v,z_v)$. Note that to obtain these, it suffices to consider the perturbative states. Therefore, the gauge group of the symplectic orientifold of type 0B string theory is
\begin{eq}
   \frac{Sp\left(\frac{k_1}{2}\right)\times Sp\left(\frac{k_1}{2}\right)\times Sp\left(\frac{k_2}{2}\right)\times Sp\left(\frac{k_2}{2}\right)}{\Z_2}\,,
\end{eq}
where $\Z_2$ is the diagonal center. The common $\Z_2$ remains the unbroken center even when one of the pairs vanishes. As in type 0A/$\Omega_s$, the group is connected, which is consistent with the absence of tachyon-free D-instantons.

\section{Conclusions}

We have provided a systematic scan of global topologies of gauge groups in the orthogonal and symplectic type 0 orientifolds. Our results are summarized in \cref{sec:summary}. Among our results, we corrected some misprints regarding type 0 one-loop amplitudes that were present in the literature, also providing a dictionary between the algebraic orientifold language of~\cite{Dudas:2001wd} and the K-theory classification of~\cite{Kaidi:2019tyf}. 

We found two surprising features of the type 0A orientifolds. In the orthogonal orientifold, contrary to what is often stated, not every D9 configuration is allowed. The number of D9-branes of type $\eta=-1$ is constrained to be even to cancel global worldvolume anomalies or, equivalently, K-theory tadpoles. In addition, the uncharged D0-brane forces the number of D9-branes of type $\eta=+1$ to be even as well. This last constraint can be relaxed in the presence of worldvolume Chern--Simons terms or tadpoles, as is the case when a non-vanishing Romans mass is present. It would be interesting to investigate this setting in more detail, studying the gauge group topology in massive deformations of type 0A.

In the symplectic type 0A/$\Omega_s$ orientifold, we identified a possible non-trivial contribution to the invariant center arising from an uncharged tachyonic D0-brane. The decay of this object is particularly puzzling, as its gauge-invariant states carry a center class that is absent from the known perturbative and non-perturbative tachyon-free brane spectrum. 
Hence, understanding the decay of the D0 requires new physical input and might signal the existence of an unknown defect in this type 0A orientifold. 

Our work has implications for string dualities in the non-supersymmetric landscape~\cite{Bergman:1997rf,Blum:1997cs,Blum:1997gw,Bergman:1999km,Blumenhagen:1999ad,Angelantonj:2007ts,Faraggi:2007tj,Acharya:2022shu,Basile:2022zee,Larotonda:2024thv,Fraiman:2025yrx,Baykara:2026gem,Altavista:2026edv,Altavista:2026evd,Tachikawa:2026jaj,Baykara:2026vdc,Altavista:2026brr,Dasgupta:2026maq,Basile:2026trt,Kamal:2026msr,Altavista:2026muc,Kan:2026sea,Cribiori:2026kpk}, which have recently seen a resurgence of interest. In fact, this was the main topic of the two related works~\cite{Larotonda:2024thv,Fraiman:2026ltu}, in which the global gauge group structures have been used to explore or assess proposed dualities of non-supersymmetric strings~\cite{Bergman:1997rf}. 

An explicit application of the results of this paper is the duality proposed in~\cite{Dudas:2001wd}, and discussed in~\cite{Baykara:2026vdc}, between type 0A/$\Omega$ with $n_1=32$ and $n_2=0$ and an orientifold of bosonic string theory on $T^{16}$ at a special $\mathfrak{so}(32)\oplus \mathfrak{so}(32)$ enhancement point, in which the orientifold action includes the exchange of the two $\mathfrak{so}(32)$ factors. Our result for this case is the gauge group $G=O(32)/\{\pm1\}$; see \cref{eq:type0A_orthogonal_special_n20}. 
This global form of the gauge group can be used to constrain the $\mathfrak{so}(32)\oplus \mathfrak{so}(32)$ lattice on which the bosonic theory is compactified.
We leave a detailed analysis of the duality for future work; here, we only stress that the massless and tachyonic spectra (see~\cite[Table~2]{Baykara:2026vdc}), together with the fact that bosonic string theory has no fermions in the perturbative and D-brane spectra, suggest that one can take the quotient by the full center, consistently with the gauge group $O(32)/\{\pm1\}$ of the type 0A orientifold.

In this work, special emphasis has been given to the symplectic orientifolds of type 0 strings. From the point of view of string dualities, these are much less explored (and considered) than their orthogonal counterparts. The most famous example of this is the Sugimoto model \cite{Sugimoto:1999tx}, which is the symplectic version of type I string theory. Very little is known about its non-perturbative fate, and some dualities have been proposed only for T-dual versions involving orientifold planes of lower dimensionality~\cite{Angelantonj:2007ts}. Investigating if and how the symplectic orientifolds fit in the duality web of string theory, or if they have an M-theory origin as in~\cite{Baykara:2026gem,Altavista:2026evd}, remains a key question for addressing the non-perturbative behavior of these models.

Another aspect of our work is tied to string universality \cite{Kumar:2009us, Adams:2010zy,Kim:2019ths,Kim:2019vuc,Cvetic:2020kuw,Montero:2020icj,Hamada:2021bbz,Bedroya:2021fbu,Hamada:2023zol} in the string landscape without supersymmetry.
As shown in \cite{Larotonda:2024thv}, knowing the gauge group topology can serve as the starting point to implement bottom-up anomaly constraints in QFTs with analogous symmetry structures, thus shedding light on the landscape of non-supersymmetric gravitational EFTs and testing universality in this regime.

Finding the global structure of the gauge groups has further implications for our understanding of non-supersymmetric strings. In the absence of the dynamical control granted by supersymmetry, other protection mechanisms, such as topology, become our only way to access non-perturbative results (see, e.g.,~\cite{Tachikawa:2023lwf,Basile:2025mnj,Wada:2026dge}). Our work is a step in this direction. Many unexplored features of string theory may not be visible in the supersymmetric corners that we have mostly explored, such as the parity condition on the D9-branes that we have found or the subtlety with the D0 in the symplectic type 0A orientifold.
While we cannot replace supersymmetry as our guiding principle, there have been some recent attempts to build ``protections'' in non-supersymmetric settings (see, e.g.,~\cite{Kachru:1998hd,Harvey:1998rc,Shiu:1998he,Blumenhagen:1999ad,Angelantonj:1999gm,Angelantonj:2003hr,GrootNibbelink:2017luf,Aoyama:2020aaw,Satoh:2021nfu,Raucci:2023xgx,Larotonda:2026hxy,Fraiman:2026irf}). Work along these lines would bring string theory closer to our non-supersymmetric universe. 

\section*{Acknowledgments}
We would like to thank Ling Lin for helpful discussions.
S.R.~is supported by the ERC Starting Grant QGuide 101042568 - StG 2021.
This publication has been funded within the framework of the R\&D\&I Projects PID2024-156043NB-I00 and CEX2025-001574-S, funded by MICIU\slash AEI\slash10.13039\slash501100011033. The research presented in this publication falls within the research line Strings and Quantum Gravity.
This work was performed in part at the Aspen Center for Physics, which is supported by a grant from the Simons Foundation (1161654, Troyer).
V.L.~would like to thank IFT Madrid for its hospitality during the workshop Pro(v/b)ing the Swampland and the Visitor Program, where part of this work was carried out.

\paragraph*{AI assistance.}
Large language models were used to assist with editing and checking calculations in this manuscript. The authors take full responsibility for the content, accuracy, and conclusions of the work.

\appendix

\section{Centers and representation classes of \texorpdfstring{$Spin(n)$}{Spin(n)} and \texorpdfstring{$Sp(n)$}{Sp(n)}}\label{app:centers}

In this appendix, we set up our conventions for the centers of spin and symplectic groups and for the representation classes.

For the center of $Spin(n)$, we have to distinguish between three cases: $n\equiv0\pmod{4}$, $n\equiv2\pmod{4}$, and $n\equiv1\pmod{2}$.

The center of $Spin(4k)$ is ${\cal Z}(Spin(4k))=\Z_2\times \Z_2$. We denote its elements by $\left\{1,z_v,z_s,z_c\right\}$, with the conditions $z_i^2 = 1$ and $z_v=z_s z_c = z_c z_s$. The irreducible representations $\mathbf{R}$ fall into equivalence classes $[\mathbf{R}]$ labeled by irreducible representations of the center. There are four classes for $Spin(4k)$: the scalar, vector, spinor, and co-spinor classes
\begin{equation}
    o = [{\bf O}] \, , \qquad v = [{\bf V}] \, , \qquad s = [{\bf S}] \, , \qquad c = [{\bf C}] \,.
\end{equation}
Our conventions for the action of the center elements on them can be found in \cref{tab:actions_of_Z_4k}.
\begin{table}[!ht]
    \centering
    \begin{tabular}{c|c|c|c}
    irrep & $z_v$ & $z_s$ & $z_c$ \\ \hline
    $o = [{\bf O}]$ & $+$ & $+$ & $+$ \\
    $v = [{\bf V}]$ & $+$ & $-$ & $-$ \\
    $s = [{\bf S}]$ & $-$ & $+$ & $-$ \\
    $c = [{\bf C}]$ & $-$ & $-$ & $+$
    \end{tabular}
    \caption{Representation classes of $Spin(4k)$ and action of the central elements on them.}
    \label{tab:actions_of_Z_4k}
\end{table}

The center of $Spin(4k+2)$ is ${\cal Z}(Spin(4k+2))=\Z_4$.\footnote{As we stress in the main text, we consider $k>0$. The $k=0$ unitary case will be the subject of a future work.} As in the previous case, we denote its elements by $\left\{1,z_v,z_s,z_c\right\}$, now with the conditions $z_s^4=z_c^4 = 1$, $z_c = z_s^3$, and $z_v=z_s^2= z_c^2$, so that $z_s$ can be taken to be the generator of $\Z_4$. The actions on the center equivalence classes can be found in \cref{tab:actions_of_Z_4k+2}.
\begin{table}[!ht]
    \centering
    \begin{tabular}{c|c|c|c}
    irrep & $z_v$ & $z_s$ & $z_c$ \\ \hline
    $o = [{\bf O}]$ & $+$ & $+$ & $+$ \\
    $v = [{\bf V}]$ & $+$ & $-$ & $-$ \\
    $s = [{\bf S}]$ & $-$ & $+i$ & $-i$ \\
    $c = [{\bf C}]$ & $-$ & $-i$ & $+i$
    \end{tabular}
    \caption{Representation classes of $Spin(4k+2)$ and action of the central elements on them.}
    \label{tab:actions_of_Z_4k+2}
\end{table}

The center of $Spin(2k+1)$ is ${\cal Z}(Spin(2k+1))=\Z_2$. We denote its elements by $\left\{1,z_v\right\}$, with the condition $z_v^2=1$. In this case, all tensor representations belong to the same class and there is a single spinor class:
\begin{equation}
    o = [{\bf O}] = [{\bf V}] \, , \qquad s' = [{\bf S'}] \,.
\end{equation}
The actions on the center equivalence classes are in \cref{tab:actions_of_Z_2k+1}.
\begin{table}[!ht]
    \centering
    \begin{tabular}{c|c}
    irrep & $z_v$ \\ \hline
    $o = [{\bf O}]$ & $+$ \\
    $s' = [{\bf S}']$ & $-$
    \end{tabular}
    \caption{Representation classes of $Spin(2k+1)$ and action of the central elements on them.}
    \label{tab:actions_of_Z_2k+1}
\end{table}

Symplectic groups have a simpler structure. Their center is always $\mathcal{Z}\left(Sp(k)\right) = \Z_2$.
We denote the elements of the center by $\{1,z_v\}$, with the condition $z_v^2=1$. The irreducible representations fall into two equivalence classes,
\begin{equation}
    o = [\mathbf{O}] \qquad \text{and} \qquad v = [\mathbf{V}]\,.
\end{equation}
Representations arising from even tensor powers of the defining representation---which we call \emph{fundamental representation} in the text, following the physics convention---fall into the singlet class, while odd tensor powers are in the class represented by $\mathbf{V}$. The actions of the central elements on the classes are shown in \cref{tab:actions_of_Z_Sp}. 
\begin{table}[!ht]
    \centering
    \begin{tabular}{c|c}
    irrep & $z_v$  \\ \hline
    $o = [{\bf O}]$ & $+$  \\
    $v = [{\bf V}]$ & $-$  \\
    \end{tabular}
    \caption{Representation classes of $Sp(k)$ and action of the central elements on them.}
    \label{tab:actions_of_Z_Sp}
\end{table}

We collect the results of this appendix in \cref{tab:centers}.
\begin{table}[!ht]
    \centering
    \begin{tabular}{c|c|c}
    $G$ & $\mathcal{Z}(G)$ & $\mathcal{Z}(G)$ generators\\ \hline
    $Spin(4k)$ & $\Z_2\times\Z_2$ & $z_s,z_c$ \\
    $Spin(4k+2)$ & $ \Z_4$ & $z_s$\\
    $Spin(2k+1)$ & $\Z_2$ & $z_v$\\
    $Sp(k)$ & $\Z_2$ & $z_v$
    \end{tabular}
    \caption{Summary of the groups of interest, their centers, and the respective generators.}
    \label{tab:centers}
\end{table}

\section{Brane spectra of type 0 orthogonal orientifolds}\label{app:Branes_orthogonal}

In this appendix and in the next, we adopt the following conventions to denote representations (that are not necessarily irreducible) of unitary, orthogonal, and symplectic groups. The Young tableau $\text{\tiny\yng(1)}$ denotes the defining representations, namely the vector representation for $\mathfrak{so}(n)$ and the fundamental representations for $\mathfrak{sp}(n)$ and $\mathfrak{u}(n)$. The symbols $\text{\tiny\yng(2)}$ and $\text{\tiny\yng(1,1)}$ are used for the symmetric and antisymmetric tensor products of vector representations of $\mathfrak{so}(n)$ or of fundamental representations of $\mathfrak{sp}(n)$ and $\mathfrak{u}(n)$.
For $\mathfrak{u}(n)$, we use the bar to denote the conjugate representations; e.g., $\ybar{\text{\tiny\yng(1)}}$ is the antifundamental representation.
In the tables, we use $\mathbf{n}_{s'}$, with $\mathbf{n = 2^{\lfloor\frac{9-p}{2}\rfloor}}$, for the irreducible Dirac spinor under the transverse rotation algebra $\mathfrak{so}(9-p)_T$, while $\mathbf{(9-p)}_v$ denotes the vector.

\subsection{Type 0A orthogonal orientifold}\label{app:0A/Omega_branes}

In this appendix, we display the tables from \cite{Fraiman:2026ltu} that contain the light open-string spectra from D-branes in the type 0A orthogonal orientifold.\footnote{In an early version of~\cite{Fraiman:2026ltu}, there were some typos in these tables, originating from the typos of~\cite{Dudas:2001wd} that we are correcting in this work.} A detailed discussion of the spectra can be found in the appendices of~\cite{Fraiman:2026ltu}.

We summarize the spectrum of type 0A/$\Omega$ arising from the D9 sector in table~\ref{tab:0A/Omega_open_spectrum}.
\begin{table}[!htbp]
\centering
\renewcommand{\arraystretch}{1.2}
\begin{tabular}{l|c|c|c}
State & character & $\mathfrak{so}(n_1)$ & $\mathfrak{so}(n_2)$ \\
\hline
Gauge vector & $V_8$ & $\mathbf{Adj}$ & $\mathbf{1}$ \\
Gauge vector & $V_8$ & $\mathbf{1}$ & $\mathbf{Adj}$ \\
Fermion & $S_8$ & \tiny\yng(1) & \tiny\yng(1) \\
Fermion & $C_8$ & \tiny\yng(1) & \tiny\yng(1) \\
Tachyon & $O_8$ & \tiny\yng(2) & $\mathbf{1}$ \\
Tachyon & $O_8$ & $\mathbf{1}$ & $\mathbf{Adj}$
\end{tabular}
\caption{Open-string light spectrum from D9-branes in the type 0A/$\Omega$ orientifold.}
\label{tab:0A/Omega_open_spectrum}
\end{table}

\FloatBarrier

The light spectra from D$p$-branes with $p=0,2,4,6,8$ are in \cref{tab:0A/Omega_D0_spectrum,tab:0A/Omega_D2_spectrum,tab:0A/Omega_D4_spectrum,tab:0A/Omega_D6_spectrum,tab:0A/Omega_D8_spectrum}. For branes with $p=-1,1,3,5,7$, we collect their worldvolume gauge structure in table~\ref{tab:0A/Omega_gauge_structures_uncharged_branes} and the light spectra in \cref{tab:0A/Omega_D1_spectrum,tab:0A/Omega_D3_spectrum,tab:0A/Omega_D5_spectrum,tab:0A/Omega_D7_spectrum}.

\begin{table}[!htbp]
\centering
\renewcommand{\arraystretch}{1.2}
\textbf{D0}
\centerresize{}{
\begin{tabular}{l|c|c|c}
Worldvolume state  & $\mathfrak{so}(9)_T$ & $\mathfrak{so}(n_1) \oplus \mathfrak{so}(n_2)$  & $\mathfrak{so}(d_1) \oplus \mathfrak{so}(d_2) \oplus \mathfrak{u}(m)$ \\
\hline
Gauge vector  & $\mathbf{1}$ & $(\mathbf{1}, \mathbf{1})$  & $(\mathbf{Adj},  \mathbf{1}, \mathbf{1}) \oplus(\mathbf{1}, \mathbf{Adj}, \mathbf{1}) \oplus (\mathbf{1},  \mathbf{1}, \mathbf{Adj})$ \\
Scalar & $\mathbf{9}_v$ & $(\mathbf{1}, \mathbf{1})$ & $\left(\text{\tiny{\yng(2)}},  \mathbf{1}, \mathbf{1}\right) \oplus\left(\mathbf{1}, \text{\tiny{\yng(2)}},\mathbf{1}\right)\oplus (\mathbf{1},  \mathbf{1}, \mathbf{Adj})$\\
Tachyon & $\mathbf{1}$ & $(\mathbf{1}, \mathbf{1})$ & $ \left(\text{\tiny{\yng(1)}},\text{\tiny{\yng(1)}}, \mathbf{1} \right) \oplus \left(\mathbf{1},  \mathbf{1}, \text{\tiny{\yng(1,1)}}\right)\oplus \left(\mathbf{1},  \mathbf{1}, \ybar{\text{\tiny{\yng(1,1)}}}\right)$ \\
Fermion & $\mathbf{16}_{s'}$ & $(\mathbf{1}, \mathbf{1})$ & $\left(\text{\tiny{\yng(1)}}, \mathbf{1},\text{\tiny{\yng(1)}}\right) \oplus \left(\mathbf{1},\text{\tiny{\yng(1)}},\text{\tiny{\yng(1)}}\right) \oplus \left(\text{\tiny{\yng(1)}}, \mathbf{1},\ybar{\text{\tiny{\yng(1)}}}\right) \oplus \left(\mathbf{1},\text{\tiny{\yng(1)}},\ybar{\text{\tiny{\yng(1)}}}\right)$ \\
Fermion & $\mathbf{1}$ & $\left(\mathbf{1},\text{\tiny\yng(1)}\right)$ & $\left(\text{\tiny\yng(1)}, \mathbf{1},\mathbf{1}\right) \oplus \left( \mathbf{1},\text{\tiny{\yng(1)}},\mathbf{1}\right)$ \\
Fermion & $\mathbf{1}$ & $\left(\text{\tiny\yng(1)},\mathbf{1}\right)$ & $\left(\mathbf{1},\mathbf{1},\text{\tiny\yng(1)}\right)\oplus\left(\mathbf{1},\mathbf{1},\ybar{\text{\tiny\yng(1)}}\right)$ \\
\end{tabular}}
\caption{Open-string light spectrum for D0-branes in the type 0A orthogonal orientifold.}
\label{tab:0A/Omega_D0_spectrum}
\end{table}
\begin{table}[!htbp]
\centering
\renewcommand{\arraystretch}{1.2}
\textbf{D2}
\centerresize{}{
\begin{tabular}{l|c|c|c}
Worldvolume state  & $\mathfrak{so}(7)_T$ & $\mathfrak{so}(n_1) \oplus \mathfrak{so}(n_2)$  & $\mathfrak{so}(d_1) \oplus \mathfrak{so}(d_2) \oplus \mathfrak{u}(m)$ \\
\hline
Gauge vector  & $\mathbf{1}$ & $(\mathbf{1}, \mathbf{1})$  & $(\mathbf{Adj},  \mathbf{1}, \mathbf{1}) \oplus(\mathbf{1}, \mathbf{Adj}, \mathbf{1}) \oplus (\mathbf{1},  \mathbf{1}, \mathbf{Adj})$ \\
Scalar & $\mathbf{7}_{v}$ & $(\mathbf{1}, \mathbf{1})$ & $\left(\text{\tiny{\yng(2)}},  \mathbf{1}, \mathbf{1}\right) \oplus\left(\mathbf{1}, \text{\tiny{\yng(2)}},\mathbf{1}\right)\oplus (\mathbf{1},  \mathbf{1}, \mathbf{Adj})$\\
Tachyon & $\mathbf{1}$ & $(\mathbf{1}, \mathbf{1})$ & $ \left(\text{\tiny{\yng(1)}},\text{\tiny{\yng(1)}}, \mathbf{1} \right) \oplus \left(\mathbf{1},  \mathbf{1}, \text{\tiny{\yng(2)}}\right)\oplus \left(\mathbf{1},  \mathbf{1}, \ybar{\text{\tiny{\yng(2)}}}\right)$ \\
Fermion & $\mathbf{8}_{s'}$ & $(\mathbf{1}, \mathbf{1})$ & $\left(\text{\tiny{\yng(1)}}, \mathbf{1},\text{\tiny{\yng(1)}} \right) \oplus \left(\mathbf{1},\text{\tiny{\yng(1)}},\text{\tiny{\yng(1)}}\right) \oplus \left(\text{\tiny{\yng(1)}}, \mathbf{1},\ybar{\text{\tiny{\yng(1)}}}\right) \oplus \left(\mathbf{1},\text{\tiny{\yng(1)}},\ybar{\text{\tiny{\yng(1)}}} \right)$ \\
Fermion & $\mathbf{1}$ & $\left(\text{\tiny\yng(1)},\mathbf{1}\right)$ & $\left(\text{\tiny\yng(1)}, \mathbf{1},\mathbf{1}\right) \oplus \left( \mathbf{1},\text{\tiny{\yng(1)}},\mathbf{1}\right)$ \\
Fermion & $\mathbf{1}$ & $\left(\mathbf{1},\text{\tiny\yng(1)}\right)$ & $\left(\mathbf{1},\mathbf{1},\text{\tiny\yng(1)}\right)\oplus\left(\mathbf{1},\mathbf{1},\ybar{\text{\tiny\yng(1)}}\right)$ 
\end{tabular}}
\caption{Open-string light spectrum for D2-branes in the type 0A orthogonal orientifold.}
\label{tab:0A/Omega_D2_spectrum}
\end{table}
\begin{table}[!htbp]
\centering
\textbf{D4}
\renewcommand{\arraystretch}{1.2}
\centerresize{}{
\begin{tabular}{l|c|c|c}
Worldvolume state  & $\mathfrak{so}(5)_T$ & $\mathfrak{so}(n_1) \oplus \mathfrak{so}(n_2)$  & $\mathfrak{sp}(d_1/2)\oplus \mathfrak{sp}(d_2/2) \oplus \mathfrak{u}(m)$ \\
\hline
Gauge vector  & $\mathbf{1}$ & $(\mathbf{1}, \mathbf{1})$  & $(\mathbf{Adj},  \mathbf{1}, \mathbf{1}) \oplus(\mathbf{1}, \mathbf{Adj}, \mathbf{1}) \oplus (\mathbf{1},  \mathbf{1}, \mathbf{Adj})$ \\
Scalar & $\mathbf{5}_{v}$ & $(\mathbf{1}, \mathbf{1})$ & $\left(\text{\tiny{\yng(1,1)}},  \mathbf{1}, \mathbf{1}\right) \oplus\left(\mathbf{1}, \text{\tiny{\yng(1,1)}},\mathbf{1}\right)\oplus (\mathbf{1},  \mathbf{1}, \mathbf{Adj})$\\
Tachyon & $\mathbf{1}$ & $(\mathbf{1}, \mathbf{1})$ & $ \left(\text{\tiny{\yng(1)}},\text{\tiny{\yng(1)}}, \mathbf{1} \right) \oplus \left(\mathbf{1},  \mathbf{1}, \text{\tiny{\yng(2)}}\right)\oplus \left(\mathbf{1},  \mathbf{1}, \ybar{\text{\tiny{\yng(2)}}}\right)$ \\
Fermion & $\mathbf{4}_{s'}$ & $(\mathbf{1}, \mathbf{1})$ & $\left(\text{\tiny{\yng(1)}}, \mathbf{1},\text{\tiny{\yng(1)}} \right) \oplus \left(\mathbf{1},\text{\tiny{\yng(1)}},\text{\tiny{\yng(1)}}\right) \oplus \left(\text{\tiny{\yng(1)}}, \mathbf{1},\ybar{\text{\tiny{\yng(1)}}} \right) \oplus \left(\mathbf{1},\text{\tiny{\yng(1)}},\ybar{\text{\tiny{\yng(1)}}} \right)$ \\
Fermion & $\mathbf{1}$ & $\left(\mathbf{1},\text{\tiny\yng(1)}\right)$ & $\left(\text{\tiny\yng(1)}, \mathbf{1},\mathbf{1}\right) \oplus \left( \mathbf{1},\text{\tiny{\yng(1)}},\mathbf{1}\right)$ \\
Fermion & $\mathbf{1}$ & $\left(\text{\tiny\yng(1)},\mathbf{1}\right)$ & $\left(\mathbf{1},\mathbf{1},\text{\tiny\yng(1)}\right)\oplus\left(\mathbf{1},\mathbf{1},\ybar{\text{\tiny\yng(1)}}\right)$ 
\end{tabular}}
\caption{Open-string light spectrum for D4-branes in the type 0A orthogonal orientifold.}
\label{tab:0A/Omega_D4_spectrum}
\end{table}
\begin{table}[!htbp]
\centering
\textbf{D6}
\renewcommand{\arraystretch}{1.2}
\centerresize{}{
\begin{tabular}{l|c|c|c}
Worldvolume state  & $\mathfrak{so}(3)_T$ & $\mathfrak{so}(n_1) \oplus \mathfrak{so}(n_2)$  & $\mathfrak{sp}(d_1/2) \oplus \mathfrak{sp}(d_2/2) \oplus \mathfrak{u}(m)$ \\
\hline
Gauge vector  & $\mathbf{1}$ & $(\mathbf{1}, \mathbf{1})$  & $(\mathbf{Adj},  \mathbf{1}, \mathbf{1}) \oplus(\mathbf{1}, \mathbf{Adj}, \mathbf{1}) \oplus (\mathbf{1},  \mathbf{1}, \mathbf{Adj})$ \\
Scalar & $\mathbf{3}_{v}$ & $(\mathbf{1}, \mathbf{1})$ & $\left(\text{\tiny{\yng(1,1)}},  \mathbf{1}, \mathbf{1}\right) \oplus\left(\mathbf{1}, \text{\tiny{\yng(1,1)}},\mathbf{1}\right)\oplus (\mathbf{1},  \mathbf{1}, \mathbf{Adj})$\\
Tachyon & $\mathbf{1}$ & $(\mathbf{1}, \mathbf{1})$ & $ \left(\text{\tiny{\yng(1)}},\text{\tiny{\yng(1)}}, \mathbf{1} \right) \oplus \left(\mathbf{1},  \mathbf{1}, \text{\tiny{\yng(1,1)}}\right)\oplus \left(\mathbf{1},  \mathbf{1}, \ybar{\text{\tiny{\yng(1,1)}}}\right)$ \\
Fermion & $\mathbf{2}_{s'}$ & $(\mathbf{1}, \mathbf{1})$ & $\left(\text{\tiny{\yng(1)}}, \mathbf{1},\text{\tiny{\yng(1)}}\right) \oplus \left(\mathbf{1},\text{\tiny{\yng(1)}},\text{\tiny{\yng(1)}}\right) \oplus \left(\text{\tiny{\yng(1)}}, \mathbf{1},\ybar{\text{\tiny{\yng(1)}}} \right) \oplus \left(\mathbf{1},\text{\tiny{\yng(1)}},\ybar{\text{\tiny{\yng(1)}}}\right)$ \\
Fermion & $\mathbf{1}$ & $\left(\text{\tiny\yng(1)},\mathbf{1}\right)$ & $\left(\text{\tiny\yng(1)}, \mathbf{1},\mathbf{1}\right) \oplus \left( \mathbf{1},\text{\tiny{\yng(1)}},\mathbf{1}\right)$ \\
Fermion & $\mathbf{1}$ & $\left(\mathbf{1},\text{\tiny\yng(1)}\right)$ & $\left(\mathbf{1},\mathbf{1},\text{\tiny\yng(1)}\right)\oplus\left(\mathbf{1},\mathbf{1},\ybar{\text{\tiny\yng(1)}}\right)$ \\
Tachyon & $\mathbf{2}_{s'}$ & $\left(\mathbf{1},\text{\tiny\yng(1)}\right)$ & $\left(\text{\tiny\yng(1)}, \mathbf{1},\mathbf{1}\right) \oplus \left( \mathbf{1},\text{\tiny{\yng(1)}},\mathbf{1}\right) $ \\
Tachyon & $\mathbf{2}_{s'}$ & $\left(\text{\tiny\yng(1)},\mathbf{1}\right)$ & $\left(\mathbf{1},\mathbf{1},\text{\tiny\yng(1)}\right)\oplus \left(\mathbf{1},\mathbf{1},\ybar{\text{\tiny\yng(1)}}\right)$ \\
\end{tabular}}
\caption{Open-string light spectrum for D6-branes in the type 0A orthogonal orientifold.}
\label{tab:0A/Omega_D6_spectrum}
\end{table}
\begin{table}[!htbp]
\centering
\renewcommand{\arraystretch}{1.2}
\textbf{D8}
\centerresize{}{
\begin{tabular}{l|c|c|c}
Worldvolume state  & $\mathfrak{so}(1)_T$ & $\mathfrak{so}(n_1) \oplus \mathfrak{so}(n_2)$  & $\mathfrak{so}(d_1) \oplus \mathfrak{so}(d_2) \oplus \mathfrak{u}(m)$ \\
\hline
Gauge vector  & $\mathbf{1}$ & $(\mathbf{1}, \mathbf{1})$  & $(\mathbf{Adj},  \mathbf{1}, \mathbf{1}) \oplus(\mathbf{1}, \mathbf{Adj}, \mathbf{1}) \oplus (\mathbf{1},  \mathbf{1}, \mathbf{Adj})$ \\
Scalar & $\mathbf{1}_v$ & $(\mathbf{1}, \mathbf{1})$ & $\left(\text{\tiny{\yng(2)}},  \mathbf{1}, \mathbf{1}\right) \oplus\left(\mathbf{1}, \text{\tiny{\yng(2)}},\mathbf{1}\right)\oplus (\mathbf{1},  \mathbf{1}, \mathbf{Adj})$\\
Tachyon & $\mathbf{1}$ & $(\mathbf{1}, \mathbf{1})$ & $ \left(\text{\tiny{\yng(1)}},\text{\tiny{\yng(1)}}, \mathbf{1} \right) \oplus \left(\mathbf{1},  \mathbf{1}, \text{\tiny{\yng(1,1)}}\right)\oplus \left(\mathbf{1},  \mathbf{1}, \ybar{\text{\tiny{\yng(1,1)}}}\right)$ \\
Fermion & $\mathbf{1}_{s'}$ & $(\mathbf{1}, \mathbf{1})$ & $\left(\text{\tiny{\yng(1)}}, \mathbf{1},\text{\tiny{\yng(1)}}\right) \oplus \left(\mathbf{1},\text{\tiny{\yng(1)}},\text{\tiny{\yng(1)}}\right) \oplus \left(\text{\tiny{\yng(1)}}, \mathbf{1},\ybar{\text{\tiny{\yng(1)}}}\right) \oplus \left(\mathbf{1},\text{\tiny{\yng(1)}},\ybar{\text{\tiny{\yng(1)}}}\right)$ \\
Fermion & $\mathbf{1}$ & $\left(\mathbf{1},\text{\tiny\yng(1)}\right)$ & $\left(\text{\tiny\yng(1)}, \mathbf{1},\mathbf{1}\right) \oplus \left( \mathbf{1},\text{\tiny{\yng(1)}},\mathbf{1}\right)$ \\
Fermion & $\mathbf{1}$ & $\left(\text{\tiny\yng(1)},\mathbf{1}\right)$ & $\left(\mathbf{1},\mathbf{1},\text{\tiny\yng(1)}\right)\oplus\left(\mathbf{1},\mathbf{1},\ybar{\text{\tiny\yng(1)}}\right)$ \\
Tachyon & $\mathbf{1}_{s'}$ & $\left(\text{\tiny\yng(1)}, \mathbf{1}\right)$ & $\left(\text{\tiny\yng(1)}, \mathbf{1},\mathbf{1}\right) \oplus \left( \mathbf{1},\text{\tiny{\yng(1)}},\mathbf{1}\right) $ \\
Tachyon & $\mathbf{1}_{s'}$ & $\left(\mathbf{1},\text{\tiny\yng(1)}\right)$ & $\left(\mathbf{1},\mathbf{1},\text{\tiny\yng(1)}\right)\oplus \left(\mathbf{1},\mathbf{1},\ybar{\text{\tiny\yng(1)}}\right)$ \\
\end{tabular}}
\caption{Open-string light spectrum for D8-branes in the type 0A orthogonal orientifold.}
\label{tab:0A/Omega_D8_spectrum}
\end{table}
\begin{table}[!htbp]
\centering
\renewcommand{\arraystretch}{1.4}
\centerresize{1,1}{
\begin{tabular}{c|c|c|c|c|c}
     & D$(-1)$ & D1 & D3 & D5 & D7 \\
    \hline
    $(\epsilon,\epsilon')$ & $(-1,-1)$ & $(1,1)$ & $(1,-1)$ & $(-1,1)$ & $(-1,-1)$  \\
    vector degeneracy &  $\frac{d_1(d_1 - 1)}{2} + \frac{d_2(d_2 + 1)}{2}$ & $\frac{d_1(d_1 - 1)}{2} + \frac{d_2(d_2 - 1)}{2}$ & $\frac{d_1(d_1 + 1)}{2} + \frac{d_2(d_2 - 1)}{2}$ & $\frac{d_1(d_1 + 1)}{2} + \frac{d_2(d_2 + 1)}{2}$ & $\frac{d_1(d_1 - 1)}{2} + \frac{d_2(d_2 + 1)}{2}$   \\
    gauge structure & $\mathfrak{so}(d_1) \oplus \mathfrak{sp}(d_2/2)$ & $\mathfrak{so}(d_1) \oplus \mathfrak{so}(d_2)$ & $\mathfrak{sp}(d_1/2) \oplus \mathfrak{so}(d_2)$ & $\mathfrak{sp}(d_1/2) \oplus \mathfrak{sp}(d_2/2)$ & $\mathfrak{so}(d_1) \oplus \mathfrak{sp}(d_2/2)$
\end{tabular}}
\caption{Worldvolume gauge structures of the uncharged D$p$-branes with $p$ odd in type 0A/$\Omega$.}
\label{tab:0A/Omega_gauge_structures_uncharged_branes}
\end{table}
\begin{table}[!htbp]
\centering
\textbf{D1}
\renewcommand{\arraystretch}{1.2}
\centerresize{0.85}{
\begin{tabular}{l|c|c|c}
Worldvolume state  & $\mathfrak{so}(8)_T$ & $\mathfrak{so}(n_1) \oplus \mathfrak{so}(n_2)$  & $\mathfrak{so}(d_1) \oplus \mathfrak{so}(d_2)$ \\
\hline
Gauge vector  & $\mathbf{1}$ & $(\mathbf{1}, \mathbf{1})$  & $(\mathbf{Adj},  \mathbf{1}) \oplus(\mathbf{1}, \mathbf{Adj})$ \\
Scalar & $\mathbf{8}_{v}$ & $(\mathbf{1}, \mathbf{1})$ & $\left(\text{\tiny{\yng(2)}},  \mathbf{1}\right) \oplus \left(\mathbf{1}, \text{\tiny{\yng(2)}}\right)$\\
Tachyon & $\mathbf{1}$ & $(\mathbf{1}, \mathbf{1})$ & $ \left(\text{\tiny{\yng(2)}}, \mathbf{1} \right) \oplus \left(\mathbf{1},\mathbf{Adj}\right)$ \\
Fermion & $\mathbf{16}$ & $(\mathbf{1}, \mathbf{1})$ & $\left(\text{\tiny{\yng(1)}},\text{\tiny{\yng(1)}}\right)$ \\
Fermion & $\mathbf{1}$ & $\left(\text{\tiny\yng(1)},\mathbf{1}\right)$ & $\left(\mathbf{1},\text{\tiny\yng(1)}\right)$ \\
Fermion & $\mathbf{1}$ & $\left(\mathbf{1},\text{\tiny\yng(1)}\right)$  & $\left(\text{\tiny\yng(1)},\mathbf{1}\right)$\\
\end{tabular}}
\caption{Open-string light spectrum for D1-branes in type 0A/$\Omega$. Here, $\mathbf{16}=\mathbf{8}_s\oplus \mathbf{8}_c$.}
\label{tab:0A/Omega_D1_spectrum}
\end{table}
\begin{table}[!htbp]
\centering
\textbf{D3}
\renewcommand{\arraystretch}{1.2}
\centerresize{0.85}{
\begin{tabular}{l|c|c|c}
Worldvolume state  & $\mathfrak{so}(6)_T$ & $\mathfrak{so}(n_1) \oplus \mathfrak{so}(n_2)$  & $\mathfrak{sp}(d_1/2) \oplus \mathfrak{so}(d_2)$ \\
\hline
Gauge vector  & $\mathbf{1}$ & $(\mathbf{1}, \mathbf{1})$  & $(\mathbf{Adj},  \mathbf{1}) \oplus(\mathbf{1}, \mathbf{Adj})$ \\
Scalar & $\mathbf{6}_{v}$ & $(\mathbf{1}, \mathbf{1})$ & $\left(\text{\tiny{\yng(1,1)}},  \mathbf{1}\right) \oplus \left(\mathbf{1}, \text{\tiny{\yng(2)}}\right)$\\
Tachyon & $\mathbf{1}$ & $(\mathbf{1}, \mathbf{1})$ & $ \left(\mathbf{Adj}, \mathbf{1} \right) \oplus \left(\mathbf{1},\text{\tiny{\yng(2)}}\right)$ \\
Fermion & $\mathbf{8}$ & $(\mathbf{1}, \mathbf{1})$ & $\left(\text{\tiny{\yng(1)}},\text{\tiny{\yng(1)}}\right)$ \\
Fermion & $\mathbf{1}$ & $\left(\text{\tiny\yng(1)},\mathbf{1}\right)$ & $\left(\mathbf{1},\text{\tiny\yng(1)}\right)$ \\
Fermion & $\mathbf{1}$ & $\left(\mathbf{1},\text{\tiny\yng(1)}\right)$  & $\left(\text{\tiny\yng(1)},\mathbf{1}\right)$\\
\end{tabular}}
\caption{Open-string light spectrum for D3-branes in type 0A/$\Omega$. Here, $\mathbf{8} = \mathbf{4}_s\oplus\mathbf{4}_c$.}
\label{tab:0A/Omega_D3_spectrum}
\end{table}
\begin{table}[!htbp]
\centering
\textbf{D5}
\renewcommand{\arraystretch}{1.2}
\centerresize{0.85}{
\begin{tabular}{l|c|c|c}
Worldvolume state  & $\mathfrak{so}(4)_T$ & $\mathfrak{so}(n_1) \oplus \mathfrak{so}(n_2)$  & $\mathfrak{sp}(d_1/2) \oplus \mathfrak{sp}(d_2/2)$ \\
\hline
Gauge vector  & $\mathbf{1}$ & $(\mathbf{1}, \mathbf{1})$  & $(\mathbf{Adj},  \mathbf{1}) \oplus(\mathbf{1}, \mathbf{Adj})$ \\
Scalar & $\mathbf{4}_{v}$ & $(\mathbf{1}, \mathbf{1})$ & $\left(\text{\tiny{\yng(1,1)}},  \mathbf{1}\right) \oplus \left(\mathbf{1}, \text{\tiny{\yng(1,1)}}\right)$\\
Tachyon & $\mathbf{1}$ & $(\mathbf{1}, \mathbf{1})$ & $ \left(\text{\tiny{\yng(1,1)}}, \mathbf{1} \right) \oplus \left(\mathbf{1},\mathbf{Adj}\right)$ \\
Fermion & $\mathbf{4}$ & $(\mathbf{1}, \mathbf{1})$ & $\left(\text{\tiny{\yng(1)}},\text{\tiny{\yng(1)}}\right)$ \\
Fermion & $\mathbf{1}$ & $\left(\text{\tiny\yng(1)},\mathbf{1}\right)$ & $\left(\mathbf{1},\text{\tiny\yng(1)}\right)$ \\
Fermion & $\mathbf{1}$ & $\left(\mathbf{1},\text{\tiny\yng(1)}\right)$  & $\left(\text{\tiny\yng(1)},\mathbf{1}\right)$\\
Scalar & $\mathbf{4}$ & $\left(\mathbf{1},\text{\tiny\yng(1)}\right)$  & $\left(\mathbf{1},\text{\tiny\yng(1)}\right)$\\
Scalar & $\mathbf{4}$ & $\left(\text{\tiny\yng(1)},\mathbf{1}\right)$  & $\left(\text{\tiny\yng(1)},\mathbf{1}\right)$\\
\end{tabular}}
\caption{Open-string light spectrum for D5-branes in type 0A/$\Omega$. Here, $\mathbf{4} = \mathbf{2}_s\oplus\mathbf{2}_c$.}
\label{tab:0A/Omega_D5_spectrum}
\end{table}
\begin{table}[!htbp]
\centering
\textbf{D7}
\renewcommand{\arraystretch}{1.2}
\centerresize{0.85}{
\begin{tabular}{l|c|c|c}
Worldvolume state  & $\mathfrak{so}(2)_T$ & $\mathfrak{so}(n_1) \oplus \mathfrak{so}(n_2)$  & $\mathfrak{so}(d_1) \oplus \mathfrak{sp}(d_2/2)$ \\
\hline
Gauge vector  & $\mathbf{1}$ & $(\mathbf{1}, \mathbf{1})$  & $(\mathbf{Adj},  \mathbf{1}) \oplus(\mathbf{1}, \mathbf{Adj})$ \\
Scalar & $\mathbf{2}_{v}$ & $(\mathbf{1}, \mathbf{1})$ & $\left(\text{\tiny{\yng(2)}},  \mathbf{1}\right) \oplus (\mathbf{1}, \text{\tiny{\yng(1,1)}})$\\
Tachyon & $\mathbf{1}$ & $(\mathbf{1}, \mathbf{1})$ & $ \left(\mathbf{Adj}, \mathbf{1} \right) \oplus \left(\mathbf{1},\text{\tiny{\yng(1,1)}}\right)$ \\
Fermion & $\mathbf{2}$ & $(\mathbf{1}, \mathbf{1})$ & $\left(\text{\tiny{\yng(1)}},\text{\tiny{\yng(1)}}\right)$ \\
Fermion & $\mathbf{1}$ & $\left(\text{\tiny\yng(1)},\mathbf{1}\right)$ & $\left(\mathbf{1},\text{\tiny\yng(1)}\right)$ \\
Fermion & $\mathbf{1}$ & $\left(\mathbf{1},\text{\tiny\yng(1)}\right)$  & $\left(\text{\tiny\yng(1)},\mathbf{1}\right)$\\
Tachyon & $\mathbf{2}$ & $\left(\mathbf{1},\text{\tiny\yng(1)}\right)$  & $\left(\mathbf{1},\text{\tiny\yng(1)}\right)$\\
Tachyon & $\mathbf{2}$ & $\left(\text{\tiny\yng(1)},\mathbf{1}\right)$  & $\left(\text{\tiny\yng(1)},\mathbf{1}\right)$\\
\end{tabular}}
\caption{Open-string light spectrum for D7-branes in type 0A/$\Omega$. Here, $\mathbf{2} = \mathbf{1}_s\oplus\mathbf{1}_c$. The D$(-1)$ case is analogous to this with the D$p$--D9 tachyons being massive scalars.}
\label{tab:0A/Omega_D7_spectrum}
\end{table}

\FloatBarrier

\subsection{Type 0B orthogonal orientifold}\label{app:0B/Omega_branes}

This appendix contains tables from \cite{Fraiman:2026ltu} showing the light open-string spectra from D-branes in the type 0B orthogonal orientifold. A detailed discussion of the spectra can be found in the appendices of~\cite{Fraiman:2026ltu}.

We summarize the spectrum arising from the D9 sector in \cref{tab:0B/Omega_spacetime_spectrum}.

\begin{table}[!htbp]
\centering
\renewcommand{\arraystretch}{1.2}
\centerresize{}{
\begin{tabular}{l|c|c}
State  & character & $\mathfrak{so}(n_1)\oplus\mathfrak{so}(n_2)\oplus\mathfrak{so}(n_3)\oplus\mathfrak{so}(n_4)$ \\
\hline
Gauge vector  & $V_8$ & $(\mathbf{Adj}, \mathbf{1}, \mathbf{1}, \mathbf{1}) \oplus (\mathbf{1}, \mathbf{Adj}, \mathbf{1},  \mathbf{1})\oplus(\mathbf{1}, \mathbf{1},\mathbf{Adj}, \mathbf{1})\oplus(\mathbf{1}, \mathbf{1}, \mathbf{1},\mathbf{Adj})$ \\
Tachyon & $O_8$ & $(\text{\tiny$\yng(1)$},\text{\tiny$\yng(1)$},\mathbf{1},\mathbf{1}) \oplus (\mathbf{1},\mathbf{1},\text{\tiny$\yng(1)$},\text{\tiny$\yng(1)$})$ \\
Fermion $(+)$ & $S_8$ & $(\text{\tiny$\yng(1)$},\mathbf{1},\text{\tiny$\yng(1)$},\mathbf{1})\oplus(\mathbf{1},\text{\tiny$\yng(1)$},\mathbf{1},\text{\tiny$\yng(1)$})$\\
Fermion $(-)$ & $C_8$ & $(\text{\tiny$\yng(1)$},\mathbf{1},\mathbf{1},\text{\tiny$\yng(1)$})\oplus(\mathbf{1},\text{\tiny$\yng(1)$},\text{\tiny$\yng(1)$},\mathbf{1})$
\end{tabular}}
\caption{Open-string light spectrum from D9-branes in the type 0B/$\Omega$ orientifold.}
\label{tab:0B/Omega_spacetime_spectrum}
\end{table}

The presence of two RR 2-forms in this orientifold signals the existence of charged D1- and D5-branes. Their light spectra are collected in \cref{tab:0B/Omega_D1_spectrum,tab:0B/Omega_D5_spectrum}.
There are also uncharged D$p$-branes for the remaining odd values of $p$ and for even $p$. The cases with $p=-1,3,7$ can be found in \cref{tab:0B/Omega_D3_spectrum,tab:0B/Omega_D7_spectrum}, and those with $p=0,2,4,6,8$ are in \cref{tab:0B/Omega_D0_spectrum,tab:0B/Omega_D2_spectrum,tab:0B/Omega_D4_spectrum,tab:0B/Omega_D6_spectrum,tab:0B/Omega_D8_spectrum}.

\begin{table}[!htbp]
\centering
\textbf{D1}
\renewcommand{\arraystretch}{1.2}
\centerresize{}{
\begin{tabular}{l|c|c|c}
Worldvolume state  & $\mathfrak{so}(8)_T$ & $\mathfrak{so}(n_1)\oplus\mathfrak{so}(n_2)\oplus\mathfrak{so}(n_3)\oplus\mathfrak{so}(n_4)$ & $\mathfrak{so}(d_1)\oplus\mathfrak{so}(d_2)\oplus\mathfrak{so}(d_3)\oplus\mathfrak{so}(d_4)$ \\
\hline
Gauge vector  & $\mathbf{1}$ & $(\mathbf{1}, \mathbf{1}, \mathbf{1}, \mathbf{1})$ & $(\mathbf{Adj}, \mathbf{1}, \mathbf{1}, \mathbf{1}) \oplus (\mathbf{1}, \mathbf{Adj}, \mathbf{1},  \mathbf{1})\oplus(\mathbf{1}, \mathbf{1},\mathbf{Adj}, \mathbf{1})\oplus(\mathbf{1}, \mathbf{1}, \mathbf{1},\mathbf{Adj})$ \\
Scalar & $\mathbf{8}_v$ & $(\mathbf{1}, \mathbf{1}, \mathbf{1}, \mathbf{1})$ & $(\text{\tiny$\yng(2)$},\mathbf{1}, \mathbf{1},\mathbf{1}) \oplus (\mathbf{1},\text{\tiny$\yng(2)$}, \mathbf{1},\mathbf{1}) \oplus(\mathbf{1}, \mathbf{1}, \text{\tiny$\yng(2)$}, \mathbf{1}) \oplus(\mathbf{1}, \mathbf{1},\mathbf{1}, \text{\tiny$\yng(2)$})$ \\
Tachyon & $\mathbf{1}$ & $(\mathbf{1}, \mathbf{1}, \mathbf{1}, \mathbf{1})$ & $(\text{\tiny$\yng(1)$},\text{\tiny$\yng(1)$},\mathbf{1},\mathbf{1})\oplus (\mathbf{1},\mathbf{1},\text{\tiny$\yng(1)$},\text{\tiny$\yng(1)$})$\\
Fermion $(+)$ & $\mathbf{8}_{s}$ & $(\mathbf{1}, \mathbf{1}, \mathbf{1}, \mathbf{1})$ & $(\text{\tiny$\yng(1)$},\mathbf{1},\text{\tiny$\yng(1)$},\mathbf{1})\oplus(\mathbf{1},\text{\tiny$\yng(1)$},\mathbf{1},\text{\tiny$\yng(1)$})$ \\
Fermion $(-)$ & $\mathbf{8}_{c}$ & $(\mathbf{1}, \mathbf{1}, \mathbf{1}, \mathbf{1})$ & $(\text{\tiny$\yng(1)$},\mathbf{1},\text{\tiny$\yng(1)$},\mathbf{1})\oplus(\mathbf{1},\text{\tiny$\yng(1)$},\mathbf{1},\text{\tiny$\yng(1)$})$ \\
Fermion $(+)$ & $\mathbf{8}_{c}$ & $(\mathbf{1}, \mathbf{1}, \mathbf{1}, \mathbf{1})$ & $(\text{\tiny$\yng(1)$},\mathbf{1},\mathbf{1},\text{\tiny$\yng(1)$})\oplus(\mathbf{1},\text{\tiny$\yng(1)$},\text{\tiny$\yng(1)$},\mathbf{1})$ \\
Fermion $(-)$ & $\mathbf{8}_{s}$ & $(\mathbf{1}, \mathbf{1}, \mathbf{1}, \mathbf{1})$ & $(\text{\tiny$\yng(1)$},\mathbf{1},\mathbf{1},\text{\tiny$\yng(1)$})\oplus(\mathbf{1},\text{\tiny$\yng(1)$},\text{\tiny$\yng(1)$},\mathbf{1})$ \\
\hline
\hline
Fermion ($+$) & $\mathbf{1}$ & $\begin{array}{l}
    (\text{\tiny$\yng(1)$}, \mathbf{1}, \mathbf{1}, \mathbf{1})\\(\mathbf{1}, \text{\tiny$\yng(1)$}, \mathbf{1}, \mathbf{1})\\(\mathbf{1}, \mathbf{1}, \text{\tiny$\yng(1)$}, \mathbf{1})\\(\mathbf{1}, \mathbf{1}, \mathbf{1}, \text{\tiny$\yng(1)$}) \end{array}$ & $\begin{array}{l} (\mathbf{1}, \mathbf{1}, \mathbf{1}, \text{\tiny$\yng(1)$}) \\ (\mathbf{1}, \mathbf{1}, \text{\tiny$\yng(1)$}, \mathbf{1}) \\ (\mathbf{1}, \text{\tiny$\yng(1)$}, \mathbf{1}, \mathbf{1}) \\ (\text{\tiny$\yng(1)$}, \mathbf{1}, \mathbf{1}, \mathbf{1}) \end{array}$ \\
\hline
Fermion ($-$) & $\mathbf{1}$ & $\begin{array}{l}
    (\text{\tiny$\yng(1)$}, \mathbf{1}, \mathbf{1}, \mathbf{1})\\(\mathbf{1}, \text{\tiny$\yng(1)$}, \mathbf{1}, \mathbf{1})\\(\mathbf{1}, \mathbf{1}, \text{\tiny$\yng(1)$}, \mathbf{1})\\(\mathbf{1}, \mathbf{1}, \mathbf{1}, \text{\tiny$\yng(1)$}) \end{array}$ & $\begin{array}{l} (\mathbf{1}, \mathbf{1}, \text{\tiny$\yng(1)$}, \mathbf{1}) \\ (\mathbf{1}, \mathbf{1}, \mathbf{1}, \text{\tiny$\yng(1)$}) \\ (\text{\tiny$\yng(1)$}, \mathbf{1}, \mathbf{1}, \mathbf{1}) \\ (\mathbf{1}, \text{\tiny$\yng(1)$}, \mathbf{1}, \mathbf{1}) \end{array}$ \\
\end{tabular}}
\caption{Open-string light spectrum for D1-branes in type 0B/$\Omega$.}
\label{tab:0B/Omega_D1_spectrum}
\end{table}
\begin{table}[!htbp]
\centering
\textbf{D5}
\renewcommand{\arraystretch}{1.2}
\centerresize{}{
\begin{tabular}{l|c|c|c}
Worldvolume state  & $\mathfrak{so}(4)_T$ & $\mathfrak{so}(n_1)\oplus\mathfrak{so}(n_2)\oplus\mathfrak{so}(n_3)\oplus\mathfrak{so}(n_4)$ & $\mathfrak{sp}(d_1/2)\oplus\mathfrak{sp}(d_2/2)\oplus\mathfrak{sp}(d_3/2)\oplus\mathfrak{sp}(d_4/2)$ \\
\hline
Gauge vector  & $\mathbf{1}$ & $(\mathbf{1}, \mathbf{1}, \mathbf{1}, \mathbf{1})$ & $(\mathbf{Adj}, \mathbf{1}, \mathbf{1}, \mathbf{1}) \oplus (\mathbf{1}, \mathbf{Adj}, \mathbf{1},  \mathbf{1})\oplus(\mathbf{1}, \mathbf{1},\mathbf{Adj}, \mathbf{1})\oplus(\mathbf{1}, \mathbf{1}, \mathbf{1},\mathbf{Adj})$ \\
Scalar & $\mathbf{4}_{v}$ & $(\mathbf{1}, \mathbf{1}, \mathbf{1}, \mathbf{1})$ & $\left(\text{\tiny$\yng(1,1)$},\mathbf{1}, \mathbf{1},\mathbf{1}\right) \oplus \left(\mathbf{1},\text{\tiny$\yng(1,1)$}, \mathbf{1},\mathbf{1}\right) \oplus\left(\mathbf{1}, \mathbf{1}, \text{\tiny$\yng(1,1)$}, \mathbf{1}\right) \oplus\left(\mathbf{1}, \mathbf{1},\mathbf{1}, \text{\tiny$\yng(1,1)$}\right)$ \\
Tachyon & $\mathbf{1}$ & $(\mathbf{1}, \mathbf{1}, \mathbf{1}, \mathbf{1})$ & $(\text{\tiny$\yng(1)$},\text{\tiny$\yng(1)$},\mathbf{1},\mathbf{1})\oplus (\mathbf{1},\mathbf{1},\text{\tiny$\yng(1)$},\text{\tiny$\yng(1)$})$\\
Fermion $(+)$ & $\mathbf{2}_{s}$ & $(\mathbf{1}, \mathbf{1}, \mathbf{1}, \mathbf{1})$ & $(\text{\tiny$\yng(1)$},\mathbf{1},\text{\tiny$\yng(1)$},\mathbf{1})\oplus(\mathbf{1},\text{\tiny$\yng(1)$},\mathbf{1},\text{\tiny$\yng(1)$})$ \\
Fermion $(-)$ & $\mathbf{2}_{c}$ & $(\mathbf{1}, \mathbf{1}, \mathbf{1}, \mathbf{1})$ & $(\text{\tiny$\yng(1)$},\mathbf{1},\text{\tiny$\yng(1)$},\mathbf{1})\oplus(\mathbf{1},\text{\tiny$\yng(1)$},\mathbf{1},\text{\tiny$\yng(1)$})$ \\
Fermion $(+)$ & $\mathbf{2}_{c}$ & $(\mathbf{1}, \mathbf{1}, \mathbf{1}, \mathbf{1})$ & $(\text{\tiny$\yng(1)$},\mathbf{1},\mathbf{1},\text{\tiny$\yng(1)$})\oplus(\mathbf{1},\text{\tiny$\yng(1)$},\text{\tiny$\yng(1)$},\mathbf{1})$ \\
Fermion $(-)$ & $\mathbf{2}_{s}$ & $(\mathbf{1}, \mathbf{1}, \mathbf{1}, \mathbf{1})$ & $(\text{\tiny$\yng(1)$},\mathbf{1},\mathbf{1},\text{\tiny$\yng(1)$})\oplus(\mathbf{1},\text{\tiny$\yng(1)$},\text{\tiny$\yng(1)$},\mathbf{1})$ \\
\hline
\hline
Various scalars & & & \\
\hline
Fermion ($-$) & $\mathbf{1}$ & $\begin{array}{l}
    (\text{\tiny$\yng(1)$}, \mathbf{1}, \mathbf{1}, \mathbf{1})\\(\mathbf{1}, \text{\tiny$\yng(1)$}, \mathbf{1}, \mathbf{1})\\(\mathbf{1}, \mathbf{1}, \text{\tiny$\yng(1)$}, \mathbf{1})\\(\mathbf{1}, \mathbf{1}, \mathbf{1}, \text{\tiny$\yng(1)$}) \end{array}$ & $\begin{array}{l} (\mathbf{1}, \mathbf{1}, \mathbf{1}, \text{\tiny$\yng(1)$}) \\ (\mathbf{1}, \mathbf{1}, \text{\tiny$\yng(1)$}, \mathbf{1}) \\ (\mathbf{1}, \text{\tiny$\yng(1)$}, \mathbf{1}, \mathbf{1}) \\ (\text{\tiny$\yng(1)$}, \mathbf{1}, \mathbf{1}, \mathbf{1}) \end{array}$ \\
\hline
Fermion ($+$) & $\mathbf{1}$ & $\begin{array}{l}
    (\text{\tiny$\yng(1)$}, \mathbf{1}, \mathbf{1}, \mathbf{1})\\(\mathbf{1}, \text{\tiny$\yng(1)$}, \mathbf{1}, \mathbf{1})\\(\mathbf{1}, \mathbf{1}, \text{\tiny$\yng(1)$}, \mathbf{1})\\(\mathbf{1}, \mathbf{1}, \mathbf{1}, \text{\tiny$\yng(1)$}) \end{array}$ & $\begin{array}{l} (\mathbf{1}, \mathbf{1}, \text{\tiny$\yng(1)$}, \mathbf{1}) \\ (\mathbf{1}, \mathbf{1}, \mathbf{1}, \text{\tiny$\yng(1)$}) \\ (\text{\tiny$\yng(1)$}, \mathbf{1}, \mathbf{1}, \mathbf{1}) \\ (\mathbf{1}, \text{\tiny$\yng(1)$}, \mathbf{1}, \mathbf{1}) \end{array}$ \\
\end{tabular}}
\caption{Open-string light spectrum for D5-branes in type 0B/$\Omega$.}
\label{tab:0B/Omega_D5_spectrum}
\end{table}
\begin{table}[!htbp]
\centering
\textbf{D3}
\renewcommand{\arraystretch}{1.2}
\centerresize{}{
\begin{tabular}{l|c|c|c}
Worldvolume state  & $\mathfrak{so}(6)_T$ & $\mathfrak{so}(n_1)\oplus\mathfrak{so}(n_2)\oplus\mathfrak{so}(n_3)\oplus\mathfrak{so}(n_4)$ & $\mathfrak{u}(d_1)\oplus\mathfrak{u}(d_2)$ \\
\hline
Gauge vector  & $\mathbf{1}$ & $(\mathbf{1}, \mathbf{1}, \mathbf{1}, \mathbf{1})$ &  $(\mathbf{Adj}, \mathbf{1}) \oplus (\mathbf{1},\mathbf{Adj})$ \\
Scalar & $\mathbf{6}_{v}$ & $(\mathbf{1}, \mathbf{1}, \mathbf{1}, \mathbf{1})$ & $(\mathbf{Adj}, \mathbf{1}) \oplus (\mathbf{1},\mathbf{Adj})$\\
Tachyon & $\mathbf{1}$ & $(\mathbf{1}, \mathbf{1}, \mathbf{1}, \mathbf{1})$ & $  \left(\text{\tiny$\yng(2)$}, \mathbf{1}\right) \oplus  \left(\mathbf{1}, \text{\tiny$\yng(2)$}\right)\oplus \left(\ybar{\text{\tiny$\yng(2)$}}, \mathbf{1}\right) \oplus  \left(\mathbf{1}, \ybar{\text{\tiny$\yng(2)$}}\right)$\\
Fermion $(+)$ & $\mathbf{4}_{s}$ & $(\mathbf{1}, \mathbf{1}, \mathbf{1}, \mathbf{1})$ & $(\text{\tiny$\yng(1)$},\text{\tiny$\ybar{\yng(1)}$})\oplus (\text{\tiny$\ybar{\yng(1)}$},\text{\tiny$\yng(1)$})$ \\
Fermion $(-)$ & $\mathbf{4}_{c}$ & $(\mathbf{1}, \mathbf{1}, \mathbf{1}, \mathbf{1})$ & $(\text{\tiny$\yng(1)$},\text{\tiny$\ybar{\yng(1)}$})\oplus (\text{\tiny$\ybar{\yng(1)}$},\text{\tiny$\yng(1)$})$ \\
Fermion $(+)$ & $\mathbf{4}_{c}$ & $(\mathbf{1}, \mathbf{1}, \mathbf{1}, \mathbf{1})$ & $(\text{\tiny$\yng(1)$},\text{\tiny$\yng(1)$}) \oplus (\text{\tiny$\ybar{\yng(1)}$},\text{\tiny$\ybar{\yng(1)}$})$ \\
Fermion $(-)$ & $\mathbf{4}_{s}$ & $(\mathbf{1}, \mathbf{1}, \mathbf{1}, \mathbf{1})$ & $(\text{\tiny$\yng(1)$},\text{\tiny$\yng(1)$}) \oplus (\text{\tiny$\ybar{\yng(1)}$},\text{\tiny$\ybar{\yng(1)}$})$\\
\hline
\hline
Fermion ($+$) & $\mathbf{1}$ & $\begin{array}{l}
    (\text{\tiny$\yng(1)$}, \mathbf{1}, \mathbf{1}, \mathbf{1})\\(\mathbf{1}, \text{\tiny$\yng(1)$}, \mathbf{1}, \mathbf{1})\\(\mathbf{1}, \mathbf{1}, \text{\tiny$\yng(1)$}, \mathbf{1})\\(\mathbf{1}, \mathbf{1}, \mathbf{1}, \text{\tiny$\yng(1)$}) \end{array}$ & $\begin{array}{l} (\mathbf{1}, \text{\tiny$\ybar{\yng(1)}$}) \\ (\mathbf{1}, \text{\tiny$\yng(1)$}) \\ (\text{\tiny$\ybar{\yng(1)}$},\mathbf{1}) \\ (\text{\tiny$\yng(1)$},\mathbf{1}) \end{array}$ \\
\hline
Fermion ($-$) & $\mathbf{1}$ & $\begin{array}{l}
    (\text{\tiny$\yng(1)$}, \mathbf{1}, \mathbf{1}, \mathbf{1})\\(\mathbf{1}, \text{\tiny$\yng(1)$}, \mathbf{1}, \mathbf{1})\\(\mathbf{1}, \mathbf{1}, \text{\tiny$\yng(1)$}, \mathbf{1})\\(\mathbf{1}, \mathbf{1}, \mathbf{1}, \text{\tiny$\yng(1)$}) \end{array}$ & $\begin{array}{l} (\mathbf{1}, \text{\tiny$\yng(1)$}) \\ (\mathbf{1}, \text{\tiny$\ybar{\yng(1)}$})  \\ (\text{\tiny$\yng(1)$},\mathbf{1}) \\ (\text{\tiny$\ybar{\yng(1)}$},\mathbf{1}) \end{array}$ \\
\end{tabular}}
\caption{Open-string light spectrum for D3-branes in type 0B/$\Omega$.
}
\label{tab:0B/Omega_D3_spectrum}
\end{table}
\begin{table}[!htbp]
\centering
\textbf{D7}
\renewcommand{\arraystretch}{1.2}
\centerresize{}{
\begin{tabular}{l|c|c|c}
Worldvolume state  & $\mathfrak{so}(2)_T$ & $\mathfrak{so}(n_1)\oplus\mathfrak{so}(n_2)\oplus\mathfrak{so}(n_3)\oplus\mathfrak{so}(n_4)$ & $\mathfrak{u}(d_1)\oplus\mathfrak{u}(d_2)$ \\
\hline
Gauge vector  & $\mathbf{1}$ & $(\mathbf{1}, \mathbf{1}, \mathbf{1}, \mathbf{1})$ &  $(\mathbf{Adj}, \mathbf{1}) \oplus (\mathbf{1},\mathbf{Adj})$ \\
Scalar & $\mathbf{2}_{v}$ & $(\mathbf{1}, \mathbf{1}, \mathbf{1}, \mathbf{1})$ & $(\mathbf{Adj}, \mathbf{1}) \oplus (\mathbf{1},\mathbf{Adj})$\\
Tachyon & $\mathbf{1}$ & $(\mathbf{1}, \mathbf{1}, \mathbf{1}, \mathbf{1})$ & $\left(\text{\tiny$\yng(1,1)$}, \mathbf{1}\right) \oplus \left(\mathbf{1}, \text{\tiny$\yng(1,1)$}\right) \oplus \left(\ybar{\text{\tiny$\yng(1,1)$}}, \mathbf{1}\right) \oplus \left(\mathbf{1}, \ybar{\text{\tiny$\yng(1,1)$}}\right) $\\
Fermion $(+)$ & $\mathbf{1}_{s}$ & $(\mathbf{1}, \mathbf{1}, \mathbf{1}, \mathbf{1})$ & $(\text{\tiny$\yng(1)$},\text{\tiny$\ybar{\yng(1)}$})\oplus (\text{\tiny$\ybar{\yng(1)}$},\text{\tiny$\yng(1)$})$ \\
Fermion $(-)$ & $\mathbf{1}_{c}$ & $(\mathbf{1}, \mathbf{1}, \mathbf{1}, \mathbf{1})$ & $(\text{\tiny$\yng(1)$},\text{\tiny$\ybar{\yng(1)}$})\oplus (\text{\tiny$\ybar{\yng(1)}$},\text{\tiny$\yng(1)$})$ \\
Fermion $(+)$ & $\mathbf{1}_{c}$ & $(\mathbf{1}, \mathbf{1}, \mathbf{1}, \mathbf{1})$ & $(\text{\tiny$\yng(1)$},\text{\tiny$\yng(1)$}) \oplus (\text{\tiny$\ybar{\yng(1)}$},\text{\tiny$\ybar{\yng(1)}$})$ \\
Fermion $(-)$ & $\mathbf{1}_{s}$ & $(\mathbf{1}, \mathbf{1}, \mathbf{1}, \mathbf{1})$ & $(\text{\tiny$\yng(1)$},\text{\tiny$\yng(1)$}) \oplus (\text{\tiny$\ybar{\yng(1)}$},\text{\tiny$\ybar{\yng(1)}$})$ \\
\hline
\hline
Various tachyons & & & \\
\hline
Fermion ($-$) & $\mathbf{1}$ & $\begin{array}{l}
    (\text{\tiny$\yng(1)$}, \mathbf{1}, \mathbf{1}, \mathbf{1})\\(\mathbf{1}, \text{\tiny$\yng(1)$}, \mathbf{1}, \mathbf{1})\\(\mathbf{1}, \mathbf{1}, \text{\tiny$\yng(1)$}, \mathbf{1})\\(\mathbf{1}, \mathbf{1}, \mathbf{1}, \text{\tiny$\yng(1)$}) \end{array}$ & $\begin{array}{l} (\mathbf{1}, \text{\tiny$\ybar{\yng(1)}$}) \\ (\mathbf{1}, \text{\tiny$\yng(1)$}) \\ (\text{\tiny$\ybar{\yng(1)}$},\mathbf{1}) \\ (\text{\tiny$\yng(1)$},\mathbf{1}) \end{array}$ \\
\hline
Fermion ($+$) & $\mathbf{1}$ & $\begin{array}{l}
    (\text{\tiny$\yng(1)$}, \mathbf{1}, \mathbf{1}, \mathbf{1})\\(\mathbf{1}, \text{\tiny$\yng(1)$}, \mathbf{1}, \mathbf{1})\\(\mathbf{1}, \mathbf{1}, \text{\tiny$\yng(1)$}, \mathbf{1})\\(\mathbf{1}, \mathbf{1}, \mathbf{1}, \text{\tiny$\yng(1)$}) \end{array}$ & $\begin{array}{l} (\mathbf{1}, \text{\tiny$\yng(1)$}) \\ (\mathbf{1}, \text{\tiny$\ybar{\yng(1)}$})  \\ (\text{\tiny$\yng(1)$},\mathbf{1}) \\ (\text{\tiny$\ybar{\yng(1)}$},\mathbf{1}) \end{array}$ \\
\end{tabular}}
\caption{Open-string light spectrum for D7-branes in type 0B/$\Omega$. The D$(-1)$ case is analogous to this with the D$p$--D9 tachyons being massive scalars.}
\label{tab:0B/Omega_D7_spectrum}
\end{table}
\begin{table}[!htbp]
\centering
\textbf{D0}
\renewcommand{\arraystretch}{1.2}
\centerresize{}{
\begin{tabular}{l|c|c|c}
Worldvolume state  & $\mathfrak{so}(9)_T$ & $\mathfrak{so}(n_1)\oplus\mathfrak{so}(n_2)\oplus\mathfrak{so}(n_3)\oplus\mathfrak{so}(n_4)$ & $\mathfrak{so}(d_1)\oplus\mathfrak{so}(d_2)$ \\
\hline
Gauge vector  & $\mathbf{1}$ & $(\mathbf{1}, \mathbf{1}, \mathbf{1}, \mathbf{1})$ &  $(\mathbf{Adj}, \mathbf{1}) \oplus (\mathbf{1},\mathbf{Adj})$ \\
Scalar & $\mathbf{9}_{v}$ & $(\mathbf{1}, \mathbf{1}, \mathbf{1}, \mathbf{1})$ & $\left(\text{\tiny$\yng(2)$}, \mathbf{1}\right) \oplus \left(\mathbf{1}, \text{\tiny$\yng(2)$}\right)$\\
Tachyon & $\mathbf{1}$ & $(\mathbf{1}, \mathbf{1}, \mathbf{1}, \mathbf{1})$ & $(\mathbf{Adj}, \mathbf{1}) \oplus (\mathbf{1},\mathbf{Adj})$\\
Dirac fermion  & $\mathbf{16}_{s'}$ & $(\mathbf{1}, \mathbf{1}, \mathbf{1}, \mathbf{1})$ & $(\text{\tiny$\yng(1)$},\text{\tiny$\yng(1)$})$ \\
\hline
\hline
Dirac fermion & $\mathbf{1}$ & $\begin{array}{l}
    (\text{\tiny$\yng(1)$}, \mathbf{1}, \mathbf{1}, \mathbf{1})\\(\mathbf{1}, \text{\tiny$\yng(1)$}, \mathbf{1}, \mathbf{1})\\(\mathbf{1}, \mathbf{1}, \text{\tiny$\yng(1)$}, \mathbf{1})\\(\mathbf{1}, \mathbf{1}, \mathbf{1}, \text{\tiny$\yng(1)$}) \end{array}$ & $\begin{array}{l} (\mathbf{1}, \text{\tiny$\yng(1)$}) \\ (\mathbf{1}, \text{\tiny$\yng(1)$}) \\ (\text{\tiny$\yng(1)$},\mathbf{1}) \\ (\text{\tiny$\yng(1)$},\mathbf{1}) \end{array}$ \\
\end{tabular}}
\caption{Open-string light spectrum for D0-branes in type 0B/$\Omega$.}
\label{tab:0B/Omega_D0_spectrum}
\end{table}
\begin{table}[!htbp]
\centering
\textbf{D2}
\renewcommand{\arraystretch}{1.2}
\centerresize{}{
\begin{tabular}{l|c|c|c}
Worldvolume state  & $\mathfrak{so}(7)_T$ & $\mathfrak{so}(n_1)\oplus\mathfrak{so}(n_2)\oplus\mathfrak{so}(n_3)\oplus\mathfrak{so}(n_4)$ & $\mathfrak{so}(d_1)\oplus\mathfrak{so}(d_2)$ \\
\hline
Gauge vector  & $\mathbf{1}$ & $(\mathbf{1}, \mathbf{1}, \mathbf{1}, \mathbf{1})$ &  $(\mathbf{Adj}, \mathbf{1}) \oplus (\mathbf{1},\mathbf{Adj})$ \\
Scalar & $\mathbf{7}_{v}$ & $(\mathbf{1}, \mathbf{1}, \mathbf{1}, \mathbf{1})$ & $(\text{\tiny$\yng(2)$}, \mathbf{1}) \oplus (\mathbf{1},\text{\tiny$\yng(2)$})$\\
Tachyon & $\mathbf{1}$ & $(\mathbf{1}, \mathbf{1}, \mathbf{1}, \mathbf{1})$ &  $(\text{\tiny$\yng(2)$}, \mathbf{1}) \oplus (\mathbf{1},\text{\tiny$\yng(2)$})$\\
Dirac fermion  & $\mathbf{8}_{s'}$ & $(\mathbf{1}, \mathbf{1}, \mathbf{1}, \mathbf{1})$ & $(\text{\tiny$\yng(1)$},\text{\tiny$\yng(1)$})$ \\
\hline
\hline
Dirac fermion & $\mathbf{1}$ & $\begin{array}{l}
    (\text{\tiny$\yng(1)$}, \mathbf{1}, \mathbf{1}, \mathbf{1})\\(\mathbf{1}, \text{\tiny$\yng(1)$}, \mathbf{1}, \mathbf{1})\\(\mathbf{1}, \mathbf{1}, \text{\tiny$\yng(1)$}, \mathbf{1})\\(\mathbf{1}, \mathbf{1}, \mathbf{1}, \text{\tiny$\yng(1)$}) \end{array}$ & $\begin{array}{l} (\mathbf{1}, \text{\tiny$\yng(1)$}) \\ (\mathbf{1}, \text{\tiny$\yng(1)$}) \\ (\text{\tiny$\yng(1)$},\mathbf{1}) \\ (\text{\tiny$\yng(1)$},\mathbf{1}) \end{array}$ \\
\end{tabular}}
\caption{Open-string light spectrum for D2-branes in type 0B/$\Omega$.}
\label{tab:0B/Omega_D2_spectrum}
\end{table}
\begin{table}[!htbp]
\centering
\textbf{D4}
\renewcommand{\arraystretch}{1.2}
\centerresize{}{
\begin{tabular}{l|c|c|c}
Worldvolume state  & $\mathfrak{so}(5)_T$ & $\mathfrak{so}(n_1)\oplus\mathfrak{so}(n_2)\oplus\mathfrak{so}(n_3)\oplus\mathfrak{so}(n_4)$ & $\mathfrak{sp}(d_1/2)\oplus\mathfrak{sp}(d_2/2)$ \\
\hline
Gauge vector  & $\mathbf{1}$ & $(\mathbf{1}, \mathbf{1}, \mathbf{1}, \mathbf{1})$ &  $(\mathbf{Adj}, \mathbf{1}) \oplus (\mathbf{1},\mathbf{Adj})$ \\
Scalar & $\mathbf{5}_{v}$ & $(\mathbf{1}, \mathbf{1}, \mathbf{1}, \mathbf{1})$ & $\left(\text{\tiny$\yng(1,1)$}, \mathbf{1}\right) \oplus \left(\mathbf{1}, \text{\tiny$\yng(1,1)$}\right)$\\
Tachyon & $\mathbf{1}$ & $(\mathbf{1}, \mathbf{1}, \mathbf{1}, \mathbf{1})$ &  $(\mathbf{Adj}, \mathbf{1}) \oplus (\mathbf{1},\mathbf{Adj})$\\
Dirac fermion  & $\mathbf{4}_{s'}$ & $(\mathbf{1}, \mathbf{1}, \mathbf{1}, \mathbf{1})$ & $(\text{\tiny$\yng(1)$},\text{\tiny$\yng(1)$})$ \\
\hline
\hline
Dirac fermion & $\mathbf{1}$ & $\begin{array}{l}
    (\text{\tiny$\yng(1)$}, \mathbf{1}, \mathbf{1}, \mathbf{1})\\(\mathbf{1}, \text{\tiny$\yng(1)$}, \mathbf{1}, \mathbf{1})\\(\mathbf{1}, \mathbf{1}, \text{\tiny$\yng(1)$}, \mathbf{1})\\(\mathbf{1}, \mathbf{1}, \mathbf{1}, \text{\tiny$\yng(1)$}) \end{array}$ & $\begin{array}{l} (\mathbf{1}, \text{\tiny$\yng(1)$}) \\ (\mathbf{1}, \text{\tiny$\yng(1)$}) \\ (\text{\tiny$\yng(1)$},\mathbf{1}) \\ (\text{\tiny$\yng(1)$},\mathbf{1}) \end{array}$ \\
\end{tabular}}
\caption{Open-string light spectrum for D4-branes in type 0B/$\Omega$.}
\label{tab:0B/Omega_D4_spectrum}
\end{table}
\begin{table}[!htbp]
\centering
\textbf{D6}
\renewcommand{\arraystretch}{1.2}
\centerresize{}{
\begin{tabular}{l|c|c|c}
Worldvolume state  & $\mathfrak{so}(3)_T$ & $\mathfrak{so}(n_1)\oplus\mathfrak{so}(n_2)\oplus\mathfrak{so}(n_3)\oplus\mathfrak{so}(n_4)$ & $\mathfrak{sp}(d_1/2)\oplus\mathfrak{sp}(d_2/2)$ \\
\hline
Gauge vector  & $\mathbf{1}$ & $(\mathbf{1}, \mathbf{1}, \mathbf{1}, \mathbf{1})$ &  $(\mathbf{Adj}, \mathbf{1}) \oplus (\mathbf{1},\mathbf{Adj})$ \\
Scalar & $\mathbf{3}_{v}$ & $(\mathbf{1}, \mathbf{1}, \mathbf{1}, \mathbf{1})$ & $\left(\text{\tiny$\yng(1,1)$}, \mathbf{1}\right) \oplus \left(\mathbf{1}, \text{\tiny$\yng(1,1)$}\right)$\\
Tachyon & $\mathbf{1}$ & $(\mathbf{1}, \mathbf{1}, \mathbf{1}, \mathbf{1})$ &  $\left(\text{\tiny$\yng(1,1)$}, \mathbf{1}\right) \oplus \left(\mathbf{1}, \text{\tiny$\yng(1,1)$}\right)$\\
Dirac fermion  & $\mathbf{2}_{s'}$ & $(\mathbf{1}, \mathbf{1}, \mathbf{1}, \mathbf{1})$ & $(\text{\tiny$\yng(1)$},\text{\tiny$\yng(1)$})$ \\
\hline
\hline
Various tachyons & & & \\
\hline
Dirac fermion & $\mathbf{1}$ & $\begin{array}{l}
    (\text{\tiny$\yng(1)$}, \mathbf{1}, \mathbf{1}, \mathbf{1})\\(\mathbf{1}, \text{\tiny$\yng(1)$}, \mathbf{1}, \mathbf{1})\\(\mathbf{1}, \mathbf{1}, \text{\tiny$\yng(1)$}, \mathbf{1})\\(\mathbf{1}, \mathbf{1}, \mathbf{1}, \text{\tiny$\yng(1)$}) \end{array}$ & $\begin{array}{l} (\mathbf{1}, \text{\tiny$\yng(1)$}) \\ (\mathbf{1}, \text{\tiny$\yng(1)$}) \\ (\text{\tiny$\yng(1)$},\mathbf{1}) \\ (\text{\tiny$\yng(1)$},\mathbf{1}) \end{array}$ \\
\end{tabular}}
\caption{Open-string light spectrum for D6-branes in type 0B/$\Omega$.}
\label{tab:0B/Omega_D6_spectrum}
\end{table}
\begin{table}[!htbp]
\centering
\textbf{D8}
\renewcommand{\arraystretch}{1.2}
\centerresize{}{
\begin{tabular}{l|c|c|c}
Worldvolume state  & $\mathfrak{so}(1)_T$ & $\mathfrak{so}(n_1)\oplus\mathfrak{so}(n_2)\oplus\mathfrak{so}(n_3)\oplus\mathfrak{so}(n_4)$ & $\mathfrak{so}(d_1)\oplus\mathfrak{so}(d_2)$ \\
\hline
Gauge vector  & $\mathbf{1}$ & $(\mathbf{1}, \mathbf{1}, \mathbf{1}, \mathbf{1})$ &  $(\mathbf{Adj}, \mathbf{1}) \oplus (\mathbf{1},\mathbf{Adj})$ \\
Scalar & $\mathbf{1}_{v}$ & $(\mathbf{1}, \mathbf{1}, \mathbf{1}, \mathbf{1})$ & $\left(\text{\tiny$\yng(2)$}, \mathbf{1}\right) \oplus \left(\mathbf{1}, \text{\tiny$\yng(2)$}\right)$\\
Tachyon & $\mathbf{1}$ & $(\mathbf{1}, \mathbf{1}, \mathbf{1}, \mathbf{1})$ & $(\mathbf{Adj}, \mathbf{1}) \oplus (\mathbf{1},\mathbf{Adj})$\\
Dirac fermion  & $\mathbf{1}_{s'}$ & $(\mathbf{1}, \mathbf{1}, \mathbf{1}, \mathbf{1})$ & $(\text{\tiny$\yng(1)$},\text{\tiny$\yng(1)$})$ \\
\hline
\hline
Various tachyons & & & \\
\hline
Dirac fermion & $\mathbf{1}$ & $\begin{array}{l}
    (\text{\tiny$\yng(1)$}, \mathbf{1}, \mathbf{1}, \mathbf{1})\\(\mathbf{1}, \text{\tiny$\yng(1)$}, \mathbf{1}, \mathbf{1})\\(\mathbf{1}, \mathbf{1}, \text{\tiny$\yng(1)$}, \mathbf{1})\\(\mathbf{1}, \mathbf{1}, \mathbf{1}, \text{\tiny$\yng(1)$}) \end{array}$ & $\begin{array}{l} (\mathbf{1}, \text{\tiny$\yng(1)$}) \\ (\mathbf{1}, \text{\tiny$\yng(1)$}) \\ (\text{\tiny$\yng(1)$},\mathbf{1}) \\ (\text{\tiny$\yng(1)$},\mathbf{1}) \end{array}$ \\
\end{tabular}}
\caption{Open-string light spectrum for D8-branes in type 0B/$\Omega$.}
\label{tab:0B/Omega_D8_spectrum}
\end{table}
\FloatBarrier

\section{Brane spectra of type 0 symplectic orientifolds}\label{app:Branes_symplectic}

We follow the conventions introduced in appendix~\ref{app:Branes_orthogonal}.

\subsection{Type 0A symplectic orientifold}\label{app:0A/Omega_s_branes}

From the amplitudes in \cref{eq:0A_orientifold_A-M} with $\sigma=-1$, the D9-branes generate an $\mathfrak{sp}(\tfrac{n_1}{2})\oplus\mathfrak{sp}(\tfrac{n_2}{2})$ gauge algebra, open-string tachyons in the representation $\left(\mathbf{\frac{n_1(n_1-1)}{2}},\mathbf{1}\right) \oplus \left(\mathbf{1},\mathbf{\frac{n_2(n_2+1)}{2}}\right)$, and Majorana fermions in the bifundamental representation of the spacetime gauge algebra. We summarize the spectrum from the D9-branes of this type 0A/$\Omega_s$ orientifold in \cref{tab:s0A_open_spectrum}.
\begin{table}[!ht]
\centering
\renewcommand{\arraystretch}{1.2}
\begin{tabular}{l|c|c|c}
State & character & $\mathfrak{sp}(\tfrac{n_1}{2})$ & $\mathfrak{sp}(\tfrac{n_2}{2})$ \\
\hline
Gauge vector & $V_8$ & $\mathbf{Adj}$ & $\mathbf{1}$ \\
Gauge vector & $V_8$ & $\mathbf{1}$ & $\mathbf{Adj}$ \\
Fermion & $S_8$ & \tiny\yng(1) & \tiny\yng(1) \\
Fermion & $C_8$ & \tiny\yng(1) & \tiny\yng(1) \\
Tachyon & $O_8$ & \tiny\yng(1,1) & $\mathbf{1}$ \\
Tachyon & $O_8$ & $\mathbf{1}$ & $\mathbf{Adj}$
\end{tabular}
\caption{Open-string light spectrum from D9-branes in the type 0A/$\Omega_s$ orientifold.}
\label{tab:s0A_open_spectrum}
\end{table}

We now consider the branes, deriving their spectra in more detail than in the orthogonal cases.

\subsubsection{Branes with $p$ even}

Type 0A/$\Omega_s$ has charged and uncharged D$p$-branes with $p=0,2,4,6,8$, denoted in \cref{sec:type_0_review} by the indices $d_{1,2}$ and $m,\bar{m}$. The annulus and the M\"obius amplitudes can be found in \cref{eq:0A_orientifold_charged_direct} (direct channel) and \cref{eq:0A_orientifold_charged_transverse} (transverse channel), setting $\sigma = -1$. The annulus amplitudes of open strings stretching between D$p$- and D9-branes are in \cref{eq:0A_orientifold_charged_p9}.
With the notation of the amplitudes, the multiplicity of the gauge boson is
\begin{equation}
    \frac{d_1(d_1+\epsilon)}{2} + \frac{d_2(d_2+\epsilon)}{2} + m \bar{m}\,.
\end{equation}
D0-, D2-, and D8-branes have $\epsilon = 1$, and therefore the worldvolume gauge algebras are $\mathfrak{sp}(\tfrac{d_1}{2})\oplus\mathfrak{sp}(\tfrac{d_2}{2})\oplus \mathfrak{u}(m)$; D4- and D6-branes have $\epsilon=-1$ and thus $\mathfrak{so}(d_1)\oplus\mathfrak{so}(d_2)\oplus \mathfrak{u}(m)$ worldvolume gauge algebras.

The spectra of these branes generically contain open-string tachyons, with degeneracy
\begin{equation}
    d_1 d_2 + \frac{m(m+\eta_d(p)\,\epsilon)}{2} +\frac{\bar{m}(\bar{m}+\eta_d(p)\,\epsilon)}{2}\,.
\end{equation}
This corresponds to tachyons in the bifundamental representation of the first two gauge factors for charged branes (labeled by $d_{1,2}$), in the antisymmetric representation of $\mathfrak{u}(m)$ for D2- and D4-branes, and in the symmetric representation of $\mathfrak{u}(m)$ for D0-, D6-, and D8-branes. 

The scalars that represent the position of the branes arise from the $O_{p-1}V_{9-p}$ combination, with multiplicity
\begin{equation}
    \frac{d_1(d_1 - \epsilon)}{2} + \frac{d_2(d_2 - \epsilon)}{2} + m \bar{m}\,.
\end{equation}
They are in the adjoint representation of the $\mathfrak{u}(m)$ factors, and in the $\left(\mathbf{\frac{d_1(d_1-1)}{2}},\mathbf{1}\right) \oplus \left(\mathbf{1},\mathbf{\frac{d_2(d_2-1)}{2}}\right)$ of $\mathfrak{sp}(\tfrac{d_1}{2})\oplus\mathfrak{sp}(\tfrac{d_2}{2})$ for D0-, D2-, and D8-branes, while in the $\left(\mathbf{\frac{d_1(d_1+1)}{2}},\mathbf{1}\right) \oplus \left(\mathbf{1},\mathbf{\frac{d_2(d_2+1)}{2}}\right)$ representation of $\mathfrak{so}(d_1)\oplus\mathfrak{so}(d_2)$ for D4- and D6-branes.

Worldvolume spinors from $S_{p-1}'$ transform in the bifundamental representation of the first or second factor and of the $\mathfrak{u}(m)$ $+$ conjugate part of the gauge algebra.

Additional states arise from the D$p$--D9 strings in $\mathcal{A}_{p9}$. In particular, fermions arise from $S'_{p-1} O_{9-p}$ and scalars from $O_{p-1} S'_{9-p}$, which become tachyonic when $p>5$.  Regarding the fermions, for $p=0,4,8$, two of them are in the bifundamental of $\mathfrak{sp}(\tfrac{n_2}{2})$ and of the first and second factors of the worldvolume gauge algebra, while for $p=2,6$, they are in an analogous representation with $n_2\to n_1$. Two more fermions transform in the bifundamental of $\mathfrak{sp}(\tfrac{n_1}{2})\oplus \mathfrak{u}(m)$ for $p=0,4,8$ and of $\mathfrak{sp}(\tfrac{n_2}{2})\oplus \mathfrak{u}(m)$ for $p=2,6$, together with their conjugates.

We summarize the light spectra from D$p$-branes with $p=0,2,4,6,8$ in \cref{tab:s0A_D0_spectrum,tab:s0A_D2_spectrum,tab:s0A_D4_spectrum,tab:s0A_D6_spectrum,tab:s0A_D8_spectrum}.

\begin{table}[!htbp]
\centering
\renewcommand{\arraystretch}{1.2}
\textbf{D0}
\centerresize{}{
\begin{tabular}{l|c|c|c}
Worldvolume state  & $\mathfrak{so}(9)_T$ & $\mathfrak{sp}(\tfrac{n_1}{2}) \oplus \mathfrak{sp}(\tfrac{n_2}{2})$  & $\mathfrak{sp}(\frac{d_1}{2}) \oplus \mathfrak{sp}(\frac{d_2}{2}) \oplus \mathfrak{u}(m)$ \\
\hline
Gauge vector  & $\mathbf{1}$ & $(\mathbf{1}, \mathbf{1})$  & $(\mathbf{Adj},  \mathbf{1}, \mathbf{1}) \oplus(\mathbf{1}, \mathbf{Adj}, \mathbf{1}) \oplus (\mathbf{1},  \mathbf{1}, \mathbf{Adj})$ \\
Scalar & $\mathbf{9}_v$ & $(\mathbf{1}, \mathbf{1})$ & $\left(\text{\tiny{\yng(1,1)}},  \mathbf{1}, \mathbf{1}\right) \oplus\left(\mathbf{1}, \text{\tiny{\yng(1,1)}},\mathbf{1}\right)\oplus (\mathbf{1},  \mathbf{1}, \mathbf{Adj})$\\
Tachyon & $\mathbf{1}$ & $(\mathbf{1}, \mathbf{1})$ & $ \left(\text{\tiny{\yng(1)}},\text{\tiny{\yng(1)}}, \mathbf{1} \right) \oplus \left(\mathbf{1},  \mathbf{1}, \text{\tiny{\yng(2)}}\right)\oplus \left(\mathbf{1},  \mathbf{1}, \ybar{\text{\tiny{\yng(2)}}}\right)$ \\
Fermion & $\mathbf{16}_{s'}$ & $(\mathbf{1}, \mathbf{1})$ & $\left(\text{\tiny{\yng(1)}}, \mathbf{1},\text{\tiny{\yng(1)}}\right) \oplus \left(\mathbf{1},\text{\tiny{\yng(1)}},\text{\tiny{\yng(1)}}\right) \oplus \left(\text{\tiny{\yng(1)}}, \mathbf{1},\ybar{\text{\tiny{\yng(1)}}}\right) \oplus \left(\mathbf{1},\text{\tiny{\yng(1)}},\ybar{\text{\tiny{\yng(1)}}}\right)$ \\
Fermion & $\mathbf{1}$ & $\left(\mathbf{1},\text{\tiny\yng(1)}\right)$ & $\left(\text{\tiny\yng(1)}, \mathbf{1},\mathbf{1}\right) \oplus \left( \mathbf{1},\text{\tiny{\yng(1)}},\mathbf{1}\right)$ \\
Fermion & $\mathbf{1}$ & $\left(\text{\tiny\yng(1)},\mathbf{1}\right)$ & $\left(\mathbf{1},\mathbf{1},\text{\tiny\yng(1)}\right)\oplus\left(\mathbf{1},\mathbf{1},\ybar{\text{\tiny\yng(1)}}\right)$
\end{tabular}}
\caption{Open-string light spectrum for D0-branes in type 0A/$\Omega_s$.}
\label{tab:s0A_D0_spectrum}
\end{table}
\begin{table}[!htbp]
\centering
\renewcommand{\arraystretch}{1.2}
\textbf{D2}
\centerresize{}{
\begin{tabular}{l|c|c|c}
Worldvolume state  & $\mathfrak{so}(7)_T$ & $\mathfrak{sp}(\tfrac{n_1}{2}) \oplus \mathfrak{sp}(\tfrac{n_2}{2})$  & $\mathfrak{sp}(\frac{d_1}{2}) \oplus \mathfrak{sp}(\frac{d_2}{2}) \oplus \mathfrak{u}(m)$ \\
\hline
Gauge vector  & $\mathbf{1}$ & $(\mathbf{1}, \mathbf{1})$  & $(\mathbf{Adj},  \mathbf{1}, \mathbf{1}) \oplus(\mathbf{1}, \mathbf{Adj}, \mathbf{1}) \oplus (\mathbf{1},  \mathbf{1}, \mathbf{Adj})$ \\
Scalar & $\mathbf{7}_{v}$ & $(\mathbf{1}, \mathbf{1})$ & $\left(\text{\tiny{\yng(1,1)}},  \mathbf{1}, \mathbf{1}\right) \oplus\left(\mathbf{1}, \text{\tiny{\yng(1,1)}},\mathbf{1}\right)\oplus (\mathbf{1},  \mathbf{1}, \mathbf{Adj})$\\
Tachyon & $\mathbf{1}$ & $(\mathbf{1}, \mathbf{1})$ & $ \left(\text{\tiny{\yng(1)}},\text{\tiny{\yng(1)}}, \mathbf{1} \right) \oplus \left(\mathbf{1},  \mathbf{1}, \text{\tiny{\yng(1,1)}}\right)\oplus \left(\mathbf{1},  \mathbf{1}, \ybar{\text{\tiny{\yng(1,1)}}}\right)$ \\
Fermion & $\mathbf{8}_{s'}$ & $(\mathbf{1}, \mathbf{1})$ & $\left(\text{\tiny{\yng(1)}}, \mathbf{1},\text{\tiny{\yng(1)}} \right) \oplus \left(\mathbf{1},\text{\tiny{\yng(1)}},\text{\tiny{\yng(1)}}\right) \oplus \left(\text{\tiny{\yng(1)}}, \mathbf{1},\ybar{\text{\tiny{\yng(1)}}}\right) \oplus \left(\mathbf{1},\text{\tiny{\yng(1)}},\ybar{\text{\tiny{\yng(1)}}} \right)$ \\
Fermion & $\mathbf{1}$ & $\left(\text{\tiny\yng(1)},\mathbf{1}\right)$ & $\left(\text{\tiny\yng(1)}, \mathbf{1},\mathbf{1}\right) \oplus \left( \mathbf{1},\text{\tiny{\yng(1)}},\mathbf{1}\right)$ \\
Fermion & $\mathbf{1}$ & $\left(\mathbf{1},\text{\tiny\yng(1)}\right)$ & $\left(\mathbf{1},\mathbf{1},\text{\tiny\yng(1)}\right)\oplus\left(\mathbf{1},\mathbf{1},\ybar{\text{\tiny\yng(1)}}\right)$ 
\end{tabular}}
\caption{Open-string light spectrum for D2-branes in type 0A/$\Omega_s$.}
\label{tab:s0A_D2_spectrum}
\end{table}
\begin{table}[!htbp]
\centering
\textbf{D4}
\renewcommand{\arraystretch}{1.2}
\centerresize{}{
\begin{tabular}{l|c|c|c}
Worldvolume state  & $\mathfrak{so}(5)_T$ & $\mathfrak{sp}(\tfrac{n_1}{2}) \oplus \mathfrak{sp}(\tfrac{n_2}{2})$  & $\mathfrak{so}(d_1)\oplus \mathfrak{so}(d_2) \oplus \mathfrak{u}(m)$ \\
\hline
Gauge vector  & $\mathbf{1}$ & $(\mathbf{1}, \mathbf{1})$  & $(\mathbf{Adj},  \mathbf{1}, \mathbf{1}) \oplus(\mathbf{1}, \mathbf{Adj}, \mathbf{1}) \oplus (\mathbf{1},  \mathbf{1}, \mathbf{Adj})$ \\
Scalar & $\mathbf{5}_{v}$ & $(\mathbf{1}, \mathbf{1})$ & $\left(\text{\tiny{\yng(2)}},  \mathbf{1}, \mathbf{1}\right) \oplus\left(\mathbf{1}, \text{\tiny{\yng(2)}},\mathbf{1}\right)\oplus (\mathbf{1},  \mathbf{1}, \mathbf{Adj})$\\
Tachyon & $\mathbf{1}$ & $(\mathbf{1}, \mathbf{1})$ & $ \left(\text{\tiny{\yng(1)}},\text{\tiny{\yng(1)}}, \mathbf{1} \right) \oplus \left(\mathbf{1},  \mathbf{1}, \text{\tiny{\yng(1,1)}}\right)\oplus \left(\mathbf{1},  \mathbf{1}, \ybar{\text{\tiny{\yng(1,1)}}}\right)$ \\
Fermion & $\mathbf{4}_{s'}$ & $(\mathbf{1}, \mathbf{1})$ & $\left(\text{\tiny{\yng(1)}}, \mathbf{1},\text{\tiny{\yng(1)}} \right) \oplus \left(\mathbf{1},\text{\tiny{\yng(1)}},\text{\tiny{\yng(1)}}\right) \oplus \left(\text{\tiny{\yng(1)}}, \mathbf{1},\ybar{\text{\tiny{\yng(1)}}} \right) \oplus \left(\mathbf{1},\text{\tiny{\yng(1)}},\ybar{\text{\tiny{\yng(1)}}} \right)$ \\
Fermion & $\mathbf{1}$ & $\left(\mathbf{1},\text{\tiny\yng(1)}\right)$ & $\left(\text{\tiny\yng(1)}, \mathbf{1},\mathbf{1}\right) \oplus \left( \mathbf{1},\text{\tiny{\yng(1)}},\mathbf{1}\right)$ \\
Fermion & $\mathbf{1}$ & $\left(\text{\tiny\yng(1)},\mathbf{1}\right)$ & $\left(\mathbf{1},\mathbf{1},\text{\tiny\yng(1)}\right)\oplus\left(\mathbf{1},\mathbf{1},\ybar{\text{\tiny\yng(1)}}\right)$
\end{tabular}}
\caption{Open-string light spectrum for D4-branes in type 0A/$\Omega_s$.}
\label{tab:s0A_D4_spectrum}
\end{table}
\begin{table}[!htbp]
\centering
\textbf{D6}
\renewcommand{\arraystretch}{1.2}
\centerresize{}{
\begin{tabular}{l|c|c|c}
Worldvolume state  & $\mathfrak{so}(3)_T$ & $\mathfrak{sp}(\frac{n_1}{2}) \oplus \mathfrak{sp}(\frac{n_2}{2})$  & $\mathfrak{so}(d_1) \oplus \mathfrak{so}(d_2) \oplus \mathfrak{u}(m)$ \\
\hline
Gauge vector  & $\mathbf{1}$ & $(\mathbf{1}, \mathbf{1})$  & $(\mathbf{Adj},  \mathbf{1}, \mathbf{1}) \oplus(\mathbf{1}, \mathbf{Adj}, \mathbf{1}) \oplus (\mathbf{1},  \mathbf{1}, \mathbf{Adj})$ \\
Scalar & $\mathbf{3}_{v}$ & $(\mathbf{1}, \mathbf{1})$ & $\left(\text{\tiny{\yng(2)}},  \mathbf{1}, \mathbf{1}\right) \oplus\left(\mathbf{1}, \text{\tiny{\yng(2)}},\mathbf{1}\right)\oplus (\mathbf{1},  \mathbf{1}, \mathbf{Adj})$\\
Tachyon & $\mathbf{1}$ & $(\mathbf{1}, \mathbf{1})$ & $ \left(\text{\tiny{\yng(1)}},\text{\tiny{\yng(1)}}, \mathbf{1} \right) \oplus \left(\mathbf{1},  \mathbf{1}, \text{\tiny{\yng(2)}}\right)\oplus \left(\mathbf{1},  \mathbf{1}, \ybar{\text{\tiny{\yng(2)}}}\right)$ \\
Fermion & $\mathbf{2}_{s'}$ & $(\mathbf{1}, \mathbf{1})$ & $\left(\text{\tiny{\yng(1)}}, \mathbf{1},\text{\tiny{\yng(1)}}\right) \oplus \left(\mathbf{1},\text{\tiny{\yng(1)}},\text{\tiny{\yng(1)}}\right) \oplus \left(\text{\tiny{\yng(1)}}, \mathbf{1},\ybar{\text{\tiny{\yng(1)}}} \right) \oplus \left(\mathbf{1},\text{\tiny{\yng(1)}},\ybar{\text{\tiny{\yng(1)}}}\right)$ \\
Fermion & $\mathbf{1}$ & $\left(\text{\tiny\yng(1)},\mathbf{1}\right)$ & $\left(\text{\tiny\yng(1)}, \mathbf{1},\mathbf{1}\right) \oplus \left( \mathbf{1},\text{\tiny{\yng(1)}},\mathbf{1}\right)$ \\
Fermion & $\mathbf{1}$ & $\left(\mathbf{1},\text{\tiny\yng(1)}\right)$ & $\left(\mathbf{1},\mathbf{1},\text{\tiny\yng(1)}\right)\oplus\left(\mathbf{1},\mathbf{1},\ybar{\text{\tiny\yng(1)}}\right)$  \\
Tachyon & $\mathbf{2}_{s'}$ & $\left(\mathbf{1},\text{\tiny\yng(1)}\right)$ & $\left(\text{\tiny\yng(1)}, \mathbf{1},\mathbf{1}\right) \oplus \left( \mathbf{1},\text{\tiny{\yng(1)}},\mathbf{1}\right) $ \\
Tachyon & $\mathbf{2}_{s'}$ & $\left(\text{\tiny\yng(1)},\mathbf{1}\right)$ & $\left(\mathbf{1},\mathbf{1},\text{\tiny\yng(1)}\right)\oplus \left(\mathbf{1},\mathbf{1},\ybar{\text{\tiny\yng(1)}}\right)$ \\
\end{tabular}}
\caption{Open-string light spectrum for D6-branes in type 0A/$\Omega_s$.}
\label{tab:s0A_D6_spectrum}
\end{table}
\begin{table}[!htbp]
\centering
\renewcommand{\arraystretch}{1.2}
\textbf{D8}
\centerresize{}{
\begin{tabular}{l|c|c|c}
Worldvolume state  & $\mathfrak{so}(1)_T$ & $\mathfrak{sp}(\tfrac{n_1}{2}) \oplus \mathfrak{sp}(\tfrac{n_2}{2})$  & $\mathfrak{sp}(\frac{d_1}{2}) \oplus \mathfrak{sp}(\frac{d_2}{2}) \oplus \mathfrak{u}(m)$ \\
\hline
Gauge vector  & $\mathbf{1}$ & $(\mathbf{1}, \mathbf{1})$  & $(\mathbf{Adj},  \mathbf{1}, \mathbf{1}) \oplus(\mathbf{1}, \mathbf{Adj}, \mathbf{1}) \oplus (\mathbf{1},  \mathbf{1}, \mathbf{Adj})$ \\
Scalar & $\mathbf{1}_v$ & $(\mathbf{1}, \mathbf{1})$ & $\left(\text{\tiny{\yng(1,1)}},  \mathbf{1}, \mathbf{1}\right) \oplus\left(\mathbf{1}, \text{\tiny{\yng(1,1)}},\mathbf{1}\right)\oplus (\mathbf{1},  \mathbf{1}, \mathbf{Adj})$\\
Tachyon & $\mathbf{1}$ & $(\mathbf{1}, \mathbf{1})$ & $ \left(\text{\tiny{\yng(1)}},\text{\tiny{\yng(1)}}, \mathbf{1} \right) \oplus \left(\mathbf{1},  \mathbf{1}, \text{\tiny{\yng(2)}}\right)\oplus \left(\mathbf{1},  \mathbf{1}, \ybar{\text{\tiny{\yng(2)}}}\right)$ \\
Fermion & $\mathbf{1}_{s'}$ & $(\mathbf{1}, \mathbf{1})$ & $\left(\text{\tiny{\yng(1)}}, \mathbf{1},\text{\tiny{\yng(1)}}\right) \oplus \left(\mathbf{1},\text{\tiny{\yng(1)}},\text{\tiny{\yng(1)}}\right) \oplus \left(\text{\tiny{\yng(1)}}, \mathbf{1},\ybar{\text{\tiny{\yng(1)}}}\right) \oplus \left(\mathbf{1},\text{\tiny{\yng(1)}},\ybar{\text{\tiny{\yng(1)}}}\right)$ \\
Fermion & $\mathbf{1}$ & $\left(\mathbf{1},\text{\tiny\yng(1)}\right)$ & $\left(\text{\tiny\yng(1)}, \mathbf{1},\mathbf{1}\right) \oplus \left( \mathbf{1},\text{\tiny{\yng(1)}},\mathbf{1}\right)$ \\
Fermion & $\mathbf{1}$ & $\left(\text{\tiny\yng(1)},\mathbf{1}\right)$ & $\left(\mathbf{1},\mathbf{1},\text{\tiny\yng(1)}\right)\oplus\left(\mathbf{1},\mathbf{1},\ybar{\text{\tiny\yng(1)}}\right)$ \\
Tachyon & $\mathbf{1}_{s'}$ & $\left(\text{\tiny\yng(1)}, \mathbf{1}\right)$ & $\left(\text{\tiny\yng(1)}, \mathbf{1},\mathbf{1}\right) \oplus \left( \mathbf{1},\text{\tiny{\yng(1)}},\mathbf{1}\right) $ \\
Tachyon & $\mathbf{1}_{s'}$ & $\left(\mathbf{1},\text{\tiny\yng(1)}\right)$ & $\left(\mathbf{1},\mathbf{1},\text{\tiny\yng(1)}\right)\oplus \left(\mathbf{1},\mathbf{1},\ybar{\text{\tiny\yng(1)}}\right)$ \\
\end{tabular}}
\caption{Open-string light spectrum for D8-branes in type 0A/$\Omega_s$.}
\label{tab:s0A_D8_spectrum}
\end{table}

\FloatBarrier

\subsubsection{Branes with $p$ odd}

For the D$p$-branes of type 0A/$\Omega_s$ with $p=-1,1,3,5,7$, the annulus and the M\"obius amplitudes are in \cref{eq:0A_orientifold_uncharged_direct,eq:0A_orientifold_uncharged_transverse}, after setting $\sigma = -1$.
The coefficient of $V_{p-1}O_{9-p}$ fixes the worldvolume gauge algebras: $\mathfrak{sp}(\tfrac{d_1}{2})\oplus \mathfrak{sp}(\tfrac{d_2}{2})$ for D1-branes, $\mathfrak{so}(d_1)\oplus \mathfrak{sp}(\tfrac{d_2}{2}) $ for D3-branes, $\mathfrak{so}(d_1) \oplus \mathfrak{so}(d_2)$ for D5-branes, and $\mathfrak{sp}(\tfrac{d_1}{2})\oplus \mathfrak{so}(d_2)$ for D7-branes.
The corresponding degeneracies and gauge structures are collected in \cref{tab:0A/Omega_s_gauge_structures_uncharged_branes}.
\begin{table}[!htbp]
\centering
\renewcommand{\arraystretch}{1.4}
\centerresize{1,1}{
\begin{tabular}{c|c|c|c|c|c}
     & D$(-1)$ & D1 & D3 & D5 & D7 \\
    \hline
    $(\epsilon,\epsilon')$ & $(-1,-1)$ & $(1,1)$ & $(1,-1)$ & $(-1,1)$ & $(-1,-1)$  \\
    vector degeneracy &  $\frac{d_1(d_1 + 1)}{2} + \frac{d_2(d_2 - 1)}{2}$ & $\frac{d_1(d_1 + 1)}{2} + \frac{d_2(d_2 + 1)}{2}$ & $\frac{d_1(d_1 - 1)}{2} + \frac{d_2(d_2 + 1)}{2}$ & $\frac{d_1(d_1 - 1)}{2} + \frac{d_2(d_2 - 1)}{2}$ & $\frac{d_1(d_1 + 1)}{2} + \frac{d_2(d_2 - 1)}{2}$   \\
    gauge structure & $\mathfrak{sp}(d_1/2) \oplus \mathfrak{so}(d_2)$ & $\mathfrak{sp}(d_1/2) \oplus \mathfrak{sp}(d_2/2)$ & $\mathfrak{so}(d_1) \oplus \mathfrak{sp}(d_2/2)$ & $\mathfrak{so}(d_1) \oplus \mathfrak{so}(d_2)$ & $\mathfrak{sp}(d_1/2) \oplus \mathfrak{so}(d_2)$
\end{tabular}}
\caption{Worldvolume gauge structures of D$p$-branes with $p$ odd in type 0A/$\Omega_s$.}
\label{tab:0A/Omega_s_gauge_structures_uncharged_branes}
\end{table}

Most of these branes have open-string tachyons, as can be seen from the coefficient of the $O_{p-1}O_{9-p}$ combination. For the D1, there are tachyons in the antisymmetric and symmetric representations of the first and second factors of the gauge algebra, respectively. For the D3, they are in the antisymmetric representations of both factors. For the D5, they are in the symmetric and antisymmetric representations of the first and second factors, respectively. For the D7, the tachyons are in the symmetric representations of both factors. Therefore, one obtains the configurations that are free of open-string tachyons as in \cref{tab:sOA/Omega_tachyon_free_branes}: a single D3 with $d_1 = 1$ and $d_2 = 0$, and a single D5 with $d_1 = 0$ and $d_2 = 1$.

From the coefficient of $O_{p-1}V_{9-p}$, we obtain the representations of brane-position modes: for the D1, they are in the antisymmetric representations of both worldvolume gauge factors; for the D5, they are in the symmetric; for the D3, they are in the symmetric and antisymmetric representations of the first and second gauge factors; for the D7, they are in the antisymmetric and symmetric representations of the two gauge factors.

Dirac fermions transform in the bifundamental representation of the gauge algebra factors. These are non-chiral fermions also under the transverse rotation algebra.

Other scalar modes arise from the D$p$--D9 strings: these are massive for $p<5$, massless when $p=5$, and tachyonic for $p>5$. Finally, from $\mathcal{A}_{p9}$, there are additional fermions in the bifundamental of the first spacetime gauge factor and the second worldvolume gauge factor, and vice versa.

We summarize these results in \cref{tab:s0A_D1_spectrum,tab:s0A_D3_spectrum,tab:s0A_D5_spectrum,tab:s0A_D7_spectrum}.

\begin{table}[!htbp]
\centering
\textbf{D1}
\renewcommand{\arraystretch}{1.2}
\centerresize{0.85}{
\begin{tabular}{l|c|c|c}
Worldvolume state  & $\mathfrak{so}(8)_T$ & $\mathfrak{sp}(\tfrac{n_1}{2}) \oplus \mathfrak{sp}(\tfrac{n_2}{2})$  & $\mathfrak{sp}(\tfrac{d_1}{2}) \oplus \mathfrak{sp}(\tfrac{d_2}{2})$ \\
\hline
Gauge vector  & $\mathbf{1}$ & $(\mathbf{1}, \mathbf{1})$  & $(\mathbf{Adj},  \mathbf{1}) \oplus(\mathbf{1}, \mathbf{Adj})$ \\
Scalar & $\mathbf{8}_{v}$ & $(\mathbf{1}, \mathbf{1})$ & $\left(\text{\tiny{\yng(1,1)}},  \mathbf{1}\right) \oplus \left(\mathbf{1}, \text{\tiny{\yng(1,1)}}\right)$\\
Tachyon & $\mathbf{1}$ & $(\mathbf{1}, \mathbf{1})$ & $ \left(\text{\tiny{\yng(1,1)}}, \mathbf{1} \right) \oplus \left(\mathbf{1},\mathbf{Adj}\right)$ \\
Fermion & $\mathbf{16}$ & $(\mathbf{1}, \mathbf{1})$ & $\left(\text{\tiny{\yng(1)}},\text{\tiny{\yng(1)}}\right)$ \\
Fermion & $\mathbf{1}$ & $\left(\text{\tiny\yng(1)},\mathbf{1}\right)$ & $\left(\mathbf{1},\text{\tiny\yng(1)}\right)$ \\
Fermion & $\mathbf{1}$ & $\left(\mathbf{1},\text{\tiny\yng(1)}\right)$  & $\left(\text{\tiny\yng(1)},\mathbf{1}\right)$\\
\end{tabular}}
\caption{Open-string light spectrum for D1-branes in type 0A/$\Omega_s$. Here, $\mathbf{16}=\mathbf{8}_s\oplus \mathbf{8}_c$.}
\label{tab:s0A_D1_spectrum}
\end{table}
\begin{table}[!htbp]
\centering
\textbf{D3}
\renewcommand{\arraystretch}{1.2}
\centerresize{0.85}{
\begin{tabular}{l|c|c|c}
Worldvolume state  & $\mathfrak{so}(6)_T$ & $\mathfrak{sp}(\tfrac{n_1}{2}) \oplus \mathfrak{sp}(\tfrac{n_2}{2})$  & $\mathfrak{so}(d_1)\oplus \mathfrak{sp}(\tfrac{d_2}{2})$ \\
\hline
Gauge vector  & $\mathbf{1}$ & $(\mathbf{1}, \mathbf{1})$  & $(\mathbf{Adj},  \mathbf{1}) \oplus(\mathbf{1}, \mathbf{Adj})$ \\
Scalar & $\mathbf{6}_{v}$ & $(\mathbf{1}, \mathbf{1})$ & $\left(\text{\tiny{\yng(2)}},  \mathbf{1}\right) \oplus \left(\mathbf{1}, \text{\tiny{\yng(1,1)}}\right)$\\
Tachyon & $\mathbf{1}$ & $(\mathbf{1}, \mathbf{1})$ & $ \left(\mathbf{Adj}, \mathbf{1} \right) \oplus \left(\mathbf{1},\text{\tiny{\yng(1,1)}}\right)$ \\
Fermion & $\mathbf{8}$ & $(\mathbf{1}, \mathbf{1})$ & $\left(\text{\tiny{\yng(1)}},\text{\tiny{\yng(1)}}\right)$ \\
Fermion & $\mathbf{1}$ & $\left(\text{\tiny\yng(1)},\mathbf{1}\right)$ & $\left(\mathbf{1},\text{\tiny\yng(1)}\right)$ \\
Fermion & $\mathbf{1}$ & $\left(\mathbf{1},\text{\tiny\yng(1)}\right)$  & $\left(\text{\tiny\yng(1)},\mathbf{1}\right)$\\
\end{tabular}}
\caption{Open-string light spectrum for D3-branes in type 0A/$\Omega_s$. Here, $\mathbf{8} = \mathbf{4}_s\oplus\mathbf{4}_c$.}
\label{tab:s0A_D3_spectrum}
\end{table}
\begin{table}[!htbp]
\centering
\textbf{D5}
\renewcommand{\arraystretch}{1.2}
\centerresize{0.85}{
\begin{tabular}{l|c|c|c}
Worldvolume state  & $\mathfrak{so}(4)_T$ & $\mathfrak{sp}(\tfrac{n_1}{2}) \oplus \mathfrak{sp}(\tfrac{n_2}{2})$  & $\mathfrak{so}(d_1) \oplus \mathfrak{so}(d_2)$ \\
\hline
Gauge vector  & $\mathbf{1}$ & $(\mathbf{1}, \mathbf{1})$  & $(\mathbf{Adj},  \mathbf{1}) \oplus(\mathbf{1}, \mathbf{Adj})$ \\
Scalar & $\mathbf{4}_{v}$ & $(\mathbf{1}, \mathbf{1})$ & $\left(\text{\tiny{\yng(2)}},  \mathbf{1}\right) \oplus \left(\mathbf{1}, \text{\tiny{\yng(2)}}\right)$\\
Tachyon & $\mathbf{1}$ & $(\mathbf{1}, \mathbf{1})$ & $ \left(\text{\tiny{\yng(2)}}, \mathbf{1} \right) \oplus \left(\mathbf{1},\mathbf{Adj}\right)$ \\
Fermion & $\mathbf{4}$ & $(\mathbf{1}, \mathbf{1})$ & $\left(\text{\tiny{\yng(1)}},\text{\tiny{\yng(1)}}\right)$ \\
Fermion & $\mathbf{1}$ & $\left(\text{\tiny\yng(1)},\mathbf{1}\right)$ & $\left(\mathbf{1},\text{\tiny\yng(1)}\right)$ \\
Fermion & $\mathbf{1}$ & $\left(\mathbf{1},\text{\tiny\yng(1)}\right)$  & $\left(\text{\tiny\yng(1)},\mathbf{1}\right)$\\
Scalar & $\mathbf{4}$ & $\left(\mathbf{1},\text{\tiny\yng(1)}\right)$  & $\left(\mathbf{1},\text{\tiny\yng(1)}\right)$\\
Scalar & $\mathbf{4}$ & $\left(\text{\tiny\yng(1)},\mathbf{1}\right)$  & $\left(\text{\tiny\yng(1)},\mathbf{1}\right)$\\
\end{tabular}}
\caption{Open-string light spectrum for D5-branes in type 0A/$\Omega_s$. Here, $\mathbf{4} = \mathbf{2}_s\oplus\mathbf{2}_c$.}
\label{tab:s0A_D5_spectrum}
\end{table}
\begin{table}[!htbp]
\centering
\textbf{D7}
\renewcommand{\arraystretch}{1.2}
\centerresize{0.85}{
\begin{tabular}{l|c|c|c}
Worldvolume state  & $\mathfrak{so}(2)_T$ & $\mathfrak{sp}(\tfrac{n_1}{2}) \oplus \mathfrak{sp}(\tfrac{n_2}{2})$  & $\mathfrak{sp}(\tfrac{d_1}{2})\oplus \mathfrak{so}(d_2)$ \\
\hline
Gauge vector  & $\mathbf{1}$ & $(\mathbf{1}, \mathbf{1})$  & $(\mathbf{Adj},  \mathbf{1}) \oplus(\mathbf{1}, \mathbf{Adj})$ \\
Scalar & $\mathbf{2}_{v}$ & $(\mathbf{1}, \mathbf{1})$ & $\left(\text{\tiny{\yng(1,1)}},  \mathbf{1}\right) \oplus (\mathbf{1}, \text{\tiny{\yng(2)}})$\\
Tachyon & $\mathbf{1}$ & $(\mathbf{1}, \mathbf{1})$ & $ \left(\textbf{Adj}, \mathbf{1} \right) \oplus (\mathbf{1},\text{\tiny{\yng(2)}})$ \\
Fermion & $\mathbf{2}$ & $(\mathbf{1}, \mathbf{1})$ & $\left(\text{\tiny{\yng(1)}},\text{\tiny{\yng(1)}}\right)$ \\
Fermion & $\mathbf{1}$ & $\left(\text{\tiny\yng(1)},\mathbf{1}\right)$ & $\left(\mathbf{1},\text{\tiny\yng(1)}\right)$ \\
Fermion & $\mathbf{1}$ & $\left(\mathbf{1},\text{\tiny\yng(1)}\right)$  & $\left(\text{\tiny\yng(1)},\mathbf{1}\right)$\\
Tachyon & $\mathbf{2}$ & $\left(\mathbf{1},\text{\tiny\yng(1)}\right)$  & $\left(\mathbf{1},\text{\tiny\yng(1)}\right)$\\
Tachyon & $\mathbf{2}$ & $\left(\text{\tiny\yng(1)},\mathbf{1}\right)$  & $\left(\text{\tiny\yng(1)},\mathbf{1}\right)$\\
\end{tabular}}
\caption{Open-string light spectrum for D7-branes in type 0A/$\Omega_s$. Here, $\mathbf{2} = \mathbf{1}_s\oplus\mathbf{1}_c$. The D$(-1)$ case is analogous to this with the D$p$--D9 tachyons being massive scalars.}
\label{tab:s0A_D7_spectrum}
\end{table}

\FloatBarrier

\subsection{Type 0B symplectic orientifold}\label{app:0B/Omega_s_branes}

From the amplitudes in \cref{eq:0B_orientifold_A-M} with $\sigma=-1$, the D9-branes generate an $\mathfrak{sp}(\tfrac{n_1}{2})\oplus\mathfrak{sp}(\tfrac{n_2}{2})\oplus\mathfrak{sp}(\tfrac{n_3}{2})\oplus\mathfrak{sp}(\tfrac{n_4}{2})$ gauge algebra, open-string tachyons in the representation $\left(\mathbf{n_1},\mathbf{n_2},\mathbf{1},\mathbf{1}\right)  \oplus\left(\mathbf{1},\mathbf{1},\mathbf{n_3},\mathbf{n_4}\right)$, and Majorana fermions in the bifundamental representation of the spacetime gauge algebra. We summarize the spectrum from the D9-branes of this type 0B/$\Omega_s$ orientifold in \cref{tab:s0B_open_spectrum}.
\begin{table}[!htbp]
\centering
\renewcommand{\arraystretch}{1.2}
\begin{tabular}{l|c|c|c|c|c}
State & character & $\mathfrak{sp}(\tfrac{n_1}{2})$ & $\mathfrak{sp}(\tfrac{n_2}{2})$ & $\mathfrak{sp}(\tfrac{n_3}{2})$ & $\mathfrak{sp}(\tfrac{n_4}{2})$ \\
\hline
Gauge vector & $V_8$ & $\mathbf{Adj}$ & $\mathbf{1}$ & $\mathbf{1}$ & $\mathbf{1}$ \\
Gauge vector & $V_8$ & $\mathbf{1}$ &  $\mathbf{Adj}$& $\mathbf{1}$& $\mathbf{1}$  \\
Gauge vector & $V_8$ & $\mathbf{1}$ & $\mathbf{1}$  & $\mathbf{Adj}$& $\mathbf{1}$ \\
Gauge vector & $V_8$ & $\mathbf{1}$ & $\mathbf{1}$  & $\mathbf{1}$ & $\mathbf{Adj}$\\
Fermion & $S_8$ & \tiny\yng(1) &  $\mathbf{1}$ & \tiny\yng(1) & $\mathbf{1}$ \\
Fermion & $S_8$ &  $\mathbf{1}$ & \tiny\yng(1) & $\mathbf{1}$ & \tiny\yng(1)  \\
Fermion & $C_8$ & \tiny\yng(1) & $\mathbf{1}$ & $\mathbf{1}$ & \tiny\yng(1) \\
Fermion & $C_8$ & $\mathbf{1}$ & \tiny\yng(1) & \tiny\yng(1) & $\mathbf{1}$ \\
Tachyon & $O_8$ & \tiny\yng(1) & \tiny\yng(1) & $\mathbf{1}$ & $\mathbf{1}$ \\
Tachyon & $O_8$ & $\mathbf{1}$ & $\mathbf{1}$ & \tiny\yng(1) & \tiny\yng(1)\\
\end{tabular}
\caption{Open-string light spectrum from D9-branes in the type 0B/$\Omega_s$ orientifold.}
\label{tab:s0B_open_spectrum}
\end{table}

The RR tadpole is canceled by the choice $n_1 = n_2=k_1$ and $n_3 = n_4=k_2$, thus fixing the gauge algebra to be $\mathfrak{sp}(\tfrac{k_1}{2})^{\oplus 2}\oplus\mathfrak{sp}(\tfrac{k_2}{2})^{\oplus 2}$.

\subsubsection{Branes with $p$ odd}

\paragraph{Charged branes}
This orientifold has charged D1- and D5-branes. The annulus and the M\"obius amplitudes can be found in \cref{eq:0B_orientifold_charged_direct}, setting $\sigma = -1$.

From the amplitudes, one obtains that D1-branes have a worldvolume $\mathfrak{sp}(\tfrac{d_1}{2})\oplus \mathfrak{sp}(\tfrac{d_2}{2})\oplus \mathfrak{sp}(\tfrac{d_3}{2})\oplus \mathfrak{sp}(\tfrac{d_4}{2})$ gauge algebra, while D5-branes have an $\mathfrak{so}(d_1)\oplus \mathfrak{so}(d_2)\oplus \mathfrak{so}(d_3)\oplus \mathfrak{so}(d_4)$ gauge algebra.
For both branes, there are tachyons in the bifundamental representations of the first and second gauge factors, and of the third and fourth factors. 

The scalars denoting the brane position are in the antisymmetric representation of each gauge factor for the D1-brane and in the symmetric representation for the D5.
Several chiral fermions arise in the bifundamental representations, as can be seen from $\mathcal{A}_{pp}$ in \cref{eq:0B_orientifold_charged_direct}.

Moreover, from the D$p$--D9 strings, we obtain additional massless chiral fermions, together with massless scalars for the D5 and massive ones for the D1-branes. They all transform in bifundamental representations of the spacetime and worldvolume gauge algebras.

The light spectra from the D1- and D5-branes are summarized in \cref{tab:s0B/Omega_D1_spectrum,tab:s0B/Omega_D5_spectrum}.

\begin{table}[!htbp]
\centering
\textbf{D1}
\renewcommand{\arraystretch}{1.2}
\centerresize{}{
\begin{tabular}{l|c|c|c}
Worldvolume state  & $\mathfrak{so}(8)_T$ & $\mathfrak{sp}(\tfrac{n_1}{2})\oplus\mathfrak{sp}(\tfrac{n_2}{2})\oplus\mathfrak{sp}(\tfrac{n_3}{2})\oplus\mathfrak{sp}(\tfrac{n_4}{2})$ & $\mathfrak{sp}(d_1/2)\oplus\mathfrak{sp}(d_2/2)\oplus\mathfrak{sp}(d_3/2)\oplus\mathfrak{sp}(d_4/2)$ \\
\hline
Gauge vector  & $\mathbf{1}$ & $(\mathbf{1}, \mathbf{1}, \mathbf{1}, \mathbf{1})$ & $(\mathbf{Adj}, \mathbf{1}, \mathbf{1}, \mathbf{1}) \oplus (\mathbf{1}, \mathbf{Adj}, \mathbf{1},  \mathbf{1})\oplus(\mathbf{1}, \mathbf{1},\mathbf{Adj}, \mathbf{1})\oplus(\mathbf{1}, \mathbf{1}, \mathbf{1},\mathbf{Adj})$ \\
Scalar & $\mathbf{8}_v$ & $(\mathbf{1}, \mathbf{1}, \mathbf{1}, \mathbf{1})$ & $\left(\text{\tiny$\yng(1,1)$},\mathbf{1}, \mathbf{1},\mathbf{1}\right) \oplus \left(\mathbf{1},\text{\tiny$\yng(1,1)$}, \mathbf{1},\mathbf{1}\right) \oplus\left(\mathbf{1}, \mathbf{1}, \text{\tiny$\yng(1,1)$}, \mathbf{1}\right) \oplus\left(\mathbf{1}, \mathbf{1},\mathbf{1}, \text{\tiny$\yng(1,1)$}\right)$ \\
Tachyon & $\mathbf{1}$ & $(\mathbf{1}, \mathbf{1}, \mathbf{1}, \mathbf{1})$ & $(\text{\tiny$\yng(1)$},\text{\tiny$\yng(1)$},\mathbf{1},\mathbf{1})\oplus (\mathbf{1},\mathbf{1},\text{\tiny$\yng(1)$},\text{\tiny$\yng(1)$})$\\
Fermion $(+)$ & $\mathbf{8}_{s}$ & $(\mathbf{1}, \mathbf{1}, \mathbf{1}, \mathbf{1})$ & $(\text{\tiny$\yng(1)$},\mathbf{1},\text{\tiny$\yng(1)$},\mathbf{1})\oplus(\mathbf{1},\text{\tiny$\yng(1)$},\mathbf{1},\text{\tiny$\yng(1)$})$ \\
Fermion $(-)$ & $\mathbf{8}_{c}$ & $(\mathbf{1}, \mathbf{1}, \mathbf{1}, \mathbf{1})$ & $(\text{\tiny$\yng(1)$},\mathbf{1},\text{\tiny$\yng(1)$},\mathbf{1})\oplus(\mathbf{1},\text{\tiny$\yng(1)$},\mathbf{1},\text{\tiny$\yng(1)$})$ \\
Fermion $(+)$ & $\mathbf{8}_{c}$ & $(\mathbf{1}, \mathbf{1}, \mathbf{1}, \mathbf{1})$ & $(\text{\tiny$\yng(1)$},\mathbf{1},\mathbf{1},\text{\tiny$\yng(1)$})\oplus(\mathbf{1},\text{\tiny$\yng(1)$},\text{\tiny$\yng(1)$},\mathbf{1})$ \\
Fermion $(-)$ & $\mathbf{8}_{s}$ & $(\mathbf{1}, \mathbf{1}, \mathbf{1}, \mathbf{1})$ & $(\text{\tiny$\yng(1)$},\mathbf{1},\mathbf{1},\text{\tiny$\yng(1)$})\oplus(\mathbf{1},\text{\tiny$\yng(1)$},\text{\tiny$\yng(1)$},\mathbf{1})$ \\
\hline
\hline
Fermion ($+$) & $\mathbf{1}$ & $\begin{array}{l}
    (\text{\tiny$\yng(1)$}, \mathbf{1}, \mathbf{1}, \mathbf{1})\\(\mathbf{1}, \text{\tiny$\yng(1)$}, \mathbf{1}, \mathbf{1})\\(\mathbf{1}, \mathbf{1}, \text{\tiny$\yng(1)$}, \mathbf{1})\\(\mathbf{1}, \mathbf{1}, \mathbf{1}, \text{\tiny$\yng(1)$}) \end{array}$ & $\begin{array}{l} (\mathbf{1}, \mathbf{1}, \mathbf{1}, \text{\tiny$\yng(1)$}) \\ (\mathbf{1}, \mathbf{1}, \text{\tiny$\yng(1)$}, \mathbf{1}) \\ (\mathbf{1}, \text{\tiny$\yng(1)$}, \mathbf{1}, \mathbf{1}) \\ (\text{\tiny$\yng(1)$}, \mathbf{1}, \mathbf{1}, \mathbf{1}) \end{array}$ \\
\hline
Fermion ($-$) & $\mathbf{1}$ & $\begin{array}{l}
    (\text{\tiny$\yng(1)$}, \mathbf{1}, \mathbf{1}, \mathbf{1})\\(\mathbf{1}, \text{\tiny$\yng(1)$}, \mathbf{1}, \mathbf{1})\\(\mathbf{1}, \mathbf{1}, \text{\tiny$\yng(1)$}, \mathbf{1})\\(\mathbf{1}, \mathbf{1}, \mathbf{1}, \text{\tiny$\yng(1)$}) \end{array}$ & $\begin{array}{l} (\mathbf{1}, \mathbf{1}, \text{\tiny$\yng(1)$}, \mathbf{1}) \\ (\mathbf{1}, \mathbf{1}, \mathbf{1}, \text{\tiny$\yng(1)$}) \\ (\text{\tiny$\yng(1)$}, \mathbf{1}, \mathbf{1}, \mathbf{1}) \\ (\mathbf{1}, \text{\tiny$\yng(1)$}, \mathbf{1}, \mathbf{1}) \end{array}$ \\
\end{tabular}}
\caption{Open-string light spectrum for D1-branes in type 0B/$\Omega_s$. Recall that $n_1=n_2=k_1$ and $n_3=n_4=k_2$.}
\label{tab:s0B/Omega_D1_spectrum}
\end{table}
\begin{table}[!htbp]
\centering
\textbf{D5}
\renewcommand{\arraystretch}{1.2}
\centerresize{}{
\begin{tabular}{l|c|c|c}
Worldvolume state  & $\mathfrak{so}(4)_T$ & $\mathfrak{sp}(\tfrac{n_1}{2})\oplus\mathfrak{sp}(\tfrac{n_2}{2})\oplus\mathfrak{sp}(\tfrac{n_3}{2})\oplus\mathfrak{sp}(\tfrac{n_4}{2})$ & $\mathfrak{so}(d_1)\oplus\mathfrak{so}(d_2)\oplus\mathfrak{so}(d_3)\oplus\mathfrak{so}(d_4)$ \\
\hline
Gauge vector  & $\mathbf{1}$ & $(\mathbf{1}, \mathbf{1}, \mathbf{1}, \mathbf{1})$ & $(\mathbf{Adj}, \mathbf{1}, \mathbf{1}, \mathbf{1}) \oplus (\mathbf{1}, \mathbf{Adj}, \mathbf{1},  \mathbf{1})\oplus(\mathbf{1}, \mathbf{1},\mathbf{Adj}, \mathbf{1})\oplus(\mathbf{1}, \mathbf{1}, \mathbf{1},\mathbf{Adj})$ \\
Scalar & $\mathbf{4}_{v}$ & $(\mathbf{1}, \mathbf{1}, \mathbf{1}, \mathbf{1})$ & $\left(\text{\tiny$\yng(2)$},\mathbf{1}, \mathbf{1},\mathbf{1}\right) \oplus \left(\mathbf{1},\text{\tiny$\yng(2)$}, \mathbf{1},\mathbf{1}\right) \oplus\left(\mathbf{1}, \mathbf{1}, \text{\tiny$\yng(2)$}, \mathbf{1}\right) \oplus\left(\mathbf{1}, \mathbf{1},\mathbf{1}, \text{\tiny$\yng(2)$}\right)$ \\
Tachyon & $\mathbf{1}$ & $(\mathbf{1}, \mathbf{1}, \mathbf{1}, \mathbf{1})$ & $(\text{\tiny$\yng(1)$},\text{\tiny$\yng(1)$},\mathbf{1},\mathbf{1})\oplus (\mathbf{1},\mathbf{1},\text{\tiny$\yng(1)$},\text{\tiny$\yng(1)$})$\\
Fermion $(+)$ & $\mathbf{2}_{s}$ & $(\mathbf{1}, \mathbf{1}, \mathbf{1}, \mathbf{1})$ & $(\text{\tiny$\yng(1)$},\mathbf{1},\text{\tiny$\yng(1)$},\mathbf{1})\oplus(\mathbf{1},\text{\tiny$\yng(1)$},\mathbf{1},\text{\tiny$\yng(1)$})$ \\
Fermion $(-)$ & $\mathbf{2}_{c}$ & $(\mathbf{1}, \mathbf{1}, \mathbf{1}, \mathbf{1})$ & $(\text{\tiny$\yng(1)$},\mathbf{1},\text{\tiny$\yng(1)$},\mathbf{1})\oplus(\mathbf{1},\text{\tiny$\yng(1)$},\mathbf{1},\text{\tiny$\yng(1)$})$ \\
Fermion $(+)$ & $\mathbf{2}_{c}$ & $(\mathbf{1}, \mathbf{1}, \mathbf{1}, \mathbf{1})$ & $(\text{\tiny$\yng(1)$},\mathbf{1},\mathbf{1},\text{\tiny$\yng(1)$})\oplus(\mathbf{1},\text{\tiny$\yng(1)$},\text{\tiny$\yng(1)$},\mathbf{1})$ \\
Fermion $(-)$ & $\mathbf{2}_{s}$ & $(\mathbf{1}, \mathbf{1}, \mathbf{1}, \mathbf{1})$ & $(\text{\tiny$\yng(1)$},\mathbf{1},\mathbf{1},\text{\tiny$\yng(1)$})\oplus(\mathbf{1},\text{\tiny$\yng(1)$},\text{\tiny$\yng(1)$},\mathbf{1})$ \\
\hline
\hline
Various scalars & & & \\
\hline
Fermion ($-$) & $\mathbf{1}$ & $\begin{array}{l}
    (\text{\tiny$\yng(1)$}, \mathbf{1}, \mathbf{1}, \mathbf{1})\\(\mathbf{1}, \text{\tiny$\yng(1)$}, \mathbf{1}, \mathbf{1})\\(\mathbf{1}, \mathbf{1}, \text{\tiny$\yng(1)$}, \mathbf{1})\\(\mathbf{1}, \mathbf{1}, \mathbf{1}, \text{\tiny$\yng(1)$}) \end{array}$ & $\begin{array}{l} (\mathbf{1}, \mathbf{1}, \mathbf{1}, \text{\tiny$\yng(1)$}) \\ (\mathbf{1}, \mathbf{1}, \text{\tiny$\yng(1)$}, \mathbf{1}) \\ (\mathbf{1}, \text{\tiny$\yng(1)$}, \mathbf{1}, \mathbf{1}) \\ (\text{\tiny$\yng(1)$}, \mathbf{1}, \mathbf{1}, \mathbf{1}) \end{array}$ \\
\hline
Fermion ($+$) & $\mathbf{1}$ & $\begin{array}{l}
    (\text{\tiny$\yng(1)$}, \mathbf{1}, \mathbf{1}, \mathbf{1})\\(\mathbf{1}, \text{\tiny$\yng(1)$}, \mathbf{1}, \mathbf{1})\\(\mathbf{1}, \mathbf{1}, \text{\tiny$\yng(1)$}, \mathbf{1})\\(\mathbf{1}, \mathbf{1}, \mathbf{1}, \text{\tiny$\yng(1)$}) \end{array}$ & $\begin{array}{l} (\mathbf{1}, \mathbf{1}, \text{\tiny$\yng(1)$}, \mathbf{1}) \\ (\mathbf{1}, \mathbf{1}, \mathbf{1}, \text{\tiny$\yng(1)$}) \\ (\text{\tiny$\yng(1)$}, \mathbf{1}, \mathbf{1}, \mathbf{1}) \\ (\mathbf{1}, \text{\tiny$\yng(1)$}, \mathbf{1}, \mathbf{1}) \end{array}$ \\
\end{tabular}}
\caption{Open-string light spectrum for D5-branes in type 0B/$\Omega_s$. Recall that $n_1=n_2=k_1$ and $n_3=n_4=k_2$.}
\label{tab:s0B/Omega_D5_spectrum}
\end{table}

\FloatBarrier

\paragraph{Uncharged branes}

For the uncharged branes with $p=-1,3,7$, whose amplitudes are in \cref{eq:0B_orientifold_uncharged_odd} with $\sigma=-1$, we obtain unitary worldvolume gauge algebras, $\mathfrak{u}(d_1)\oplus\mathfrak{u}(d_2)$.
Tachyons appear in the antisymmetric representations of the gauge factors for D3-branes and in the symmetric representations for D$(-1)$ and D7-branes, together with the conjugate representations.
The scalars denoting the position of the branes are in the adjoint representations.
In addition, there are fermions in bi(anti)fundamental representations.

The D$p$--D9 open-string sector contains fermions in the bifundamental representation or in the fundamental-antifundamental representations of a spacetime and a worldvolume gauge factor. These have opposite chiralities between D3- and D7-branes. The latter also contain tachyons in analogous representations, which become massive scalars for the D3.

We collect the results for the D3- and D7-branes in \cref{tab:s0B/Omega_D3_spectrum,tab:s0B/Omega_D7_spectrum}. The case of the D$(-1)$ is analogous to the D7.

\begin{table}[!htbp]
\centering
\textbf{D3}
\renewcommand{\arraystretch}{1.2}
\centerresize{}{
\begin{tabular}{l|c|c|c}
Worldvolume state  & $\mathfrak{so}(6)_T$ & $\mathfrak{sp}(\tfrac{n_1}{2})\oplus\mathfrak{sp}(\tfrac{n_2}{2})\oplus\mathfrak{sp}(\tfrac{n_3}{2})\oplus\mathfrak{sp}(\tfrac{n_4}{2})$ & $\mathfrak{u}(d_1)\oplus\mathfrak{u}(d_2)$ \\
\hline
Gauge vector  & $\mathbf{1}$ & $(\mathbf{1}, \mathbf{1}, \mathbf{1}, \mathbf{1})$ &  $(\mathbf{Adj}, \mathbf{1}) \oplus (\mathbf{1},\mathbf{Adj})$ \\
Scalar & $\mathbf{6}_{v}$ & $(\mathbf{1}, \mathbf{1}, \mathbf{1}, \mathbf{1})$ & $(\mathbf{Adj}, \mathbf{1}) \oplus (\mathbf{1},\mathbf{Adj})$\\
Tachyon & $\mathbf{1}$ & $(\mathbf{1}, \mathbf{1}, \mathbf{1}, \mathbf{1})$ & $  \left(\text{\tiny$\yng(1,1)$}, \mathbf{1}\right) \oplus  \left(\mathbf{1}, \text{\tiny$\yng(1,1)$}\right) \oplus \left(\ybar{\text{\tiny$\yng(1,1)$}}, \mathbf{1}\right) \oplus  \left(\mathbf{1}, \ybar{\text{\tiny$\yng(1,1)$}}\right)$\\
Fermion $(+)$ & $\mathbf{4}_{s}$ & $(\mathbf{1}, \mathbf{1}, \mathbf{1}, \mathbf{1})$ & $(\text{\tiny$\yng(1)$},\text{\tiny$\ybar{\yng(1)}$})\oplus (\text{\tiny$\ybar{\yng(1)}$},\text{\tiny$\yng(1)$})$ \\
Fermion $(-)$ & $\mathbf{4}_{c}$ & $(\mathbf{1}, \mathbf{1}, \mathbf{1}, \mathbf{1})$ & $(\text{\tiny$\yng(1)$},\text{\tiny$\ybar{\yng(1)}$})\oplus (\text{\tiny$\ybar{\yng(1)}$},\text{\tiny$\yng(1)$})$ \\
Fermion $(+)$ & $\mathbf{4}_{c}$ & $(\mathbf{1}, \mathbf{1}, \mathbf{1}, \mathbf{1})$ & $(\text{\tiny$\yng(1)$},\text{\tiny$\yng(1)$})\oplus (\text{\tiny$\ybar{\yng(1)}$},\text{\tiny$\ybar{\yng(1)}$})$ \\
Fermion $(-)$ & $\mathbf{4}_{s}$ & $(\mathbf{1}, \mathbf{1}, \mathbf{1}, \mathbf{1})$ & $(\text{\tiny$\yng(1)$},\text{\tiny$\yng(1)$})\oplus (\text{\tiny$\ybar{\yng(1)}$},\text{\tiny$\ybar{\yng(1)}$})$  \\
\hline
\hline
Fermion ($+$) & $\mathbf{1}$ & $\begin{array}{l}
    (\text{\tiny$\yng(1)$}, \mathbf{1}, \mathbf{1}, \mathbf{1})\\(\mathbf{1}, \text{\tiny$\yng(1)$}, \mathbf{1}, \mathbf{1})\\(\mathbf{1}, \mathbf{1}, \text{\tiny$\yng(1)$}, \mathbf{1})\\(\mathbf{1}, \mathbf{1}, \mathbf{1}, \text{\tiny$\yng(1)$}) \end{array}$ & $\begin{array}{l} (\mathbf{1}, \text{\tiny$\ybar{\yng(1)}$}) \\ (\mathbf{1}, \text{\tiny$\yng(1)$}) \\ (\text{\tiny$\ybar{\yng(1)}$},\mathbf{1}) \\ (\text{\tiny$\yng(1)$},\mathbf{1}) \end{array}$ \\
\hline
Fermion ($-$) & $\mathbf{1}$ & $\begin{array}{l}
    (\text{\tiny$\yng(1)$}, \mathbf{1}, \mathbf{1}, \mathbf{1})\\(\mathbf{1}, \text{\tiny$\yng(1)$}, \mathbf{1}, \mathbf{1})\\(\mathbf{1}, \mathbf{1}, \text{\tiny$\yng(1)$}, \mathbf{1})\\(\mathbf{1}, \mathbf{1}, \mathbf{1}, \text{\tiny$\yng(1)$}) \end{array}$ & $\begin{array}{l} (\mathbf{1}, \text{\tiny$\yng(1)$}) \\ (\mathbf{1}, \text{\tiny$\ybar{\yng(1)}$})  \\ (\text{\tiny$\yng(1)$},\mathbf{1}) \\ (\text{\tiny$\ybar{\yng(1)}$},\mathbf{1}) \end{array}$ \\
\end{tabular}}
\caption{Open-string light spectrum for D3-branes in type 0B/$\Omega_s$. Recall that $n_1=n_2=k_1$ and $n_3=n_4=k_2$.}
\label{tab:s0B/Omega_D3_spectrum}
\end{table}
\begin{table}[!htbp]
\centering
\textbf{D7}
\renewcommand{\arraystretch}{1.2}
\centerresize{}{
\begin{tabular}{l|c|c|c}
Worldvolume state  & $\mathfrak{so}(2)_T$ & $\mathfrak{sp}(\tfrac{n_1}{2})\oplus\mathfrak{sp}(\tfrac{n_2}{2})\oplus\mathfrak{sp}(\tfrac{n_3}{2})\oplus\mathfrak{sp}(\tfrac{n_4}{2})$ & $\mathfrak{u}(d_1)\oplus\mathfrak{u}(d_2)$ \\
\hline
Gauge vector  & $\mathbf{1}$ & $(\mathbf{1}, \mathbf{1}, \mathbf{1}, \mathbf{1})$ &  $(\mathbf{Adj}, \mathbf{1}) \oplus (\mathbf{1},\mathbf{Adj})$ \\
Scalar & $\mathbf{2}_{v}$ & $(\mathbf{1}, \mathbf{1}, \mathbf{1}, \mathbf{1})$ & $(\mathbf{Adj}, \mathbf{1}) \oplus (\mathbf{1},\mathbf{Adj})$\\
Tachyon & $\mathbf{1}$ & $(\mathbf{1}, \mathbf{1}, \mathbf{1}, \mathbf{1})$ & $(\text{\tiny$\yng(2)$}, \mathbf{1}) \oplus (\mathbf{1}, \text{\tiny$\yng(2)$})\oplus (\ybar{\text{\tiny$\yng(2)$}}, \mathbf{1}) \oplus (\mathbf{1}, \ybar{\text{\tiny$\yng(2)$}})$\\
Fermion $(+)$ & $\mathbf{1}_{s}$ & $(\mathbf{1}, \mathbf{1}, \mathbf{1}, \mathbf{1})$ & $(\text{\tiny$\yng(1)$},\text{\tiny$\ybar{\yng(1)}$})\oplus (\text{\tiny$\ybar{\yng(1)}$},\text{\tiny$\yng(1)$})$ \\
Fermion $(-)$ & $\mathbf{1}_{c}$ & $(\mathbf{1}, \mathbf{1}, \mathbf{1}, \mathbf{1})$ & $(\text{\tiny$\yng(1)$},\text{\tiny$\ybar{\yng(1)}$})\oplus (\text{\tiny$\ybar{\yng(1)}$},\text{\tiny$\yng(1)$})$ \\
Fermion $(+)$ & $\mathbf{1}_{c}$ & $(\mathbf{1}, \mathbf{1}, \mathbf{1}, \mathbf{1})$ & $(\text{\tiny$\yng(1)$},\text{\tiny$\yng(1)$})\oplus (\text{\tiny$\ybar{\yng(1)}$},\text{\tiny$\ybar{\yng(1)}$})$  \\
Fermion $(-)$ & $\mathbf{1}_{s}$ & $(\mathbf{1}, \mathbf{1}, \mathbf{1}, \mathbf{1})$ & $(\text{\tiny$\yng(1)$},\text{\tiny$\yng(1)$})\oplus (\text{\tiny$\ybar{\yng(1)}$},\text{\tiny$\ybar{\yng(1)}$})$  \\
\hline
\hline
Various tachyons & & & \\
\hline
Fermion ($-$) & $\mathbf{1}$ & $\begin{array}{l}
    (\text{\tiny$\yng(1)$}, \mathbf{1}, \mathbf{1}, \mathbf{1})\\(\mathbf{1}, \text{\tiny$\yng(1)$}, \mathbf{1}, \mathbf{1})\\(\mathbf{1}, \mathbf{1}, \text{\tiny$\yng(1)$}, \mathbf{1})\\(\mathbf{1}, \mathbf{1}, \mathbf{1}, \text{\tiny$\yng(1)$}) \end{array}$ & $\begin{array}{l} (\mathbf{1}, \text{\tiny$\ybar{\yng(1)}$}) \\ (\mathbf{1}, \text{\tiny$\yng(1)$}) \\ (\text{\tiny$\ybar{\yng(1)}$},\mathbf{1}) \\ (\text{\tiny$\yng(1)$},\mathbf{1}) \end{array}$ \\
\hline
Fermion ($+$) & $\mathbf{1}$ & $\begin{array}{l}
    (\text{\tiny$\yng(1)$}, \mathbf{1}, \mathbf{1}, \mathbf{1})\\(\mathbf{1}, \text{\tiny$\yng(1)$}, \mathbf{1}, \mathbf{1})\\(\mathbf{1}, \mathbf{1}, \text{\tiny$\yng(1)$}, \mathbf{1})\\(\mathbf{1}, \mathbf{1}, \mathbf{1}, \text{\tiny$\yng(1)$}) \end{array}$ & $\begin{array}{l} (\mathbf{1}, \text{\tiny$\yng(1)$}) \\ (\mathbf{1}, \text{\tiny$\ybar{\yng(1)}$})  \\ (\text{\tiny$\yng(1)$},\mathbf{1}) \\ (\text{\tiny$\ybar{\yng(1)}$},\mathbf{1}) \end{array}$ \\
\end{tabular}}
\caption{Open-string light spectrum for D7-branes in type 0B/$\Omega_s$. Recall that $n_1=n_2=k_1$ and $n_3=n_4=k_2$. The D$(-1)$ case is analogous to this with the D$p$--D9 tachyons being massive scalars.}
\label{tab:s0B/Omega_D7_spectrum}
\end{table}

\FloatBarrier

\subsubsection{Branes with $p$ even}

The amplitudes corresponding to the D$p$-branes with $p=0,2,4,6,8$ are in \cref{eq:0B_orientifold_uncharged_even}, after setting $\sigma=-1$.
These branes are uncharged, and their worldvolume gauge algebras are: $\mathfrak{sp}(\tfrac{d_1}{2})\oplus \mathfrak{sp}(\tfrac{d_2}{2})$ for D0-, D2-, and D8-branes, and $\mathfrak{so}(d_1)\oplus \mathfrak{so}(d_2) $ for D4- and D6-branes.
Tachyons appear in the symmetric representations of the first and second factors of the gauge algebra for D0-, D6- and D8-branes, while they are in the antisymmetric representations for D2- and D4-branes.

The scalars denoting the brane position are in the antisymmetric representations of the worldvolume gauge algebra for D0-, D2- and D8-branes, and in the symmetric representations for D4- and D6-branes.
There are also Dirac spinors, from the combination $S'_{p-1} S'_{9-p}$ in $\mathcal{A}_{pp}$, in the bifundamental representation of the worldvolume gauge algebra.

The D$p$--D9 sector from \cref{eq:0B_orientifold_uncharged_even} yields Dirac fermions in the bifundamental representation of a spacetime and a worldvolume factor, in addition to tachyons for $p>5$.

We collect the results for these branes in \cref{tab:s0B/Omega_D0_spectrum,tab:s0B/Omega_D2_spectrum,tab:s0B/Omega_D4_spectrum,tab:s0B/Omega_D6_spectrum,tab:s0B/Omega_D8_spectrum}.

\begin{table}[!htbp]
\centering
\textbf{D0}
\renewcommand{\arraystretch}{1.2}
\centerresize{}{
\begin{tabular}{l|c|c|c}
Worldvolume state  & $\mathfrak{so}(9)_T$ & $\mathfrak{sp}(\tfrac{n_1}{2})\oplus\mathfrak{sp}(\tfrac{n_2}{2})\oplus\mathfrak{sp}(\tfrac{n_3}{2})\oplus\mathfrak{sp}(\tfrac{n_4}{2})$ & $\mathfrak{sp}(\tfrac{d_1}{2})\oplus\mathfrak{sp}(\tfrac{d_2}{2})$ \\
\hline
Gauge vector  & $\mathbf{1}$ & $(\mathbf{1}, \mathbf{1}, \mathbf{1}, \mathbf{1})$ &  $(\mathbf{Adj}, \mathbf{1}) \oplus (\mathbf{1},\mathbf{Adj})$ \\
Scalar & $\mathbf{9}_{v}$ & $(\mathbf{1}, \mathbf{1}, \mathbf{1}, \mathbf{1})$ &  $  \left(\text{\tiny$\yng(1,1)$}, \mathbf{1}\right) \oplus \left(\mathbf{1}, \text{\tiny$\yng(1,1)$}\right)$\\
Tachyon & $\mathbf{1}$ & $(\mathbf{1}, \mathbf{1}, \mathbf{1}, \mathbf{1})$ & $(\mathbf{Adj}, \mathbf{1}) \oplus (\mathbf{1},\mathbf{Adj})$\\
Dirac fermion  & $\mathbf{16}_{s'}$ & $(\mathbf{1}, \mathbf{1}, \mathbf{1}, \mathbf{1})$ & $(\text{\tiny$\yng(1)$},\text{\tiny$\yng(1)$})$ \\
\hline
\hline
Dirac fermion & $\mathbf{1}$ & $\begin{array}{l}
    (\text{\tiny$\yng(1)$}, \mathbf{1}, \mathbf{1}, \mathbf{1})\\(\mathbf{1}, \text{\tiny$\yng(1)$}, \mathbf{1}, \mathbf{1})\\(\mathbf{1}, \mathbf{1}, \text{\tiny$\yng(1)$}, \mathbf{1})\\(\mathbf{1}, \mathbf{1}, \mathbf{1}, \text{\tiny$\yng(1)$}) \end{array}$ & $\begin{array}{l} (\mathbf{1}, \text{\tiny$\yng(1)$}) \\ (\mathbf{1}, \text{\tiny$\yng(1)$}) \\ (\text{\tiny$\yng(1)$},\mathbf{1}) \\ (\text{\tiny$\yng(1)$},\mathbf{1}) \end{array}$ \\
\end{tabular}}
\caption{Open-string light spectrum for D0-branes in type 0B/$\Omega_s$. Recall that $n_1=n_2=k_1$ and $n_3=n_4=k_2$.}
\label{tab:s0B/Omega_D0_spectrum}
\end{table}
\begin{table}[!htbp]
\centering
\textbf{D2}
\renewcommand{\arraystretch}{1.2}
\centerresize{}{
\begin{tabular}{l|c|c|c}
Worldvolume state  & $\mathfrak{so}(7)_T$ & $\mathfrak{sp}(\tfrac{n_1}{2})\oplus\mathfrak{sp}(\tfrac{n_2}{2})\oplus\mathfrak{sp}(\tfrac{n_3}{2})\oplus\mathfrak{sp}(\tfrac{n_4}{2})$ & $\mathfrak{sp}(\tfrac{d_1}{2})\oplus\mathfrak{sp}(\tfrac{d_2}{2})$ \\
\hline
Gauge vector  & $\mathbf{1}$ & $(\mathbf{1}, \mathbf{1}, \mathbf{1}, \mathbf{1})$ &  $(\mathbf{Adj}, \mathbf{1}) \oplus (\mathbf{1},\mathbf{Adj})$ \\
Scalar & $\mathbf{7}_{v}$ & $(\mathbf{1}, \mathbf{1}, \mathbf{1}, \mathbf{1})$ & $  \left(\text{\tiny$\yng(1,1)$}, \mathbf{1}\right) \oplus \left(\mathbf{1}, \text{\tiny$\yng(1,1)$}\right)$\\
Tachyon & $\mathbf{1}$ & $(\mathbf{1}, \mathbf{1}, \mathbf{1}, \mathbf{1})$ &  $  \left(\text{\tiny$\yng(1,1)$}, \mathbf{1}\right) \oplus \left(\mathbf{1}, \text{\tiny$\yng(1,1)$}\right)$\\
Dirac fermion  & $\mathbf{8}_{s'}$ & $(\mathbf{1}, \mathbf{1}, \mathbf{1}, \mathbf{1})$ & $(\text{\tiny$\yng(1)$},\text{\tiny$\yng(1)$})$ \\
\hline
\hline
Dirac fermion & $\mathbf{1}$ & $\begin{array}{l}
    (\text{\tiny$\yng(1)$}, \mathbf{1}, \mathbf{1}, \mathbf{1})\\(\mathbf{1}, \text{\tiny$\yng(1)$}, \mathbf{1}, \mathbf{1})\\(\mathbf{1}, \mathbf{1}, \text{\tiny$\yng(1)$}, \mathbf{1})\\(\mathbf{1}, \mathbf{1}, \mathbf{1}, \text{\tiny$\yng(1)$}) \end{array}$ & $\begin{array}{l} (\mathbf{1}, \text{\tiny$\yng(1)$}) \\ (\mathbf{1}, \text{\tiny$\yng(1)$}) \\ (\text{\tiny$\yng(1)$},\mathbf{1}) \\ (\text{\tiny$\yng(1)$},\mathbf{1}) \end{array}$ \\
\end{tabular}}
\caption{Open-string light spectrum for D2-branes in type 0B/$\Omega_s$. Recall that $n_1=n_2=k_1$ and $n_3=n_4=k_2$.}
\label{tab:s0B/Omega_D2_spectrum}
\end{table}
\begin{table}[H]
\centering
\textbf{D4}
\renewcommand{\arraystretch}{1.2}
\centerresize{}{
\begin{tabular}{l|c|c|c}
Worldvolume state  & $\mathfrak{so}(5)_T$ & $\mathfrak{sp}(\tfrac{n_1}{2})\oplus\mathfrak{sp}(\tfrac{n_2}{2})\oplus\mathfrak{sp}(\tfrac{n_3}{2})\oplus\mathfrak{sp}(\tfrac{n_4}{2})$ & $\mathfrak{so}(d_1)\oplus\mathfrak{so}(d_2)$ \\
\hline
Gauge vector  & $\mathbf{1}$ & $(\mathbf{1}, \mathbf{1}, \mathbf{1}, \mathbf{1})$ &  $(\mathbf{Adj}, \mathbf{1}) \oplus (\mathbf{1},\mathbf{Adj})$ \\
Scalar & $\mathbf{5}_{v}$ & $(\mathbf{1}, \mathbf{1}, \mathbf{1}, \mathbf{1})$ & $(\text{\tiny$\yng(2)$}, \mathbf{1}) \oplus (\mathbf{1}, \text{\tiny$\yng(2)$})$\\
Tachyon & $\mathbf{1}$ & $(\mathbf{1}, \mathbf{1}, \mathbf{1}, \mathbf{1})$ &  $(\mathbf{Adj}, \mathbf{1}) \oplus (\mathbf{1},\mathbf{Adj})$\\
Dirac fermion  & $\mathbf{4}_{s'}$ & $(\mathbf{1}, \mathbf{1}, \mathbf{1}, \mathbf{1})$ & $(\text{\tiny$\yng(1)$},\text{\tiny$\yng(1)$})$ \\
\hline
\hline
Dirac fermion & $\mathbf{1}$ & $\begin{array}{l}
    (\text{\tiny$\yng(1)$}, \mathbf{1}, \mathbf{1}, \mathbf{1})\\(\mathbf{1}, \text{\tiny$\yng(1)$}, \mathbf{1}, \mathbf{1})\\(\mathbf{1}, \mathbf{1}, \text{\tiny$\yng(1)$}, \mathbf{1})\\(\mathbf{1}, \mathbf{1}, \mathbf{1}, \text{\tiny$\yng(1)$}) \end{array}$ & $\begin{array}{l} (\mathbf{1}, \text{\tiny$\yng(1)$}) \\ (\mathbf{1}, \text{\tiny$\yng(1)$}) \\ (\text{\tiny$\yng(1)$},\mathbf{1}) \\ (\text{\tiny$\yng(1)$},\mathbf{1}) \end{array}$ \\
\end{tabular}}
\caption{Open-string light spectrum for D4-branes in type 0B/$\Omega_s$. Recall that $n_1=n_2=k_1$ and $n_3=n_4=k_2$.}
\label{tab:s0B/Omega_D4_spectrum}
\end{table}
\begin{table}[!htbp]
\centering
\textbf{D6}
\renewcommand{\arraystretch}{1.2}
\centerresize{}{
\begin{tabular}{l|c|c|c}
Worldvolume state  & $\mathfrak{so}(3)_T$ & $\mathfrak{sp}(\tfrac{n_1}{2})\oplus\mathfrak{sp}(\tfrac{n_2}{2})\oplus\mathfrak{sp}(\tfrac{n_3}{2})\oplus\mathfrak{sp}(\tfrac{n_4}{2})$ & $\mathfrak{so}(d_1)\oplus\mathfrak{so}(d_2)$ \\
\hline
Gauge vector  & $\mathbf{1}$ & $(\mathbf{1}, \mathbf{1}, \mathbf{1}, \mathbf{1})$ &  $(\mathbf{Adj}, \mathbf{1}) \oplus (\mathbf{1},\mathbf{Adj})$ \\
Scalar & $\mathbf{3}_{v}$ & $(\mathbf{1}, \mathbf{1}, \mathbf{1}, \mathbf{1})$ & $(\text{\tiny$\yng(2)$}, \mathbf{1}) \oplus (\mathbf{1}, \text{\tiny$\yng(2)$})$\\
Tachyon & $\mathbf{1}$ & $(\mathbf{1}, \mathbf{1}, \mathbf{1}, \mathbf{1})$ &  $(\text{\tiny$\yng(2)$}, \mathbf{1}) \oplus (\mathbf{1}, \text{\tiny$\yng(2)$})$\\
Dirac fermion  & $\mathbf{2}_{s'}$ & $(\mathbf{1}, \mathbf{1}, \mathbf{1}, \mathbf{1})$ & $(\text{\tiny$\yng(1)$},\text{\tiny$\yng(1)$})$ \\
\hline
\hline
Various tachyons & & & \\
\hline
Dirac fermion & $\mathbf{1}$ & $\begin{array}{l}
    (\text{\tiny$\yng(1)$}, \mathbf{1}, \mathbf{1}, \mathbf{1})\\(\mathbf{1}, \text{\tiny$\yng(1)$}, \mathbf{1}, \mathbf{1})\\(\mathbf{1}, \mathbf{1}, \text{\tiny$\yng(1)$}, \mathbf{1})\\(\mathbf{1}, \mathbf{1}, \mathbf{1}, \text{\tiny$\yng(1)$}) \end{array}$ & $\begin{array}{l} (\mathbf{1}, \text{\tiny$\yng(1)$}) \\ (\mathbf{1}, \text{\tiny$\yng(1)$}) \\ (\text{\tiny$\yng(1)$},\mathbf{1}) \\ (\text{\tiny$\yng(1)$},\mathbf{1}) \end{array}$ \\
\end{tabular}}
\caption{Open-string light spectrum for D6-branes in type 0B/$\Omega_s$. Recall that $n_1=n_2=k_1$ and $n_3=n_4=k_2$.}
\label{tab:s0B/Omega_D6_spectrum}
\end{table}
\begin{table}[!htbp]
\centering
\textbf{D8}
\renewcommand{\arraystretch}{1.2}
\centerresize{}{
\begin{tabular}{l|c|c|c}
Worldvolume state  & $\mathfrak{so}(1)_T$ & $\mathfrak{sp}(\tfrac{n_1}{2})\oplus\mathfrak{sp}(\tfrac{n_2}{2})\oplus\mathfrak{sp}(\tfrac{n_3}{2})\oplus\mathfrak{sp}(\tfrac{n_4}{2})$ & $\mathfrak{sp}(\tfrac{d_1}{2})\oplus\mathfrak{sp}(\tfrac{d_2}{2})$ \\
\hline
Gauge vector  & $\mathbf{1}$ & $(\mathbf{1}, \mathbf{1}, \mathbf{1}, \mathbf{1})$ &  $(\mathbf{Adj}, \mathbf{1}) \oplus (\mathbf{1},\mathbf{Adj})$ \\
Scalar & $\mathbf{1}_{v}$ & $(\mathbf{1}, \mathbf{1}, \mathbf{1}, \mathbf{1})$ & $\left(\text{\tiny$\yng(1,1)$}, \mathbf{1}\right) \oplus \left(\mathbf{1}, \text{\tiny$\yng(1,1)$}\right)$\\
Tachyon & $\mathbf{1}$ & $(\mathbf{1}, \mathbf{1}, \mathbf{1}, \mathbf{1})$ & $(\mathbf{Adj}, \mathbf{1}) \oplus (\mathbf{1},\mathbf{Adj})$ \\
Dirac fermion  & $\mathbf{1}_{s'}$ & $(\mathbf{1}, \mathbf{1}, \mathbf{1}, \mathbf{1})$ & $(\text{\tiny$\yng(1)$},\text{\tiny$\yng(1)$})$ \\
\hline
\hline
Various tachyons & & & \\
\hline
Dirac fermion & $\mathbf{1}$ & $\begin{array}{l}
    (\text{\tiny$\yng(1)$}, \mathbf{1}, \mathbf{1}, \mathbf{1})\\(\mathbf{1}, \text{\tiny$\yng(1)$}, \mathbf{1}, \mathbf{1})\\(\mathbf{1}, \mathbf{1}, \text{\tiny$\yng(1)$}, \mathbf{1})\\(\mathbf{1}, \mathbf{1}, \mathbf{1}, \text{\tiny$\yng(1)$}) \end{array}$ & $\begin{array}{l} (\mathbf{1}, \text{\tiny$\yng(1)$}) \\ (\mathbf{1}, \text{\tiny$\yng(1)$}) \\ (\text{\tiny$\yng(1)$},\mathbf{1}) \\ (\text{\tiny$\yng(1)$},\mathbf{1}) \end{array}$ \\
\end{tabular}}
\caption{Open-string light spectrum for D8-branes in type 0B/$\Omega_s$. Recall that $n_1=n_2=k_1$ and $n_3=n_4=k_2$.}
\label{tab:s0B/Omega_D8_spectrum}
\end{table}

\FloatBarrier

\section{A proof for the symplectic orientifold}\label{app:symplectic_proof}

Consider $4mn$ real fermionic modes $\psi_I$, $I=1,\dots ,4mn$ with $\acomm{\psi_I}{\psi_J}=\delta_{IJ}$, in the $(\mathbf{2m},\mathbf{2n})$ of $\mathfrak{sp}(m)\oplus\mathfrak{sp}(n)$. Let $M=\mathbf{S}_{4mn}\oplus\mathbf{C}_{4mn}$ be the Clifford module that they generate. Then, every weight $\lambda$ of $M$ is such that
\begin{eq}\label{eq:appendix_weights}
    [\lambda]_{\mathfrak{sp}(m)}+[\lambda]_{\mathfrak{sp}(n)}\equiv mn\pmod{2}\,,
\end{eq}
where the classes $o$ and $v$ are represented as $o\to 0$ and $v\to 1$.

To prove this, let $\pm e_{i=1,\dots,m}$ and $\pm f_{a=1,\dots,n}$ be the weights of the defining representations of $\mathfrak{sp}(m)$ and $\mathfrak{sp}(n)$. The (complexification of the real) representation $(\mathbf{2m},\mathbf{2n})$ has weights $\pm e_i\pm f_a$ (there are, in fact, $4mn$ such weights). After pairing opposite weights, we can choose the representatives 
\begin{eq}
    e_{i}+f_a \qquad \text{and} \qquad e_i-f_a\,,
\end{eq}
for any pair $(i,a)$. Since we have a real orthogonal representation, if its weights are $\pm w_k$, the weights of its spinor module are $\lambda=\frac{1}{2}\sum_k \epsilon_k w_k$, where $\epsilon_k$ are signs. Hence, for the Clifford module of our $4mn$ modes,
\begin{eq}
    \lambda=\frac{1}{2}\sum_{i=1}^m \sum_{a=1}^n \left[\epsilon_{ia}(e_i+f_a)+\epsilon'_{ia}(e_i-f_a)\right]\,,
\end{eq}
where $\epsilon_{ia}$ and $\epsilon'_{ia}$ are signs. This can be written as
\begin{eq}
    \lambda=\sum_{i=1}^m \alpha_i e_i+\sum_{a=1}^n \beta_a f_a\,,\qquad \text{with}\qquad \alpha_i=\frac{1}{2}\sum_{a=1}^n\left(\epsilon_{ia}+\epsilon'_{ia}\right)\,,\quad \beta_{a}=\frac{1}{2}\sum_{i=1}^m \left(\epsilon_{ia}-\epsilon'_{ia}\right)\,.
\end{eq}

To prove our statement, note that for $\mathfrak{sp}(m)$, if one writes a weight as $\sum_i \alpha_i e_i$, the center charge is 
\begin{eq}
    [\lambda]_{\mathfrak{sp}(m)}\equiv\sum_{i=1}^m \alpha_i\pmod{2}\,.
\end{eq}
Therefore, using
\begin{eq}
    \sum_{i=1}^m\alpha_i+\sum_{a=1}^n \beta_a=\frac{1}{2}\sum_{i,a}\left(\epsilon_{ia}+\epsilon'_{ia}+\epsilon_{ia}-\epsilon'_{ia}\right)=\sum_{i,a}\epsilon_{ia}\equiv \sum_{i,a}1\equiv mn\pmod{2}\,,
\end{eq}
we obtain \cref{eq:appendix_weights}.

If there are $c$ copies of the $(\mathbf{2m},\mathbf{2n})$ fermion zero modes, the statement becomes 
\begin{eq}
    [\lambda]_{\mathfrak{sp}(m)}+[\lambda]_{\mathfrak{sp}(n)}\equiv cmn\pmod{2}\,.
\end{eq}

\bibliographystyle{utphys}
\bibliography{bibliography}

\end{document}